\documentclass[]{tADP2e}
\begin{document}
\doi{10.1080/0001873YYxxxxxxxx}
 \issn{1460-6976}
\issnp{0001-8732}  \jvol{00} \jnum{00} \jyear{2008} \jmonth{June}

\markboth{E. B. Sonin}{Advances in Physics}

\articletype{REVIEW}

\title{Spin currents and spin superfluidity}

\author{E. B. Sonin$^{\ast}$\thanks{$^\ast$Email: sonin@cc.huji.ac.il
\vspace{6pt}} \\\vspace{6pt}  {\em{Racah Institute of Physics, Hebrew University of Jerusalem,
Jerusalem 91904, Israel}}\\\vspace{6pt}\received{received}
}

\maketitle

\begin{abstract}
The present review analyzes and compares various types of dissipationless spin transport: (1) Superfluid transport, when the spin-current state is a metastable state (a local but not the absolute minimum in the parameter space). (2) Ballistic spin transport, when spin is  transported without losses simply because sources of dissipation are very weak. (3) Equilibrium spin currents, i.e., genuine persistent currents. (4) Spin currents in the spin Hall effect. Since superfluidity is frequently connected with Bose condensation, recent debates about magnon Bose condensation are also reviewed.
For any type of spin currents   simplest models were chosen for discussion in order to concentrate on concepts rather than details of numerous models. The various hurdles on the way of using the concept of spin current (absence of the spin-conservation law, ambiguity of spin current definition,  {\em etc.}) were analyzed. The final conclusion is that the spin-current concept can be developed in a fully consistent manner, and is a useful language for description of various phenomena in spin dynamics.
\bigskip

\begin{keywords}spin current; spin superfluidity; easy-plane (anti)ferromagnet; Landau criterion; spin-orbit coupling; spin Hall effect 
\end{keywords}\bigskip

\end{abstract}

\section{Introduction}
\label{intro}

The problem of spin transport occupies minds of condensed matter physicists for decades. A simple example  of spin transport is spin diffusion, which is a process accompanied with dissipation. Conceptually more complicated is ``dissipationless'' spin transport, which was also discussed  long time but was in the past and remains now to be a matter of controversy. The main source of controversy is that spin is not a conserved quantity.  This leads to many complications and ambiguities in defining such concepts as spin {\em flow, current,  or transport}.
 Sometimes   these complications are purely semantic. However, this does not make them simpler for discussion.  ``Semantic traps'' very often are a serious obstacle for understanding physics  and for deriving    proper conclusions concerning observation and practical application of the phenomenon. The best strategy in these cases is to focus not on names but on concepts hidden under these names. Only after this one may ``take sides'' in semantic disputes not forgetting, however, that choosing names is to considerable extent a matter of convention and taste.

During long history of studying the problem of ``dissipationless'' spin transport one can notice three periods, when studies in this field were especially intensive. The first period started from theoretical suggestions on possible  ``superfluidity of electron-hole pairs''  \cite{KozMax}, which were later  extended on possible spin superfluidity \cite{ES-78a,ES-78b}.  At the same period the concept of spin superfluidity was exploited \cite{Vuo,Vuo2,ES-79} for interpretation of experiments demonstrating unusually fast spin relaxation in $^3$He-A \cite{CorOsh}. The second period was marked by intensive theoretical and experimental work on spin superfluidity in $^3$He-B starting from interpretation of experiments on the so-called Homogeneously Precessing Domain (HPD) \cite{HeBex} in terms of spin supercurrents \cite{Fom-84}. Finally in these days (the third period) we observe a growing interest to dissipationless spin currents in connection with work on spintronics \cite{sptr}. The final goal of spintronics is to create devices based on spin manipulation, and transport of spin with minimal losses is crucial for this goal. Now one can find reviews summarizing the investigations done during the first \cite{ES-82} and the second \cite{Fom-91,Bun} periods of works on dissipationless spin transport. On the other hand, the work on spin transport in spintronics  is a developing story, and probably it is still premature to write summarizing reviews. Nevertheless, 
some reviews mostly addressing the spin Hall effect have already appeared \cite{ERH,Dy08}.  It looks also useful to have a glance on the current status of the field from a broader viewpoint and to find bridges between current investigations and those done in the ``last millennium''. The present review aims at this goal. The intention is to discuss mostly concepts without unnecessary deepening in details, and simplest models were chosen for this.
 
The term ``superfluidity'' is used in the literature to cover a broad range of phenomena, which have been observed in superfluid $^4$He and $^3$He, Bose-Einstein condensates of cold atoms, and, in the broader sense of this term, in superconductors.  In the present review superfluidity  means only a possibility to transport a physical quantity (mass, charge, spin, ...) without dissipation. Exactly this phenomenon  gave a rise to the terms ``superconductivity'' and ``superfluidity'', discovered  nearly 100 years  and 70 years ago respectively. It is worthwhile to stress that one should not understand  the adjective ``dissipationless'' too literally. In reality we deal with an essential suppression of dissipation due to the presence of energetic barriers of the topological origin. How essential suppression could be, is a matter of a special analysis.  In the present review we restrict discussion with the question whether activation barriers, which suppresses dissipation, can appear. 

But superfluidity is not the only reason for suppression of dissipation in the transport process, and it is important to understand the difference between various types of dissipationless transport. In the present review we shall discuss four types of them:
\begin{itemize}
\item Superfluid transport: The spin-current state is a metastable state (a local but not the absolute minimum in the parameter space).
\item Ballistic transport. Here spin is  transported without losses simply because sources of dissipation are very weak.
\item Equilibrium currents. Sometimes symmetry allows currents even at the equilibrium. A superconductor in a magnetic field is a simple example. Equilibrium spin currents are also possible, though there is a dispute on whether they have something to do with spin transport.  Equilibrium spin currents are genuine persistent currents, since no dissipation is possible at the equilibrium by definition.
\item Spin currents in the spin Hall effect. These currents are also called dissipationless since they are normal to the driving force (electric field) and therefore do not produce any work. However,  in  the spin Hall effect there is dissipation connected with a longitudinal charge current through a conducting medium. On the other hand, it was recently revealed that the spin Hall effect  is possible also in insulators where a charge current is absent. Then  spin currents are not accompanied by any dissipation becoming similar to equilibrium spin currents.  
\end{itemize}
The second type (ballistic) looks mostly trivial: dissipation is absent because sources of dissipation are absent. Still it is worth of short discussion since sometimes they confuse ballistic transport with superfluid transport (an example of it is discussed in section~\ref{SPV}). The superfluid transport does not  require the absence of dissipation mechanisms. One may expect that in an ideally clean metal at zero temperature resistance would be absent. But this would not be superconductivity. Superconductivity is the absence of resistance in a dirty metal at $T>0$. 

The first two types of spin currents are discussed in  Part I of the review, which is devoted to magnetically ordered systems. The third and the fourth  types are discussed mostly in  Part II, which addresses time-reversal-invariant systems without magnetic order, though in magnetically ordered media equilibrium spin currents are also possible (section \ref{heli}). 
Since from the very beginning of the  theory of superfluidity the relation between superfluidity and Bose condensation was permanently in the focus of attention, discussing spin superfluidity one cannot avoid to consider the concept of magnon Bose condensation, which is vividly debated nowadays. Section~\ref{MBEC} addresses this issue.

\vspace{1cm}
\centerline{\bf \Large Part I: Spin currents in magnetically ordered systems}

\section{Mass supercurrents} \label{mass}

Since the idea of spin superfluidity originated from the analogy with the more common concept of mass superfluidity let us shortly summarize the latter.  The essence of the transition to the superfluid or superconducting state is that below the critical temperature the complex order parameter $\psi =|\psi| e^{i\varphi}$, which has a meaning of the wave function of the bosons or the fermion Cooper pairs, emerges as an additional macroscopical variable of the liquid.   For the sake of simplicity, we restrict ourselves to the case of a neutral superfluid at zero temperature putting aside the two-fluid theory for finite temperatures. Then the theory of superfluidity tells that the order parameter $\psi$ determines the particle density $n=|\psi|^2$   and the velocity of the liquid is given by the standard quantum-mechanical expression
\begin{equation}
\bm v = -i {\hbar \over 2 m |\psi|^2}(\psi^* \bm \nabla \psi-\psi \bm \nabla \psi^*)= {\hbar \over m} \bm \nabla \varphi.
         \end{equation}
Thus the velocity is a gradient of a scalar, and any flow is potential. Since the phase and the particle number are a pair of canonically conjugate variables, one can write down the Hamilton equations for the pair of the canonically conjugated variables ``phase -- density'':
 \begin{eqnarray}
\hbar {d\varphi \over dt}=-{\delta {\cal E}\over \delta n},~~
{dn \over dt}= {\delta  {\cal E}\over \hbar \delta \varphi}.
     \label{IdHam} \end{eqnarray}
Here ${\cal E}=\int d^3\bm R E$ is the total liquid energy, whereas $E$ is the energy density, and  $\delta {\cal E}/ \delta n$ and $\delta {\cal E} / \delta \varphi$ are functional derivatives of the total energy:
\begin{equation}
{\delta {\cal E}\over \delta n}={\partial  E\over \partial n} -\bm \nabla \cdot {\partial  E\over \partial \bm \nabla n} \approx {\partial  E\over \partial n} =  \mu,  
         \end{equation}
\begin{equation}
{ \delta {\cal E}\over \delta \varphi}={\partial  E\over \partial \varphi} -\bm \nabla \cdot {\partial  E\over \partial \bm \nabla \varphi} 
= -\bm \nabla \cdot {\partial  E\over \partial \bm \nabla \varphi} =-\hbar \bm \nabla \cdot\bm g.
         \end{equation}
In these expressions $\mu$ is the chemical potential,
\begin{equation}
 \bm g=n\bm v ={\partial  E\over \hbar \partial \bm \nabla \varphi} 
 \end{equation} 
 is the particle current, and the dependence of the energy on the density gradient was ignored. Eventually the Hamilton equations are reduced to  the equations of hydrodynamics for an ideal liquid:
  \begin{eqnarray}
m{d\bm v \over dt}= -\bm \nabla \mu ,
\label{Eul}           \end{eqnarray}
 \begin{eqnarray}
{dn \over dt}=-\bm \nabla \cdot \bm g.
     \label{IdLiq} \end{eqnarray}

A crucial property of the system is the gauge invariance: the energy does not depend on the phase directly ($\partial E / \partial \varphi=0$) but only on its gradient. According to Noether's theorem this must lead to the conservation law for a conjugate variable, the total number of particles. The conservation law manifests itself in the continuity equation (\ref{IdLiq}), which contains the particle supercurrent.   The prefix ``super'' stresses that this current is not connected with dissipation. It is derived from the Hamiltonian or the Lagrangian but not from the dissipation function. In contrast to the diffusion current proportional to the density gradient, the supercurrent is proportional to the phase gradient. Therefore it appears only in a coherent state with broken gauge invariance.
The equations of superfluid hydrodynamics can be derived from the Gross--Pitaevskii equation for a weakly non-ideal Bose-gas. However, they are much more general than this model. They can be formulated from the most general principles of symmetry and conservation laws. Indeed, deriving the two-fluid theory of superfluidity Landau did not use the concept of Bose-condensation. 

\begin{figure}
\begin{center}
\begin{minipage}{100mm}
{\resizebox*{10cm}{!}
{\includegraphics{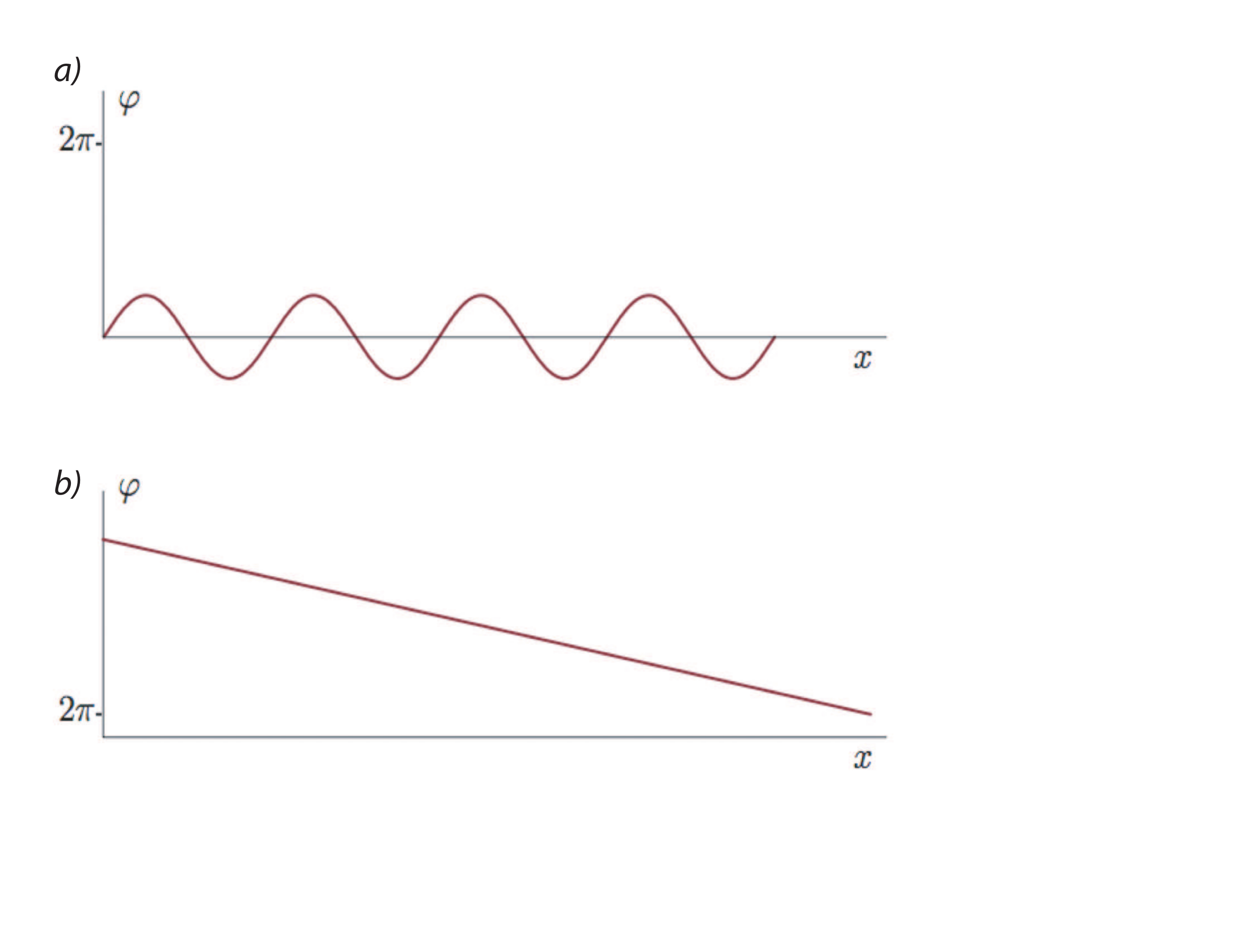}}}%
\caption{Phase (inplane rotation angle)  variation at the presence of mass (spin) supercurrents. a) Oscillations in a sound (spin) wave). b) Stationary mass (spin) supercurrent.}%
\label{fig1}
\end{minipage}
\end{center}
\end{figure}

An elementary collective mode of the ideal liquid is a sound wave. In a sound wave the phase varies in space, i.e., the wave is accompanied by mass supercurrents (figure~\ref{fig1}a). An amplitude of the time and space dependent phase variation is small, and currents transport mass on distances of the order of the wavelength. A really superfluid transport on macroscopic distances is related with stationary solutions of the hydrodynamic equations corresponding to finite constant currents with constant nonzero phase gradients (current states). In the current state the phase rotates through a large number of full 2$\pi$-rotations along streamlines of the current (figure~\ref{fig1}b).

The crucial point of the superfluidity concept  is why the supercurrent is a persistent current, which does not decay despite it is not the ground state of the system and has a larger energy. The first explanation of the supercurrent stability was given on the basis of the well known Landau criterion \cite{LC}. According to this criterion, the current state is stable as far as \emph{any quasiparticle} of the Bose-liquid in the laboratory frame has a positive energy and therefore its creation requires an energy input. Let us suppose that elementary quasiparticles of  the Bose-liquid at rest have an energy spectrum $\varepsilon(\bm p)$ where $\bm p$ is the quasiparticle momentum. If the Bose-liquid moves with the velocity $\bm v$ the quasiparticle energy in the laboratory frame is $\varepsilon(\bm p)+\bm p\cdot \bm v$. The energy cannot be negative (which would mean instability) if 
\begin{equation}
v < v_L =\mbox{min}{\varepsilon(\bm p)\over p}.
  \label{LC}       \end{equation}
In superfluid $^4$He the Landau critical velocity $v_L$ is determined by the roton part of the spectrum. But in this review we focus on the long-wavelength collective excitations, which are phonons with the spectrum $\varepsilon=u_s p$. Then according to  equation~(\ref{LC}) the supercurrent cannot be stable if the velocity $v$ exceeds the sound velocity $u_s$. 

As far as one wants to check  the Landau criterion for long-wavelength collective modes like sound  waves, it is not necessary  to solve a dynamical problem looking for the spectrum of phonons. It is enough to estimate the energy of possible static fluctuations in the stationary current state with particle density $n_0$ and velocity $\bm v_0$. Let us write down the energy of the current state taking into account possible local fluctuations of the particle density,  $n'=n-n_0$, and of the velocity, $\bm v'=\bm v-\bm v_0$, up to the terms of the second order: 
 \[
{\cal E}=\int d^3\bm R\left[\mu_0(n_0) n'+{\partial  \mu_0(n_0) \over \partial n}{n'^2\over 2}+ {m(n_0+n')(\bm v_0 + \bm v\,')^2\over 2} \right].
\]
Here $\mu_0 =\partial E_0 /\partial n$ and $E_0$ are the chemical potential and the energy density of the liquid at rest. One may neglect terms of the first order with respect to the density fluctuation $n'$ and the velocity $\bm v\,'$ since we look for an energy extremum at fixed averaged density and velocity and the first-order term must vanish after integration. Using  the thermodynamic relation $mu_s^2/n=\partial \mu_0 /\partial n=\partial^2 E_0 /\partial n^2$ and omitting the subscript 0 in $n_0$ and $\bm v_0$, one obtains
 \begin{eqnarray}
{\cal E}\approx \int d^3\bm R\left[{m u_s^2n'^2\over 2n} +mn'\bm v\cdot \bm v'+ {mn \bm v'^2\over 2} \right].
   \end{eqnarray}
The quadratic form under the integral is positive definite, i.e., the current state corresponds to the energy minimum, if the condition $u_s^2>v^2$ is satisfied. This condition is identical to the Landau criterion equation~(\ref{LC}) for the phonon spectrum $\varepsilon=u_s p$. 

The theory of superfluidity tells that the Landau criterion is a necessary but not sufficient condition for current metastability. The Landau criterion checks only small deviations from the current state. Meanwhile the current state can be destroyed via \emph{large} perturbations of the current state. In superfluids these large perturbations are vortices.   In the current state the phase rotates along the current direction. The current can relax if one can remove one  2$\pi$-turn of the phase. This requires that a singular vortex line  crossed or ``cut'' the channel  cross-section. The process is called ``phase slip''.

\begin{figure}
\begin{center}
\begin{minipage}{100mm}
{\resizebox*{7cm}{!}{\includegraphics{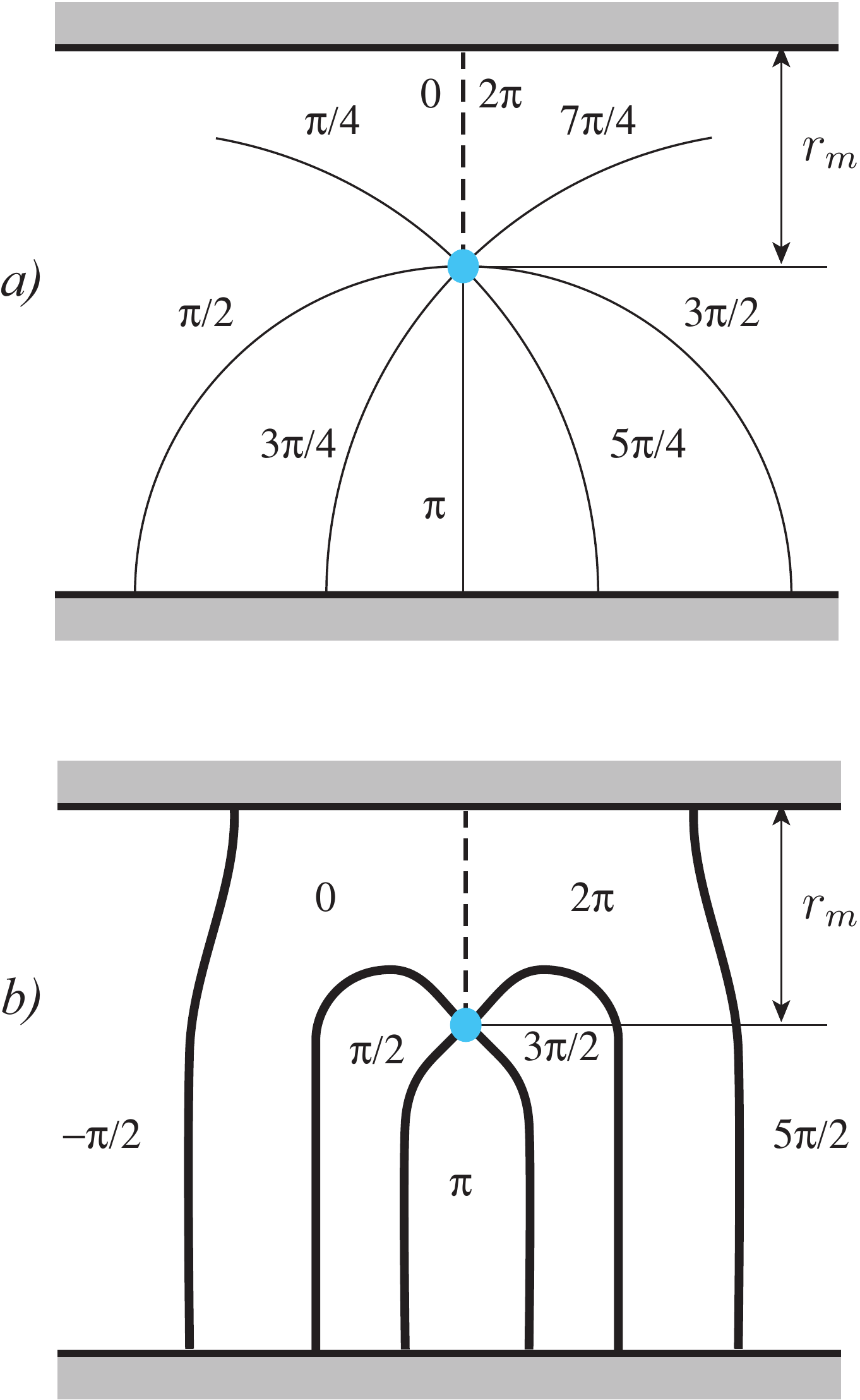}}}%
\caption{Mass and spin vortices. a) Mass vortex or spin vortex in an easy-plane ferromagnet without inplane anisotropy. b) Spin vortex at small spin currents ($\langle \nabla \varphi \rangle \ll 1/l$) for four-fold inplane symmetry. The vortex line is a confluence of four $90^\circ$ domain walls (solid lines).}%
\label{fig3}
\end{minipage}
\end{center}
\end{figure}

If the vortex axis (vortex line) coincides with the $z$ axis, the phase gradient around the vortex line  is given by 
\begin{eqnarray}
\bm \nabla \varphi_v= \frac{[\hat z \times \bm r]}{r^2},
   \label{vort}         \end{eqnarray}
where $\bm r$ is the position vector in the $xy$ plane. The phase changes by 2$\pi$ around the vortex line  (figure~\ref{fig3}a).
Creation of the vortex requires some energy. The vortex energy per unit length (line tension) is determined by the kinetic (gradient) energy: 
\begin{eqnarray}
\epsilon =\int d^2\bm r {\hbar^2 n (\bm \nabla \varphi_v)^2\over 2m} = {\pi  \hbar^2 n \over m} \ln{r_m\over r_c},
    \label{vorEnM}
          \end{eqnarray}
where the upper cut-off $r_m$ is determined by geometry. For example, for the vortex shown in figure~\ref{fig3}a it is  the distance  of the vortex line from a sample border.
The lower cut-off $r_c$ is the vortex-core radius. It determines the distance $r$ at which the phase gradient is so high that the hydrodynamic expression for the energy becomes invalid. A good estimation for  $r_c$ is $r_c \sim \kappa/u_s$, where $\kappa=h/m$ is the circulation quantum of the velocity. Inside the core the modulus of the order parameter goes down to zero  eliminating the singularity in the kinetic energy at the vortex axis. For the weakly non-ideal Bose-gas this estimation yields the coherence length.

Now suppose that a vortex appears in the current state with the constant gradient $\bm \nabla \varphi_0$: The phase gradients induced by the vortex are superimposed on the constant phase gradient related to the current: $\bm \nabla \varphi=\bm \nabla \varphi_0+\bm \nabla \varphi_v$. The total gradient energy includes that of the current, the vortex energy given by equation~(\ref{vorEnM}), and the energy from the cross terms of the two gradient fields. Only the last two contributions are connected with the vortex, and their sum determines the energy of the vortex in the current state:
\begin{eqnarray}
\tilde \epsilon = {\pi  \hbar^2 n \over m} L \ln{r_m\over r_c}-  {2\pi  \hbar^2 n \over m} S \nabla \varphi_0, 
   \label{vorCurEnM}
          \end{eqnarray}
where $L$ is the length of the vortex line and $S$ is the area of the cut, at which the phase jumps by $2\pi$. For the 2D case shown in figure \ref{fig3}a (a straight vortex in a slab of thickness $L$ normal to the picture plane) $S=Lr_m$. One can see that vortex motion across the channel (growth of $r_m$) is impeded by the barrier, which is determined by variation of the energy $\tilde \epsilon$ with respect to $r_m$. The peak of the barrier corresponds to $r_m =1/ 2 \nabla \varphi_0$. The height of the barrier is 
\begin{eqnarray}
 \epsilon_m \approx {\pi  \hbar^2 n \over m} L \ln{1\over  r_c \nabla \varphi_0}.
   \label{barrM} 
           \end{eqnarray}
Thus the barrier disappears at gradients $\nabla \varphi_0 \sim 1/r_c$, which are of the same order as the critical gradient determined from the  Landau criterion. In the 3D geometry  the phase slip is realized  with expansion of vortex rings. For the ring of radius $R$ the vortex-length and the area of the cut are $L=2\pi R$ and $S=\pi R^2$ respectively, and the barrier disappears at the same critical gradient $\sim 1/r_c$ as in the 2D case.

\begin{figure}
\begin{center}
\begin{minipage}{100mm}
{\resizebox*{9cm}{!}{\includegraphics{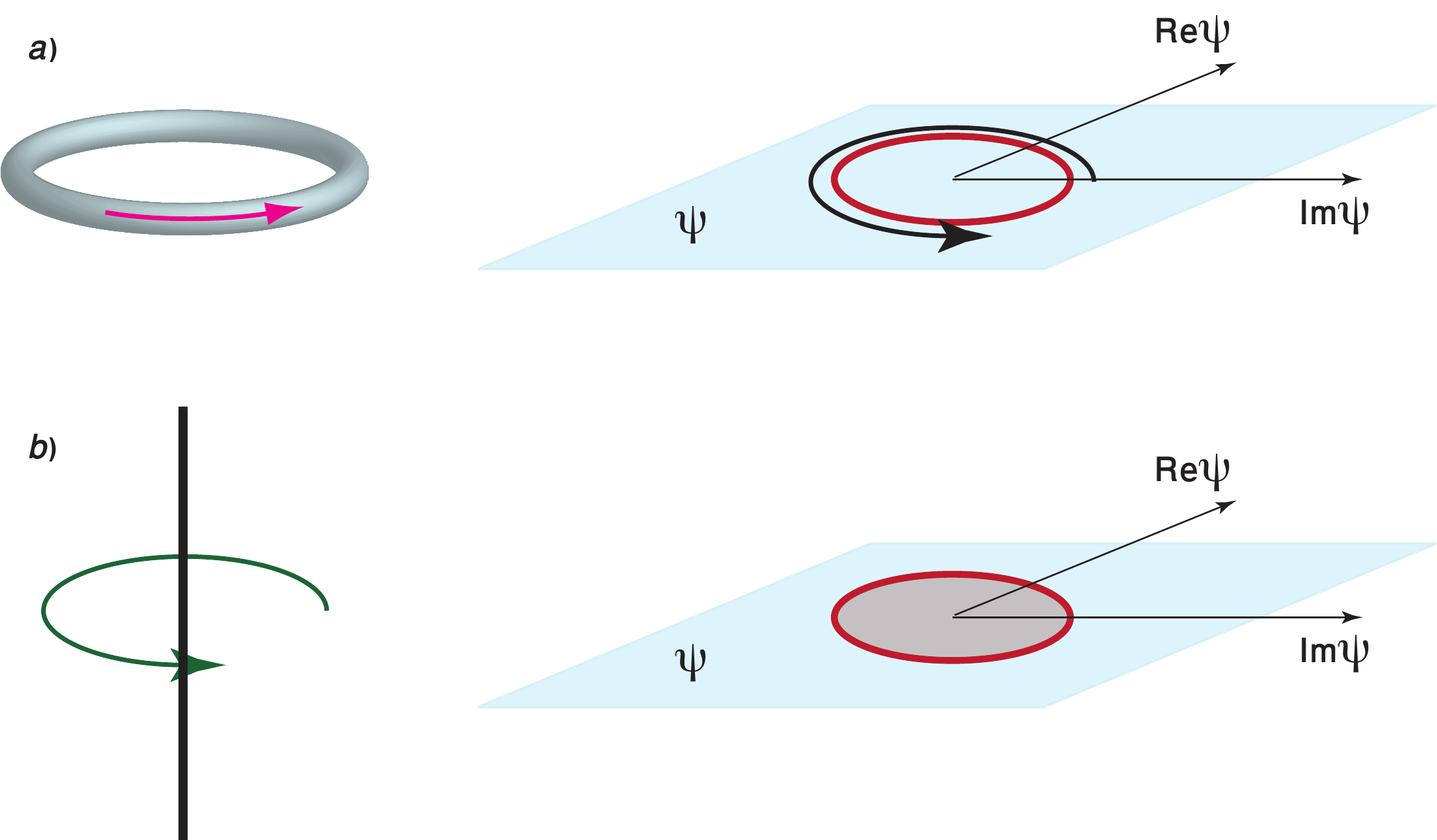}}}%
\caption{Topology of the uniform mass current and the vortex states. a) The current state in a torus maps onto the circumference $|\psi|=|\psi_0|=const$ in the complex $\psi$ - plane, where $\psi_0$ is the equilibrium order parameter wave function of the uniform state. b) The vortex state maps onto the circle $|\psi|\leq |\psi_0| $. }%
\label{fig2a}
\end{minipage}
\end{center}
\end{figure}

The barriers stabilizing metastable current states are connected with topology of the order parameter space. In a superfluid the order parameter  is a complex wave function $\psi(\bm r)$. At the equilibrium $\psi = |\psi_0|e^{i\varphi}$, where the modulus $|\psi_0|$ is a constant determined by minimization of the energy and the
 phase $\varphi$ is a degeneration parameter since the energy does not depend on  $\varphi$. Any current state
 in a closed annular channel (torus) with  the phase change $2\pi n$ around the channel maps onto a circumference $|\psi|=|\psi_0|$ in the complex plane (figure~\ref{fig2a}a) 
 winding the  circumference $n$ times. It is evident that it is impossible to change $n$ keeping the path on the circumference $|\psi|=|\psi_0|$ all the time. Thus $n$ is a {\em topological charge}. One can change it (removing, e.g., one winding around the circumference) only by leaving the circumference (the equilibrium order parameter space in the case). This should cost energy, which is spent on creation of a vortex. Figure~\ref{fig2a}b shows mapping of the vortex state onto the circle  $|\psi|\leq |\psi_0|$. 

Without such topological barriers superfluidity is ruled out.
However, barriers do not automatically provide the life-time of currents  long enough. In practice,
dissipation via phase slips is possible even in the presence of barriers due to thermal fluctuations or quantum tunneling. Here and later on we  address only ``ideal'' critical currents (the upper bound for critical currents) at which barriers disappear leaving ``practical'' critical currents beyond the scope of the present review.  

\section{Phenomenology of magnetically ordered systems and spin currents}
\label{magn}

The main interaction responsible for magnetic order is exchange interaction, which is invariant with respect to rotations of the whole spin system.  Then according to Noether's theorem the total spin must be conserved.  For ferromagnets where the order parameter is the spontaneous magnetization $\bm M$, this means that the exchange energy can depend on the absolute value of $\bm M$ but not on its direction.  Other contributions to the free energy (anisotropy energy or dipole-dipole interaction) are related with spin-orbit interaction, which does not conserve the total spin. But these interactions are relativistically small, i.e., governed by the small relativistic parameter $v/c$, where $v$ is a typical electron velocity and $c$ is the speed of light. The spin-orbit interaction does depend on $\bm M$ direction, but because of its weakness cannot affect the absolute value $M$ in slow dynamics.  This is a crucial point in the phenomenological theory of magnetism of Landau and Lifshitz \cite{LL}, which determines the form of the  equation of motion for ferromagnet magnetization known as the Landau-Lifshitz equation \cite{LP}: 
\begin{eqnarray}
{\partial \bm M\over \partial t}=\gamma \left[\bm H _{eff} \times  \bm M\right],
     \label{LLP}      \end{eqnarray}
where $\gamma$ is the gyromagnetic ratio between the magnetic and mechanical moment  ($\bm M=-\gamma \bm S$). The effective magnetic field is determined by the functional derivative of the total free energy  ${\cal F}=\int d^3\bm R \,F$ with density $F$:
\begin{eqnarray}
\bm H _{eff} =- {\delta {\cal F }\over \delta \bm M}.
           \end{eqnarray}
According to the   Landau-Lifshitz equation, the absolute value $M$ of the magnetization cannot vary.  The evolution of  $\bm M$  is a precession around the effective magnetic field $\bm H _{eff}$ .

At first let us discuss \emph{exchange approximation}, in which relativistic effects are ignored and the conservation of total spin is not violated.
In this approximation the free energy density is 
\begin{eqnarray}
F(\bm M)=F_0(M)+ {\alpha \over 2} \nabla_i \bm M \cdot \nabla_i \bm M.
       \label{exch}    \end{eqnarray}
The  first exchange-energy term $F_0(M)$, being the largest term, is crucial for determination of the equilibrium value of $M$.  But after determination of $M$ it
can be ignored as an inessential large constant. Indeed, its contribution to the effective field in Landau-Lifshitz equation (\ref{LLP}) does not produce
any effect: the contribution is parallel to $\bm M$ and vanishes in the vector product. 
In the absence of external fields, which  break invariance with respect to rotations in the spin space, 
the  Landau-Lifshitz equation reduces to the continuity equations for components of the spin density $\bm S=-\bm M/\gamma$:
\begin{eqnarray}
{\partial  S_i \over \partial t}=-{1\over \gamma}{\partial  M_i \over \partial t}=- \nabla_j J_j^i,
           \end{eqnarray}
where 
\begin{eqnarray}
 J_j^i=-\left[\bm M \times {\partial  F \over \partial \nabla_j \bm M}\right]_i= - \alpha [\bm M \times \nabla_j \bm M]_i= -\alpha\varepsilon_{ikl}  M_k  \nabla_j M_l
     \label{spCur}      \end{eqnarray}
is the $j$th component of the spin current transporting the $i$th component of spin. Thus in an isotropic ferromagnet all three components of spin are conserved.

The Landau-Lifshitz equation has plane-wave solutions describing spatially nonuniform precession of the magnetization $\bm M =\bm M_0 +\bm m$ around the ground-state magnetization $\bm M_0$:  $\bm m \propto e^{i \bm k \bm r -i\omega t}$. Here the magnetization deviation $\bm m$ is small and normal to $\bm M_0$. Linearizing with respect to $\bm m$, one obtains spin waves with the spectrum
\begin{eqnarray}
\omega = \gamma \alpha M_0 k^2.
     \label{IFsp}      \end{eqnarray}
In an  isotropic ferromagnet spin waves at $k\neq 0$  are accompanied by spin currents, but  superfluid spin transport is impossible as will be clear from section \ref{stab}.

Next we shall consider the case  when spin-rotational invariance is partially broken, and  there is uniaxial crystal magnetic anisotropy given by the third term in the phenomenological free energy:
\begin{eqnarray}
F=F_0(M)+ {\alpha\over 2} \nabla_i \bm M \cdot \nabla_i \bm M+E_A{M_z^2\over 2M^2}.
           \end{eqnarray}
If the anisotropy energy $E_A$ is positive, it is the ``easy plane'' anisotropy, which keeps the spontaneous magnetization $\bm M_0$  in the $xy$ plane (the continuous limit of the $XY$ model). In this model the $z$ component of spin is conserved, because invariance with respect to rotations in the easy plane remains unbroken. Since the absolute value of magnetization is fixed, the vector $\bm M$ of the magnetization is fully determined by the angle $\varphi $ showing the direction of $\bm M$ in the easy plane $xy$ ($M_x=M\cos \varphi$, $M_y=M\sin\varphi$) and by the $z$ component of the magnetization $m_z$. We use the notation $m_z$ instead of $M_z$ in order to  emphasize that $m_z$ is a small dynamic correction to the magnetization, which is absent at the equilibrium. In the new variables the free energy is
\begin{eqnarray}
{\cal F}=\int d^3\bm R \,F=\int d^3\bm R\left[{ m_z^2\over 2\chi} + {A(\bm \nabla \varphi)^2\over 2} \right].
  \label{Ener}    \end{eqnarray}
The constant  $A=\alpha M^2$ is stiffness of the spin system determined by exchange interaction, and
the magnetic susceptibility $\chi= M^2/E_A$ along the $z$ axis is determined by the uniaxial anisotropy energy $E_A$ keeping the magnetization in the plane.  The Landau-Lifshitz equation reduces to the Hamilton equations for a pair of canonically conjugate continuous variables ``angle--angular momentum'' (analogous to the canonically conjugate pair ``coordinate--momentum''):
   \begin{eqnarray}
{d\varphi \over dt}=-\gamma {\delta {\cal F} \over \delta m_z}=-\gamma {\partial F \over \partial m_z},
     \label{HEp} \end{eqnarray}
     \begin{eqnarray}
{1\over \gamma}{dm_z \over dt}={\delta {\cal F}\over \delta \varphi}={\partial F \over \partial \varphi}-\bm \nabla \cdot{\partial F \over \partial \bm \nabla \varphi},
 \label{HEm}      \end{eqnarray}
where functional derivatives on the right-hand sides are taken from the free energy $F$ given by equation~(\ref{Ener}). Using the expressions for functional derivatives one can write the Hamilton equations as
         \begin{eqnarray}
{d\varphi \over dt}=-\gamma { m_z\over \chi},
     \label{Ep} \end{eqnarray}
     \begin{eqnarray}
-{1\over \gamma} {dm_z \over dt}+ \bm \nabla \cdot \bm J^z=0,
 \label{Em}      \end{eqnarray}
where 
\begin{eqnarray}
\bm J^z=-{\partial F \over \partial \bm \nabla \varphi} =-A   \bm \nabla \varphi
   \label{cur}    \end{eqnarray}
is the spin current. 

There is an evident analogy of equations (\ref{Ep}) and (\ref{Em}) with the hydrodynamic equations
(\ref{Eul})    and (\ref{IdLiq})  for an ideal liquid, equation (\ref{Em})  being the continuity equation for spin. This analogy was exploited by Halperin and Hohenberg \cite{HH} in their hydrodynamic theory of spin waves. In contrast to the isotropic ferromagnet with the quadratic spin-wave spectrum, the spin wave in the easy-plane ferromagnet has a sound-like spectrum as in a superfluid: $\omega=c_s k$, where the spin-wave velocity is $c_s=\gamma \sqrt{A/\chi}$. Halperin and Hohenberg introduced the concept of 
 spin current, which appears in a propagating spin wave like a mass supercurrent appears in a sound wave (figure~\ref{fig1}a). This current transports the $z$ component of spin on distances of the order of the wavelength.
 But as well as the mass supercurrent in a sound wave, this small oscillating spin current does not lead to superfluid spin transport, which this review addresses. Spin superfluid transport on long distances is realized in current states with magnetization rotating in the plane through a large number of full 2$\pi$-rotations as shown in figure~\ref{fig1}b. 
 
Let us consider now the case of antiferromagnetic order. The simplest model of an antiferromagnet is two sublattices with magnetizations $\bm M_1$ and $\bm M _2$.   In the absence of weak ferromagnetism and external magnetic fields two magnetizations $\bm M_1=-\bm M _2$ completely compensate each other without producing any total magnetization $\bm m=\bm M_1+\bm M_2$ .  However, a small magnetization $\bm m$ does appear due to external magnetic fields or dynamical effects. The amplitudes of $\bm M_1$ and $\bm M_2$ and their mutual orientation are mostly determined by strong exchange interaction, but the latter does not fix the direction of the staggered magnetization $\bm L=\bm M_1-\bm M_2$, which is the order parameter of a two-sublattice antiferromagnet.The equations of motion for two vectors $\bm L$ and $\bm m$ can be derived from the two Landau-Lifshitz equations for $\bm M_1$ and $\bm M_2$ taking into account the exchange interaction between two sublattices. But it would be useful to present a more general version of the macroscopic phenomenological theory, which is able to describe an antiferromagnetic structure of any complexity \cite{AM,HS}.   The theory of spin dynamics in superfluid phases of $^3$He developed by Leggett and Takagi \cite{LT} also belongs to this class. Following the same principle ``exchange is the strongest interaction'' as 
in the Landau-Lifshitz theory,  macroscopic theories of this type deal with phenomena at scales essentially exceeding microscopic scales (the coherence length in the case of $^3$He), at which the exchange energy establishes the tensor structure of the order parameter. This permits to assume that the entire dynamic evolution of the order parameter reduces to rotations in the 3D spin state, which cannot change the exchange energy. Then  the dynamics of the system is described by three independent pairs of canonically conjugated variables ``angle--moment''  $\varphi_i $--$m_i/\gamma$ ($i=1,2,3$): 
\begin{eqnarray}
{\partial \varphi_i\over \partial t}=-\gamma {\delta {\cal F}\over \delta m_i},
\nonumber \\
{1\over \gamma }{\partial m_i\over \partial t}={\delta {\cal F}\over \delta \varphi_i}.
     \label{AM}    \end{eqnarray}   
Here $\varphi_i$ are the angles of spin rotations around three Cartesian axes ($i=x,y,z$). Apart from spatial dependence of the variables, these equations are similar to the equations of motion of a 3D rigid top.  In our case the top is an antiferromagnetic  spin order parameter rigidly fixed by exchange interaction. As in the case of a two-sublattice antiferromagnet,  magnetization $\bm m$ results from  deformation of the equilibrium spin structure. The approach is valid as far as this deformation is weak, i.e., $\bm m$ is smaller than the characteristic moments of the antiferromagnetic structure (staggered magnetization $\bm L$ in the case of a two-sublattice antiferromagnet).  Since rotation around the vector $\bm L$ has no effect on the state of the system the latter has only two  degrees of freedom corresponding to two pairs ``angle--moment''. Then the equations become the equations of motion of a rotator. In contrast to the spontaneous magnetization $\bm M$ in the Landau-Lifshitz equation (\ref{LLP}), the small absolute value of the magnetization $\bm m$ is not kept constant. 

Because the group of 3D rotations is non-commutative,  the state of the system depends on the order, in which rotations around different axes are performed. In practice they frequently use the Euler angles (they are introduced in section \ref{HeB}).    
For the most content of Part I (except for section~\ref{HeB}), one can choose one degree of freedom connected with the conjugate pair $\varphi_z$--$m_z$, and the problem of non-commutativity is absent (further we shall omit  the subscript $z$  of the angle $\varphi_z$).  If the energy of the ground state does not depend (or depends weakly as discussed in section \ref{Phase}) on the angle $\varphi$, the equations of motions for $\varphi$ and $m_z$ are  the same Hamilton equations (\ref{HEp})  and(\ref{HEm}), which were formulated for an easy--plane ferromagnet. In the case of a two-sublattice antiferromagnet the angle $\varphi$ is the angle of the staggered magnetization $\bm L$ in the easy plane.

The discussion of this section has not made any reference to a concrete microscopic model of magnetism. Indeed, the approach is general enough and is valid for models of magnetism based  on the concepts of either localized or itinerant electrons.  In particular, ferromagnetism  of localized electrons is described by   the Heisenberg  model  with the Hamiltonian:
\begin{eqnarray}
H= -J\sum_{i,j} \bm s_i\cdot \bm s_{i+1},
     \label{xy}       \end{eqnarray}
where  $J>0$,  $\bm s_i$ are spins at the sites $i$,  and the summation over $j$ includes only the nearest neighbors to the site $i$.   In the continuum limit, when the spin rotates very slowly at scales of the intersite distance $a$, the Hamiltonian (\ref{xy}) reduces to   the free energy (\ref{exch}) in the Landau-Lifshitz theory  with the magnetization $\bm M =-\gamma  \langle \bm s_i \rangle/a^3$ and  the stiffness constant $\alpha =Ja^5/\gamma ^2 $.
 
The debates on reliability of the general phenomenological approach to magnetism are as old as the approach itself. Nearly sixty years ago Herring and Kittel \cite{HK} argued with their opponents that their phenomenological theory of spin waves ``is not contingent upon the choice of any particular approximate model for the ferromagnetic electrons''.
Interestingly these discussions are still continuing in connection with the concept of the spin current, which originates from the general phenomenological approach. Originally they connected spin supercurrents with counterflows of particles with opposite spins,  for example, of He atoms in the A-phase of $^3$He  \cite{Vuo,Vuo2}.  Bunkov \cite{Bun} insisted that only a counterflow of particles with opposite spins would lead to superfluid spin transport, thus ruling out spin superfluidity in materials with magnetic order   resulting from exchange interaction between localized spins (see p. 93 in his review).  However,  the spin current does not require itinerant electrons for its existence \cite{ES-78a}.
The presumption that  spin transport  in insulators is impossible is still alive nowadays. According to Shi et al. \cite{Niu}, 
it is  a critical  flaw of spin-current definition if it predicts spin currents in insulators.

\section{Stability of spin-current states} \label{stab}

For the sake of simplicity further we focus on current states in an easy-plane ferromagnet, though the analysis can be easily generalized to other magnetically ordered systems discussed in the previous section. In the current state the spontaneous magnetization $\bm M(\bm r)$ rotates in the easy plane  through a large number of full $2\pi$-rotations when the position vector $\bm r$ is varying
along the direction of spin current (figure \ref{fig1}b). The spin-current state is metastable if it corresponds to a local minimum of the free energy, i.e., any transition to nearby states would require an increase of energy. This condition is an analog of the Landau criterion for mass supercurrents discussed in section~\ref{mass}.
In order to check current metastability, one should estimate the energy of possible small static fluctuations around the stationary current state. For this estimation, one should take into account that the stiffness constant $A$ is proportional to the squared inplane  component of the spontaneous magnetization  $ M_{\perp}^2 = M_0^2-m_z^2$, and in the presence of large angle gradients $A$ must be replaced with $A(1-m_z^2/M_0^2)$. So the free energy is 
\begin{eqnarray}
{\cal F}=\int d^3\bm R\left[{ m_z^2\over 2\chi} + {A(1-m_z^2/M_0^2)(\bm \nabla \varphi)^2\over 2} \right]
\nonumber \\
=\int d^3\bm R\left[{ m_z^2\over 2}\frac{E_A-A(\bm \nabla \varphi)^2}{M_0^2} + {A(\bm \nabla \varphi)^2\over 2} \right].
   \end{eqnarray}
 One can see that if $\nabla \varphi$ exceeds $\sqrt{M_0^2/\chi  A}=\sqrt{E_A/ A}$ the current state is unstable with respect to the exit of $\bm M_0$ from the easy plane. This is the Landau criterion for the stability of the spin current.

\begin{figure}
\begin{center}
\begin{minipage}{100mm}
{\resizebox*{7cm}{!}{\includegraphics{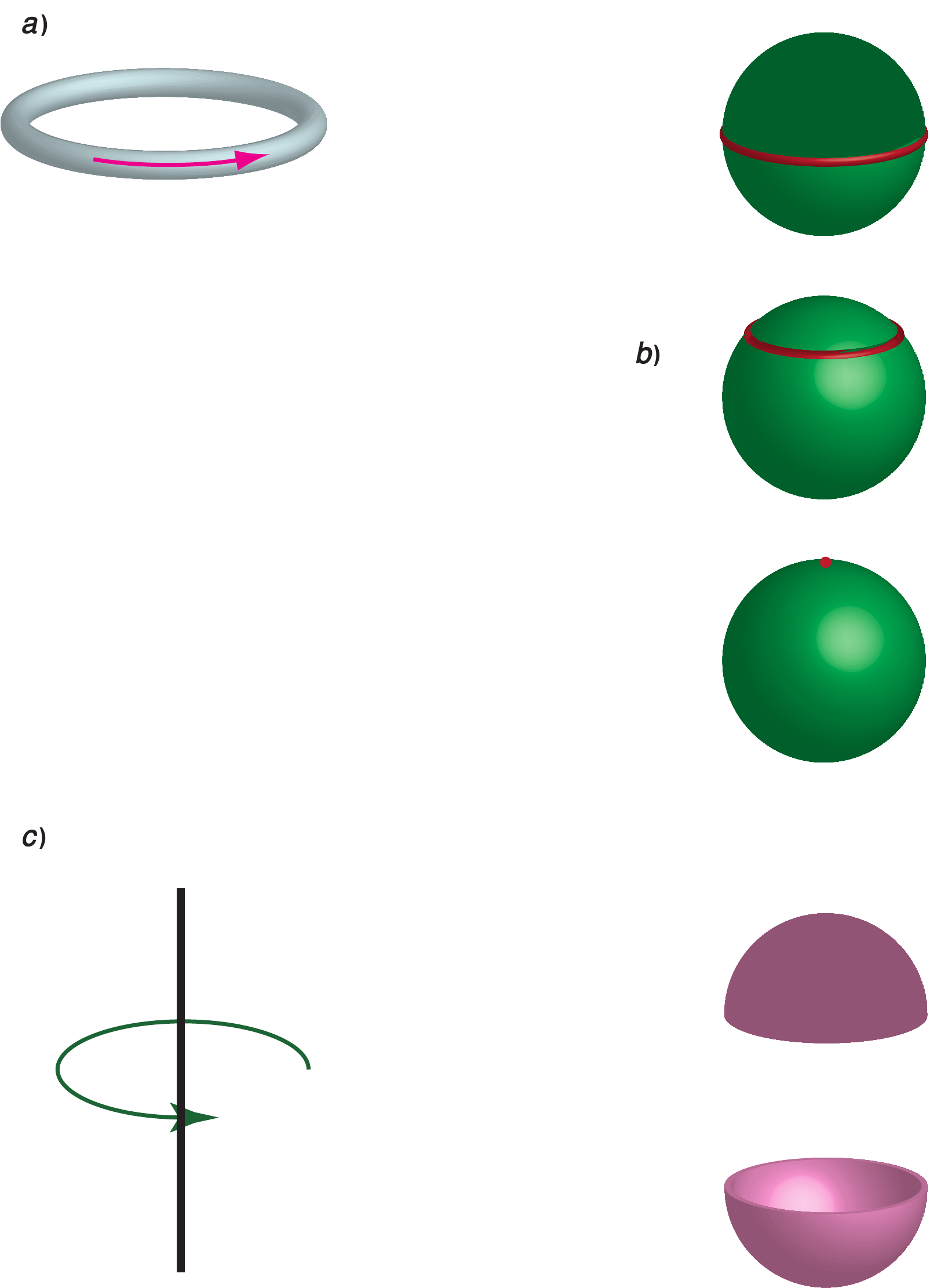}}}%
\caption{Topology of the uniform spin current and the spin vortex states. a) The current state in a torus maps onto the equatorial circumference of the order parameter sphere  $|\bm M|=const$. b) Isotropic ferromagnet: continuous deformation reduces a circumference (current state) to a point (uniform current-free state).
  c) The spin vortex state maps onto either an upper or a lower half of the sphere  $|\bm M|=const$.}%
\label{fig3b}
\end{minipage}
\end{center}
\end{figure}
 
 Like in superfluids, stability of current states is connected with topology of the order parameter space. For ferromagnets the order parameter is the magnetization vector $\bm M$. For isotropic ferromagnets the space of degenerated equilibrium states is a sphere $|\bm M|=|\bm M_0|$, whereas for an easy-plane ferromagnet this space reduced to an equatorial circumference on this sphere (figure~\ref{fig3b}a). Thus the order parameter space for an easy-plane ferromagnet is topologically equivalent to that space for superfluids (the circumference on the complex plane shown in figure~\ref{fig2a}). Spin-current states are stable because they belong to the topological classes different from the class of the uniform ground state and cannot be reduced to the latter by continuous deformation of the path. In contrast, for an isotropic ferromagnet the path around  the equatorial circumference can be continuously transformed to a point on the sphere as shown in figure~\ref{fig3b}b. In this process  the energy monotonously decreases, and topological barriers are absent. Topology of an easy-axis (anti)ferromagnet also does not allow stable spin-current states.
  
  The connection of superfluidity-like phenomena with topology of the order parameter space is universal and not restricted with the examples of mass and spin superfluidity considered here. The same arguments support possibility of exciton superfluidity, which was discussed even earlier than spin superfluidity (see the introductory section \ref{intro}). Though the whole problem of superfluid exciton transport is far from its resolution, in some special case 
 experimental evidences of this transport has already been reported. Kellogg et al.  \cite{QHsup} observed vanishing resistance in double quantum Hall layers, which was interpreted as a consequence of Bose condensation of interlayer excitons (or pseudospin ferromagnetism).  
 
 As well as in the theory of mass superfluidity, the Landau criterion is a necessary but not sufficient condition for current metastability. One  should also to check stability with respect to large 
perturbations, which are \emph{magnetic vortices}.  The magnetic vortices were well known in magnetism. Bloch lines in ferromagnetic domain walls are an example of them \cite{Maloz}.  In the spin-current state the magnetization  $\bm M$ traces a spiral at moving along the current direction. The spin current can relax if one can remove one turn of the spiral. This requires that a singular line (magnetic vortex)  crossed or ``cut'' the channel  cross-section \cite{ES-78a,ES-78b} as shown in figure \ref{fig3}b. 

The  structure of the magnetic vortex outside the vortex core is the same as of the mass superfluid vortex   given by equation~(\ref{vort}). Correspondingly, the  magnetic vortex energy is  determined by the expression similar to equation~(\ref{vorEnM}): 
\begin{eqnarray}
\epsilon =\int d^2\bm r {A(\bm \nabla \varphi_v)^2\over 2} =\pi A \ln{r_m\over r_c},
    \label{vorEn}
          \end{eqnarray}
where the upper cut-off $r_m$ depends on geometry. However, the radius $r_c$ and the structure of the magnetic vortex core are determined differently  from the mass vortex. In a magnetic system the order parameter must not vanish at the vortex axis since
there is a more effective way to eliminate the singularity in the gradient energy:  an excursion of the spontaneous magnetization out of the easy plane $xy$. This would require an increase of the uniaxial anisotropy energy, which keeps $\bm M$ in the plane, but normally this energy is much less than the exchange energy, which keeps the order-parameter amplitude $M$ constant. Finally the core size $r_c$ is determined as a distance at which the uniaxial anisotropy energy density $E_A$ is in balance with the gradient energy  $A(\nabla \varphi)^2 \sim A/r_c^2$.  This yields $r_c \sim \sqrt{A/E_A}$. Figure~\ref{fig3b}c shows mapping of the spin vortex state onto the order parameter space. In contrast to superfluid vortices mapping onto a plane circle, the spin vortex state can map onto one of two halves of the  sphere $|\bm M|=const$. Thus a magnetic (spin) vortex has an additional topological charge having two values $\pm 1$ \cite{Nik}.

The energy of the spin-current state with a vortex and the energy of the barrier, which blocks the phase slip,  i.e., the decay of the current, are determined similarly to the case of mass superfluidity [see equations~(\ref{vorCurEnM}) and (\ref{barrM})]:
\begin{eqnarray}
\tilde \epsilon=\pi L A \ln{r_m\over r_c}- 2 \pi AS \nabla \varphi_0, 
    \label{vorCurEn}
          \end{eqnarray}
\begin{eqnarray}
\epsilon_m \approx \pi L A \ln{1\over  r_c \nabla \varphi_0},
    \label{barr}        \end{eqnarray}
where $L$ is the length of the vortex line and $S$ is the area of the cut, at which the angle jumps by $2\pi$. Thus the barrier disappears at gradients $\nabla \varphi_0 \sim 1/r_c$, which are of the same order as the critical gradient determined from the  Landau criterion. This is a typical situation in the superfluidity theory. But sometimes the situation is more complicated as we shall see in section~\ref{SPV}.

\section{Spin currents without spin conservation law}    \label{Phase}

Though processes violating the conservation law for the total spin are relativistically weak, their  effect is of principal importance and in no case can be ignored. The attention to superfluid transport   in the absence of conservation law was attracted first in connection with discussions of superfluidity of electron-hole pairs. The number of electron-hole pairs can vary due to interband transitions. As was shown by Guseinov and Keldysh \cite{GK}, interband transitions lift the degeneracy with respect to the phase of the  ``pair Bose-condensate''   and make the existence of spatially \emph{homogeneous} stationary current states impossible. On the basis of it Guseinov and Keldysh concluded that there is no analogy with superfluidity. This phenomenon was called ``fixation of phase''. However some time  later it was demonstrated \cite{ES-77} that phase fixation does not rule out existence \emph{inhomogeneous} stationary current states, which admit some analogy with superfluid current states\footnote{Similar conclusions have been done with respect to possibility of supercurrents in systems with spatially separated electrons and holes \cite{LY,Sh}.}. This analysis was extended on spin currents \cite{ES-78a,ES-78b}.

In the spin system the role of the phase is played by the angle of the magnetization $\bm M$ in the easy plane, and the degeneracy with respect to the angle is lifted by magnetic anisotropy in the plane. Adding the $n$-fold inplane anisotropy energy to the total free energy (\ref{Ener})  the latter can be written as
\begin{eqnarray}
{\cal F}=\int d^3\bm R\left\{{ m_z^2\over 2\chi} + {A(\bm \nabla \varphi)^2\over 2}+K[1-\cos (n\varphi)] \right\}.
 \label{an}  \end{eqnarray}
Then the spin continuity equation (\ref{Em}) becomes
   \begin{eqnarray}
{1\over \gamma}{dm_z \over dt}=\bm \nabla \cdot \bm J^z+nK \sin(n\varphi)=- A\left[\nabla^2 \varphi -{\sin(n\varphi)\over l^2}\right] ,
 \label{EmK}      \end{eqnarray}
where 
 \begin{eqnarray}
l^2={ A\over nK}.
             \end{eqnarray}
Excluding $m_z$ from equations~ (\ref{Ep}) and (\ref{EmK})  one obtains the sine Gordon equation for the angle $\varphi$:
  \begin{eqnarray}
{\partial ^2 \varphi \over \partial t^2}- c_s^2\left[\nabla^2 \varphi -{\sin(n\varphi)\over l^2}\right]=0 ,
          \end{eqnarray}
where $c_s =\sqrt{\gamma A/\chi }$ is the spin-wave velocity. According to this equation,  the inplane anisotropy leads to a gap in the spin-wave spectrum: 
 \begin{eqnarray}
\omega^2 ={nc_s^2\over l^2}+c_s^2 k^2.
             \end{eqnarray}

There are  one-dimensional  solutions $\varphi(x-v t)$ of the sine Gordon equation  with non-zero average $\langle \nabla \varphi \rangle$, which correspond to a periodic lattice of solitons (domain walls) of the width $\sim \tilde l=l\sqrt{1-v^2/c_s^2}$ with the period $x_0=2\pi /n \langle \nabla \varphi \rangle$ moving  with  the velocity $v$. The function inverse to $\varphi(x-v t)$ is
  \begin{eqnarray}
x-vt=\sqrt{n \over 2}\tilde l \int ^\varphi \frac{d\varphi'}{\sqrt{\kappa -\cos n\varphi'}},
          \end{eqnarray}
where the constant $\kappa>1$ is determined by the equation
  \begin{eqnarray}
x_0={2\pi \over n \langle \nabla \varphi \rangle} =\sqrt{n \over 2}\tilde l\int _0^{2\pi /n}  \frac{d\varphi'}{\sqrt{\kappa-\cos n\varphi'}}. 
          \end{eqnarray}
The free energy of the soliton lattice is given by
\begin{eqnarray}
F =A \left[{1-\kappa \over n  l^2}  +  {\langle \nabla \varphi \rangle\over \pi \tilde l} \sqrt{n\over 2}\int_0^{2\pi/n } \sqrt {\kappa -\cos  n\varphi} d\varphi \right].
                                \end{eqnarray}

\begin{figure}
\begin{center}
\begin{minipage}{100mm}
{\resizebox*{7cm}{!}{\includegraphics{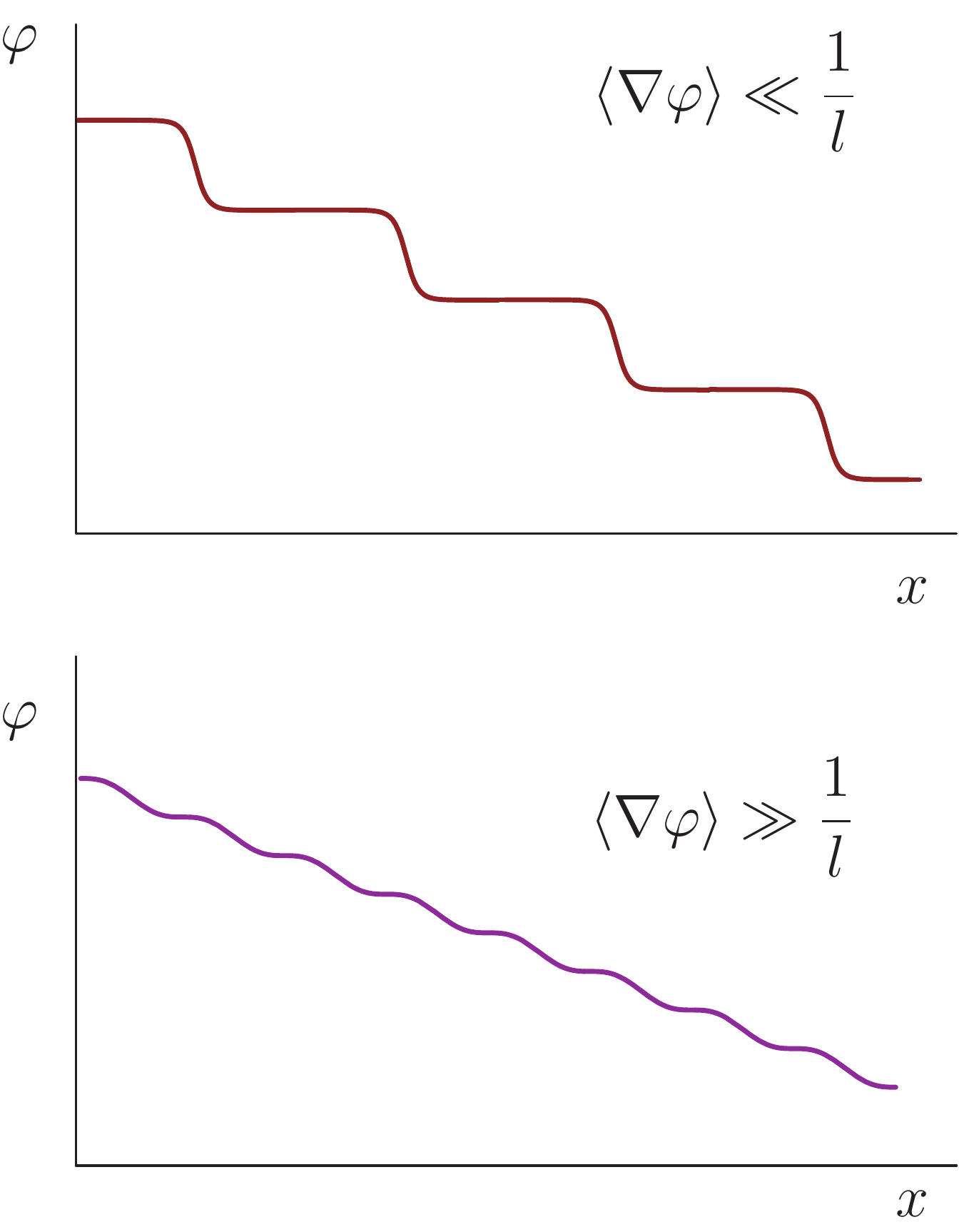}}}%
\caption{The nonuniform spin-current states with $\langle \nabla \varphi \rangle \ll 1/l$ and $\langle \nabla \varphi \rangle \gg 1/l$.}%
\label{fig2}
\end{minipage}
\end{center}
\end{figure}

It is possible to develop the hydrodynamic theory of the soliton lattice in the terms of local density and velocity of solitons \cite{ES-81}, which is able to describe deformations of the lattice slow in space and time. Here we focus on stationary current states when $dm_z/dt=0$ ($v=0$).  At small average twisting of the spontaneous magnetization $\langle \nabla \varphi \rangle \ll 1/l$ the structure constitutes domains that correspond to the $n$ equivalent easiest directions in the easy plane.  In this limit ($\kappa \to 1$)  the free energy density  is the product  of the energy of an isolated domain wall and the density of domain walls $n \langle \nabla \varphi \rangle /2\pi$: 
\begin{eqnarray}
F =A{ 4\langle \nabla \varphi \rangle\over \pi\sqrt{n}   l}.
           \label{enDW}                     \end{eqnarray}
Spin currents (gradients) inside domains are negligible but there are essential spin currents inside domain walls where  $\nabla \varphi  \sim 1/l$. This hardly reminds genuine  superfluid transport on macroscopical scales: spin is transported over distances on the order of the domain-wall width $l$. With increasing $\langle \nabla \varphi \rangle$ the density of domain walls grows, and at $\langle \nabla \varphi \rangle\gg 1/l$ they coalesce while for a displacement along the direction of the gradient $\langle \nabla \varphi \rangle$, the end point of the vector $\bm M$ describes, a line close to a helix. 
The nonuniform states with $\langle \nabla \varphi \rangle \ll 1/l$ and $\langle \nabla \varphi \rangle \gg 1/l$ are shown in figure~\ref{fig2}. Thus the processes violating spin conservation law are not important for large deformations (gradients) of the spin structure. This means that the analogy of these deformed states with current states in superfluids makes sense.

Studying stability of nonuniform current states it is possible to ignore the inplane anisotropy only  for large spin currents when $\nabla\varphi \gg 1/l$. Let us consider the opposite limit of $\langle\nabla\varphi \rangle \ll 1/l$ when the spin structure reduces to a chain of domain walls. The relaxation of the spin current,  which is proportional to the wall density,  requires that some domain walls vanish from the channel. This process is illustrated in figure \ref{fig3}b for the four-fold inplane symmetry ($n=4$). When a magnetic vortex appears, $n$ domain walls finish not at the wall but at the vortex line, around which  the angle $\varphi$ changes by 2$\pi$. The $2\pi$ angle jump occurs at the cut restricted by the vortex line. The $n$ domain walls disappear via  motion of the vortex line across the channel cross section. In the  course of this process, the change of the energy consists  of the vortex-line energy, which is proportional to the line length $L$, and of a decrease of the surface energy of the $n$ walls themselves proportional to the cut area $S$. The latter contribution is determined by the product of the free energy density (\ref{enDW}) and the volume $2 \pi S/\langle \nabla \varphi \rangle$.
Taking these two contributions into account, the energy during the process of annihilation of $n$ walls is
\begin{eqnarray}
\tilde \epsilon =\pi L A \ln{r_m\over r_c}-{8A\over \sqrt{n}}{S\over l}.
            \end{eqnarray}
Comparing it with the energy given by equation (\ref{vorCurEn}) one sees that the gradient $\nabla \varphi_0$ is replaced by the maximum gradient $\sim 1/l$ inside the domain wall. Correspondingly for the 2D case shown in figure \ref{fig3}b the expression (\ref{barr}) for the barrier energy must be replaced by
\begin{eqnarray}
\epsilon_m \approx \pi L A \ln{l \over r_c} \approx {\pi\over 2} L A \ln{E_A\over K}, 
            \end{eqnarray}
where the  two lengths $r_c \sim \sqrt{A/E_A}$ and $l\sim \sqrt{A/K}$ are determined by the uniaxial and the inplane anisotropy energies  $E_A$ and $K$.  Thus large barriers stabilizing  spin-current states are possible only if the condition $E_A\gg K$ is satisfied.  This conclusion \cite{ES-78a,ES-78b,ES-82} was recently confirmed by the analysis of K\"onig et al. \cite{McD}.

An important difference with conventional mass superfluidity is that in conventional superfluidity the barrier, which suppresses supercurrent relaxation, grows unrestrictedly when the gradient $\nabla\varphi$ decreases. In contrast, in spin superfluidity the barrier growth stops when the gradient reaches the values of the order $1/l$ (inverse width of the domain wall). Since the current relaxation time exponentially depends on the barrier (whether the barrier is overcome due to thermal fluctuations or via quantum tunneling) the life time of the current state  in conventional superfluidity diverges when the velocity (phase gradient)  decreases.  In contrast,  the life time of the spin current can be exponentially large but  always {\em finite}.  This provides an ammunition for rigorists, who  are not ready to accept the concept  ``spin superfluidity'' (or superfluidity of any non-conserved quantity) in principle. {\em In principle}, one could agree with them. But {\em in practice}, whatever we call it,  ``non-ideal superfluidity'' or ``quasi-superfluidity'', some consequences should outcome from the fact of the existence of topological  barriers suppressing relaxation of spin-current states. A key point is whether these consequences are observable. This is the topic of the next section.

\section{Is  superfluid spin transport ``real''?} \label{real}

From early days of discussions on spin supercurrents and up to now there are arguments on whether the spin supercurrent can result in ``real'' transport of spin. Partially this is a semantic problem:  One must carefully define what ``real'' transport really means. Let us suppose that one has a usual superfluid mass persistent current in a ring geometry. Nobody doubts that real mass transport occurs in this case, but how can one notice it in the experiment? In any part of the ring channel there is no accumulation (increase or decrease) of the mass. Of course, one  can detect gyroscopic effects related with persistent currents, but it is an indirect evidence. What may be a direct evidence? One could suggest a \emph{Gedanken Experiment} in which the ring channel is suddenly closed in some place. In the wake of it one can observe that the mass increases on one side from the closure and decreases on the other side. This would be  a real transport if one required a demonstration of mass accumulation as a proof of it.
Accepting this definition of transport reality  one can notice real transport only in a non-equilibrium  process, when the transported  quantity  decreases in some place and increases in another . 
Naturally one can discard these semantic exercises  as irrelevant for practice, but only as far as they refer to mass currents.  In the case of spin currents in the past and nowadays  spin accumulation sometimes  is considered as a necessary proof of real spin transport. Therefore, in old publications on spin superfluidity \cite{ES-78a,ES-78b,ES-82} much attention was paid to possible experimental demonstration of spin transport from one place to another.

\begin{figure}
\begin{center}
\begin{minipage}{100mm}
{\resizebox*{7cm}{!}{\includegraphics{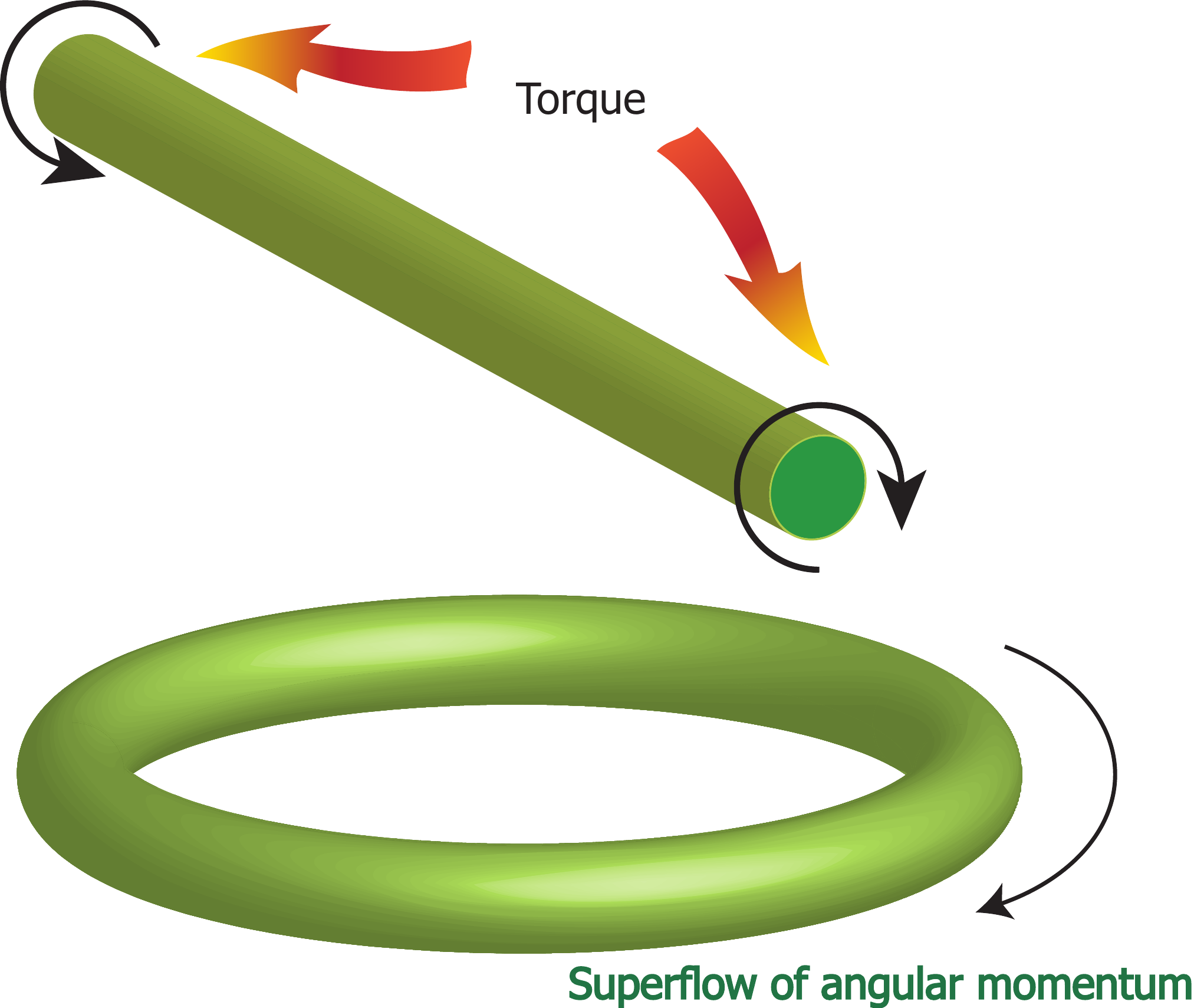}}}%
\caption{Mechanical analogue of a persistent current: A twisted elastic rod bent into a closed ring. There is a persistent angular-momentum flux around the ring.}%
\label{fig3a}
\end{minipage}
\end{center}
\end{figure}

Before starting discussion of possible spin-transport demonstration it is useful to  consider a mechanical analogue of superfluid mass or spin supercurrent \cite{ES-82}. Let  us twist a long elastic rod so  that  a twisting angle at one end of the rod with respect to an opposite end  reaches values many times $2\pi$. Bending the rod into a ring and connecting the ends rigidly, one obtains a ring with a circulating persistent angular-momentum flux (figure~\ref{fig3a}). The intensity of the flux is proportional to the gradient of twisting angle, which plays the role of the phase gradient in the mass supercurrent or the spin-rotation-angle gradient in the spin supercurrent. The analogy with spin current is especially close because spin is also a part of the angular momentum.  The deformed state of the ring is not the ground state of the ring, but it cannot relax to the ground state via any elastic process, because it is topologically stable.  The only way to relieve the strain inside the rod is {\em plastic displacements}. This means that dislocations must move across rod cross-sections. The role of dislocations in the twisted rod is the same as the role of vortices in the mass or spin current states: In both of the cases some critical deformation (gradient) is required to switch the process  on. There are various ways to detect deformations or strains in an elastically deformed body. Similarly, it is certainly possible, at least in principle, to notice deformation (angle gradient) of the spin structure in the spin-current state. It would be a legitimate evidence of the spin current, not less legitimate than a magnetic field measured around the ring as an evidence of the persistent charge current in the ring.

Of course, it is not obligatory to discuss the twisted rod in terms of angular-momentum flux. One can describe it only in terms of deformations, stresses, and elastic stiffness. So we must have in mind that there are two languages, or descriptions of the {\em same} physical phenomenon. A choice of one of them is a matter of taste and tradition. For example, in order to describe the transfer of momentum they use the momentum-flux tensor (``flux'', or ``current'' language) in hydrodynamics, while in the elasticity theory they prefer to call the  same tensor as {\em stress tensor}. In principle one can avoid the term ``superfluidity'' and speak only about the ``phase stiffness'' even in the case of mass supercurrents.

\begin{figure}
\begin{center}
\begin{minipage}{100mm}
{\resizebox*{7cm}{!}{\includegraphics{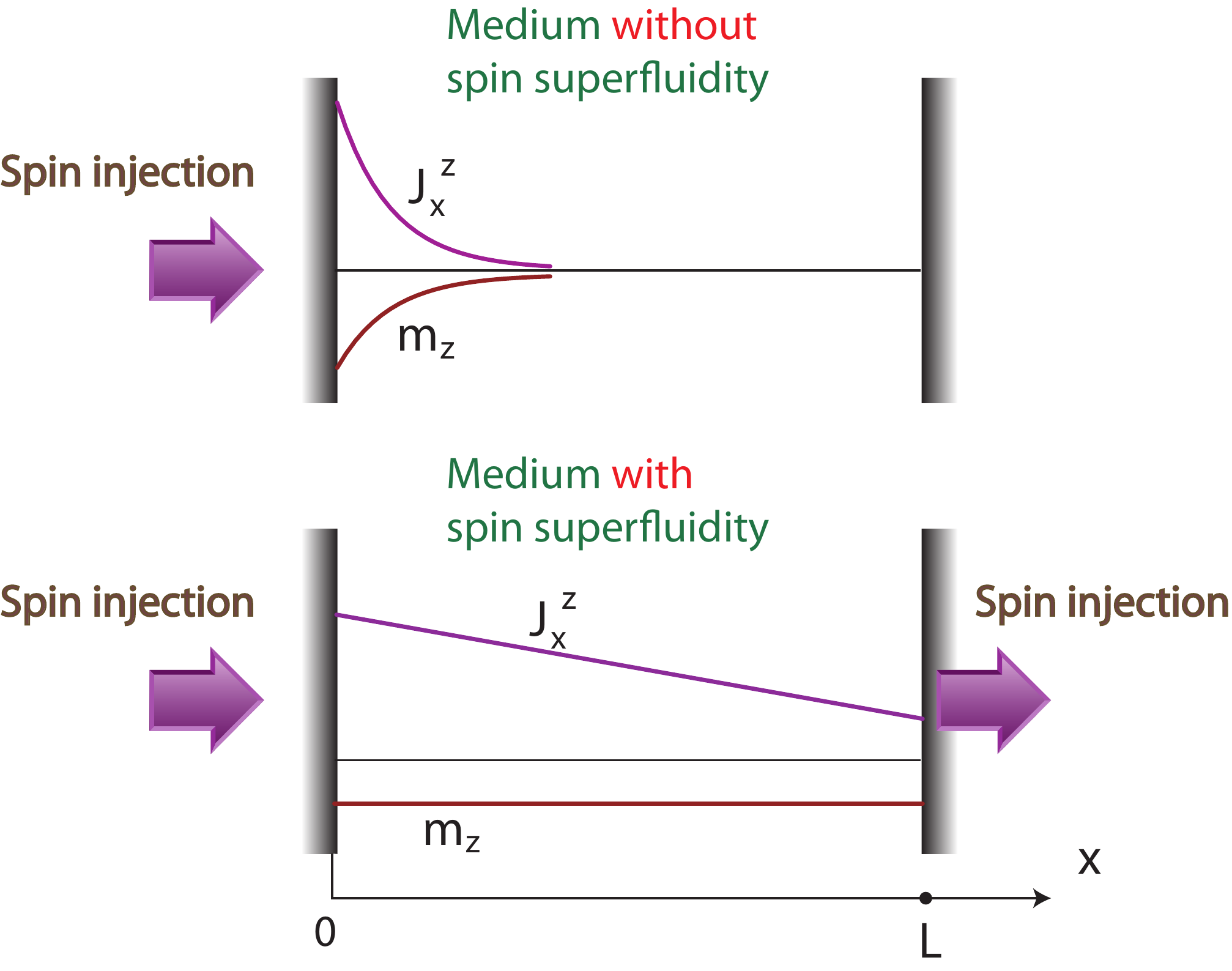}}}%
\caption{Spin injection to a spin-nonsuperfluid and a spin-superfluid medium.}%
\label{fig4}
\end{minipage}
\end{center}
\end{figure}

Let us return to possible demonstration of ``real'' spin transport. Suppose that spin is injected into a sample  at the sample boundary $x=0$ (figure \ref{fig4}). The injection can be realized  practically either with an injection of  a spin-polarized current (for the sake of simplicity we put aside the problem what happens with charge in this case), or with pumping the spin with a circularly polarized microwave irradiation.  If the medium at $x>0$ cannot support superfluid spin transport,  the only way of spin propagation is spin diffusion described by the equations
 \begin{eqnarray}
{\partial m_z \over \partial t}-\gamma \bm \nabla \cdot \bm J_d^z+ {m_z \over T_1}=0,~~ \bm J_d^z=D_s \bm \nabla m_z,
 \label{dif}      \end{eqnarray}
where $D_s$ is the spin-diffusion coefficient and  $T_1$ is the time characterizing the Bloch longitudinal  relaxation, which violates the spin-conservation law. In the stationary case $\partial m_z /\partial t=0$, and both the spin current and the nonequilibrium magnetization $m_z$ exponentially decay inside the sample: $J_d^z \propto m_z \propto e^{- x/L_s}$,  where $L_s=\sqrt{D_sT_1}$ is the spin-diffusion length. So no spin can reach the other boundary $x=L$ of the sample provided $L \gg L_s$. 

Now let us suppose that the medium at $0<x<L$ is magnetically ordered and can support superfluid spin transport. If the injection is so weak that the angle  gradient $\nabla \varphi$ is much less than $1/l$, the perturbation of the  medium by the injection can penetrate at the length not longer than the domain-wall width $l$. So the injection pushes a piece of a domain wall into the sample. The spin current exponentially decays like in the medium without superfluidity.  If the injection is so strong that the angle gradient   $\nabla \varphi $  exceeds its maximum value $1/\sqrt{n} l$ in a center of an isolated domain wall, continuity of the spin current on the boundary requires appearance of a soliton lattice with a period of the order or smaller than $l$. This means that the spin current {\em can reach} the other boundary $x=L$. In the thermodynamic limit $L\to \infty$ one may expect a stationary soliton lattice with $m_z=0$. However, at the boundary $x=L$ the spin current must be injected into the medium without spin superfluidity. This is impossible without a finite $m_z$, and the finite $m_z$ means that the soliton lattice {\em is moving}. In the limit $\nabla \varphi \gg 1/l$ the soliton lattice is rather dense and one may neglect periodical spin-current modulation caused by inplane anisotropy. Then spin transport is described by the equations 
        \begin{eqnarray}
{d\varphi \over dt}=-\gamma { m_z\over \chi},
     \label{EpB} \end{eqnarray}
     \begin{eqnarray}
{dm_z \over dt}-\gamma\bm \nabla \cdot \bm J^z+ {m_z \over T_1}=0,
 \label{EmB}      \end{eqnarray}
with the boundary conditions for the supercurrent  $\bm J^z(0) = \bm J^z_0$ at $x=0$ and $\bm J^z(L) =- f m_z (L) $ at $x=L$. The current $J^z_0$ in the first condition is the spin-injection current, while the second boundary condition takes into account that the medium at $x>L$ is not spin-superfluid and spin injection there is possible only if some non-equilibrium magnetization $m_z(L)$ is present. The coefficient $f$ can be found by solving   the spin-diffusion equations in the medium at  $x>L$ \cite{ES-78b}. It also depends on properties of the contact at $x=L$. While  the inplane anisotropy violating the spin conservation (phase fixation) was neglected, one cannot neglect irreversible dissipative processes, which also violate the spin-conservation law. The simplest example of such a process is  
the  longitudinal spin relaxation characterized by time $T_1$. 

The stationary solution of equations (\ref{EpB}) and (\ref{EmB}) is 
    \begin{eqnarray}
m_z= -{\gamma  T_1 \over L+f \gamma T_1}J^z_0\approx- {\gamma  T_1 \over L}J^z_0,~~J^z(x) = J^z_0 \left(1-{x\over L+f \gamma T_1} \right) \approx J^z_0{L-x\over L}.
     \end{eqnarray}
Though the solution is stationary in the sense that  $\partial m_z /\partial t=0$, but  $\partial \varphi /\partial t \neq 0$. We consider a non-equilibrium process (otherwise spin accumulation is impossible), which is accompanied by the precession of $\bm M$ in the easy plane. But the process is stationary  only if the precession angular velocity is constant in space. The condition $m_z=$const, which results from it, is similar to the condition of constant chemical potential in superfluids or electrochemical potential in superconductors in stationary processes. If this condition were not satisfied, there would be steady growth of the angle twist as is evident from equation (\ref{EpB}). As already mentioned above, the nonzero $m_z$ means that the soliton lattice is moving. In our case the soliton velocity is rather slow since it is inversely proportional to $L$: $v= c_s^2 T_1/L$. 

One can see that irreversible loss of spin is a more serious obstacle for superfluid spin transport than coherent phase fixation, to which most of attention was attracted in the literature. Because of spin relaxation,  the spin current inevitably decreases while moving away from the injection point, in contrast to constant superfluid mass currents. However, in a spin-superfluid medium this decrease is linear and therefore less destructive than exponential decay  of currents in non-superfluid media. So one have a good chance to notice spin accumulation in the medium at $x>L$ rather distant from the place of original spin injection.
This justifies using the term ``superfluid''.

In the presented analysis we assumed that at the boundary the entire spin-injection current is immediately transformed  into a supercurrent. Actually spin injection can also generate the diffusion current close to the boundary. However, at some distance from the boundary the diffusion current inevitably transforms into a supercurrent \cite{ES-78b}. If this distance (healing length) is much shorter than the size $L$ of the sample our  boundary condition at $x=0$ is fully justified. Similar effects take place at contacts ``normal metal - superconductor'': The current from the normal metal to the superconductor is completely transformed into the supercurrent at some finite distance from the contact. 

Spin injection is not the only method of generation of spin currents. One can generate spin currents by a rotating inplane magnetic field, which is applied to one end of the sample and  is strong enough to orient the magnetization parallel to it. Because of the stiffness of the spin system, the spin rotation at one end is transmitted to the other end of the sample, which is not subject to the direct effect of the rotating magnetic field. Transmission of the torque through the sample is {\em spin current}.  Since the rotating field is acting on the phase (angle) of the order parameter, it can be called {\em coherent} method, in contrast to the {\em incoherent} method of spin injection. The coherent  method of spin-current generation has no analog in superfluids and superconductors, since in the latter cases there is no field linked to the phase of the order parameter. Referring to the set up shown in figure \ref{fig4} with spin injection replaced by rotating magnetic field, in the coherent method the magnetization $m_z$ is fixed by the frequency of the rotating field. In this case there is no threshold for spin-current generation, and the spin current appears whatever small the frequency could be.  But if the frequency (and $m_z$ proportional to it) is low, the spin transmission is realized via generation of  a chain of well separated solitons (domain walls), which propagate to the other end of the sample. Thus, a ``moving soliton lattice'' is another synonym for spin superfluid transport. Long-distance propagation of solitons through a slab of the $A$ phase of superfluid $^3$He generated by a pulse of a radio-frequency magnetic field has already experimentally realized by Bartolac et al. \cite{SolHe}. This experiment was discussed in terms of spin transport in reference \cite{ES-81}.

\section{Spin-precession superfluidity in superfluid $^3$He-B} \label{HeB}

\subsection{Stationary uniform precession in $^3$He-B}

Now we focus on the experimental and theoretical investigations of superfluid spin transport in the $B$ phase of superfluid $^3$He. The spin superfluidity in the $B$ phase has several important features, which distinguish it from the spin superfluidity discussed previously in this review. First, in contrast to what was considered earlier, observed spin-current states in the $B$ phase are dynamical nonlinear states very far from the equilibrium, which require for their support permanent pumping of energy. Thus dissipation is always present, and speaking about ``superfluidity'', i.e., ``dissipationless'' spin transport, we have in mind the absence of additional dissipation connected with the spin current itself. Second, while the previous discussion dealt with the transport of a single spin component ($z$-component), in the $B$ phase spin vector performs a more complicated 3D rotation and the spin current refers to the transport of some combination of spin components. This combination may be called ``precession moment''  because it is a canonical conjugate of the precession rotation angle (precession phase) rather the rotation angle of genuine spin in the spin space.
So one should discern two types of spin superfluid transport: transport of spin precession and transport of spin  \cite{ES-87}. 
In the experiment \cite{HeBex} they used slightly nonuniform magnetic fields, and precession took place only inside the homogeneously precessing domain (HPD). But for discussion of superfluid spin transport it is not so important, and in the following we consider only processes inside the HPD ignoring gradients of the magnetic field.

\begin{figure}
\begin{center}
\begin{minipage}{100mm}
{\resizebox*{7cm}{!}{\includegraphics{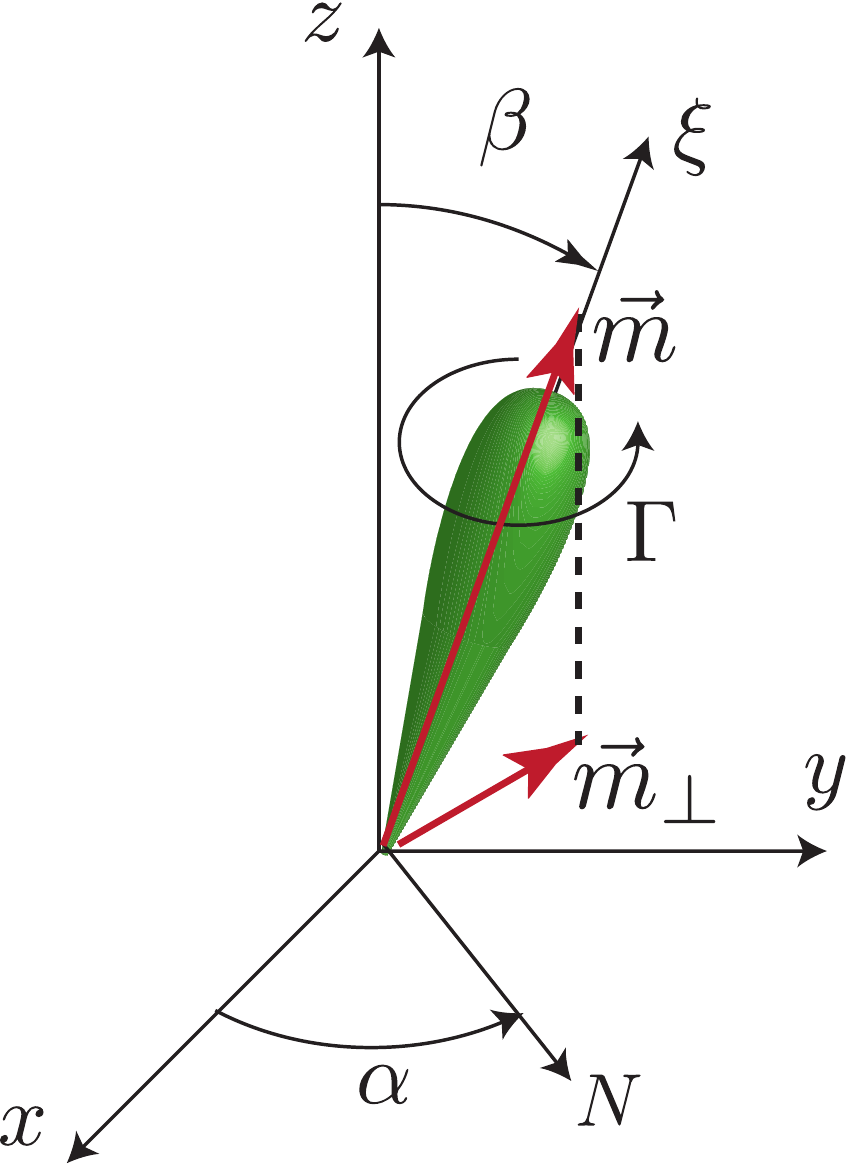}}}%
\caption{Euler angles for spin precession in $^3$He-B. The rigid top presents the rigid spin order parameter structure controlled  by large exchange energy. The rotation of the coordinate frame transforming the axis $z$ to the axis $\xi$ is performed  around the axis $N$ (line of nodes). }%
\label{fig4.0}
\end{minipage}
\end{center}
\end{figure}

The spin dynamics of superfluid phases of $^3$He is described by the theory of Leggett and Takagi \cite{LT}, which is  an example of the general phenomenological theory of magnetically ordered systems in terms of conjugate canonical variables ``angle--moment'', which was shortly discussed in section~\ref{magn}. As well as in studies of rotating solid tops, sometimes it is more convenient to describe spin rotations via the Euler angles $\alpha$, $\beta$, and $\Gamma$ (figure~\ref{fig4.0}). In $^3$He superfluid spin dynamics these angles were used by  Fomin \cite{Fom-84,Fom-91}, who, however, replaced the angle $\Gamma$ by the angle  $\Phi=\alpha+\Gamma$.
The angle $\beta$ is the precession tipping angle, and $\alpha$ is the precession phase determining the direction of the line of nodes $N$. The angle $\Phi$ characterizes the resultant rotation of the order parameter in the laboratory frame, and in the limit $\beta\to 0$ (no precession) becomes the angle of rotation around the $z$ axis. The magnetic moments canonically conjugate to the angles $\alpha$, $\beta$, and $\Phi$ are $P=m_z-m_\xi$, $m_\beta$, and $m_\xi$ respectively, where $m_z$ is the $z$ component of the magnetization $\bm m$ in the laboratory coordinate frame, $m_\xi$ is the projection of $\bm m$  on the $\xi$ axis of the rotating coordinate frame  (see figure~\ref{fig4.0}), and $m_\beta$ is the projection of $\bm m$  on the  line of nodes $N$, which is perpendicular to the axes $z$ and $\xi$. The free energy density consists of three terms, $F=F_0 +F_\nabla +V$, where
\begin{eqnarray}
F_0={m^2\over 2 \chi}-\bm m \cdot \bm H
          \end{eqnarray}   
includes the magnetization and the Zeeman energies,  and the gradient energy $F_\nabla$ and the order-parameter dependent energy $V$ (dipole energy in the case of $^3$He)  will be determined later on.  Here $\chi$ is the magnetic susceptibility.

For the phenomena observed experimentally only  one degree of freedom is essential, which is connected with precession, i.e., with the conjugate pair ``precession phase $\alpha$--precession moment $P$''.  In contrast to the mode connected to the longitudinal magnetic resonance (oscillations of the longitudinal spin component), which was discussed in previous sections, the precession mode is connected with the transverse magnetic resonance (nuclear magnetic resonance in the case of $^3$He), in which $m_z$ does not oscillate essentially.

Neglecting nutation,   the directions of the axis $\xi$ and of the moment $\bm m$ coincide. Then $\beta$ is constant, $m_\xi=m$, $m_\beta=0$, $P=m(\cos \beta -1)$, and the free energy density $F_0$ becomes 
\begin{eqnarray}
F_0={m^2\over 2 \chi}-(m+P)H ,
          \end{eqnarray}   
where $H$ is a strong constant magnetic field parallel to the $z$ axis. 
The Hamilton equations for the precession mode are: 
\begin{eqnarray}
{\partial \alpha\over \partial t}=\gamma {\delta {\cal F}\over \delta P}=-\omega_L+\gamma{\partial  (F_\nabla+V)\over \partial P} ,
\nonumber \\
{1\over \gamma } {\partial P\over \partial t}=-{\delta {\cal F}\over \delta \alpha }=-{\partial  (F_\nabla+V)\over \partial \alpha }+\nabla_i{\partial  F_\nabla\over \partial \nabla_i \alpha },
    \label{prec}      \end{eqnarray}   
where $\omega_L=\gamma H$ is the Larmor frequency and ${\cal F}=\int d^3\bm R\,F$ is the total free energy. Since this section deals with nuclear spins, here $\gamma $ is the nuclear gyromagnetic  ratio.
 In the experiment the magnetization amplitude is determined by the  magnetic field: $m=\chi H$.

Equation (\ref{prec}) describes free precession without dissipation. One  of the most important mechanisms of dissipation is the Leggett--Takagi mechanism \cite{LT} related with the process of equilibration of the magnetization of the normal component with the precessing  magnetization of the superfluid component  (for details see the reviews by  Bunkov  \cite{Bun} and Fomin \cite{Fom-91} and references therein). This mechanism becomes ineffective at low temperatures, and this leads to the Suhl instability of the uniform precession, which is discussed below. So one cannot observe uniform precession in $^3$He-B at very low temperatures.  The dissipation leads to a precession decay, and in order to support the state of uniform precession in the experiment the energy dissipation  must be compensated by the energy pumped by the rotating transverse magnetic field. Assuming that the balance of the pumped energy and the dissipated energy eventually leads to stationary precession, we may further ignore the both. The stationary precession state corresponds to the extremum of the Gibbs thermodynamic potential, which is obtained from the free energy with the Legendre transformation: $G= F+\omega_P P/\gamma$, where the precession frequency $\omega_P=-\partial \alpha/\partial t$ plays the role of the ``chemical potential'' conjugate to the precession moment density $P$.  

Up to now the theory was rather general and valid for precession in any magnetically ordered system. 
Referring  to $^3$He-B particularly,  the dipole energy  in $^3$He-B is \cite{Bun,Fom-91} 
\begin{eqnarray}
V ={2 \chi\Omega ^2\over 15 \gamma^2}\left[(1+\cos \Phi )u +\cos \Phi -{1\over 2}\right]^2,
    \label{DE}           \end{eqnarray}   
where  $\Omega$ is the longitudinal NMR frequency and $u=\cos \beta$.  At the stationary precession the angle $\Phi$ does not vary in time and can be found by minimization of the Gibbs potential.  In the state of uniform precession without spatial gradients only the dipole energy depends on $\Phi$, and the equation for $\Phi$ is
\begin{eqnarray}
{\partial V\over \partial \Phi}={4 \chi\Omega ^2\over 15 \gamma^2}\left[(1+\cos \Phi )u +\cos \Phi -{1\over 2}\right]^2(1+u)\sin\Phi=0.
     \label{dip-phi}     \end{eqnarray}   
Solution of this equation yields
 \begin{eqnarray}
\cos \Phi={1/2-u\over 1+u},~~V(u) =0  &\mbox{for}~\beta<104^\circ~(u>-1/4),
\nonumber \\
\cos \Phi=1,~~V(u) ={8\chi \Omega^2\over 15 \gamma^2}\left({1\over 4}+u\right)^2  &\mbox{for}~\beta>104^\circ~(u<-1/4).
    \label{dip}      \end{eqnarray}   
 Thus at $u>-1/4$ ($\beta<104^\circ$) and at $u<-1/4$ ($\beta>104^\circ$) one must choose two different branches of the solution of equation~(\ref{dip-phi}). The critical angle $\beta_c=104^\circ$ is called the  Leggett magic angle.
 
It has already been known about 50 years from studies of nonlinear ferromagnetic \cite{Suhl} and antiferromagnetic \cite{Heeger} resonance that the state of uniform spin precession with finite precession angle can be unstable with respect to excitation of spin waves (Suhl parametric instability). Though Suhl instability is a phenomenon of the nonlinear classical wave theory \cite{Lvov} it is easier to qualitatively explain it in terms of spin-wave quanta (magnons). The processes leading to instability are transformations of $n$ quanta  of uniform precession into two spin-wave quanta with wave vectors $\pm \bm k$:
  \begin{eqnarray}
n\omega_P =\omega(\bm k) +\omega(-\bm k). 
          \end{eqnarray}   
The precession is unstable if at least one of these processes is allowed by the laws of energy and momentum conservation. The three-magnon process ($n=1$) corresponds to the first order Suhl instability. The process is possible if a quantum of  uniform precession of frequency $\omega_P$ can dissociate into two quanta of the lower spectral branch with  frequency $\omega_P/2$. Another possibility to destabilize uniform precession is a four-magnon process of transformation of two quanta of uniform precession into  two quanta of the same spectral branch with  finite wave vectors $\pm \bm k$ (the second-order Suhl instability, $n=2$). The process becomes possible if a nonlinear correction to the frequency of uniform precession has an opposite sign with respect to the frequency dispersion $d\omega^2/dk^2$. In the theory of nonlinear waves the latter condition is called Lighthill's condition, which is necessary for modulation instability \cite{RT}. The second-order Suhl instability is an example of it. In all known examples of uniform precessions in magnetically ordered systems there are conditions for at least one type of Suhl instability.  In superfluid $^3$He Suhl instability is possible in the A \cite{Fom-79,ES-87b} and B \cite{SF,Sur} phases. But as any parametric instability, the Suhl instability can be suppressed by dissipation, which leads to a critical precession angle below which the state of uniform precession remains stable.  In the B phase stable uniform precession is possible  at temperatures $T >0.4 \,T_c$. At lower temperatures dissipation is weak and cannot block the Suhl instability. This explains a sudden transition to the regime of ``catastrophic relaxation'' observed by Bunkov et al. \cite{Cata}.\footnote{In contrast to Surovtsev and Fomin \cite{SF,Sur}, who explained catastrophic relaxation with the bulk Suhl instability, Bunkov et al. \cite{BLV} suggested another mechanism of  Suhl instability, which exists near the surface (see arguments over the two mechanisms by Fomin \cite{Fom-06} and Bunkov et al.  \cite{BLVb})}

\subsection{Stability of spin-precession supercurrents (Landau criterion)} \label{SPV}

Let us consider the state with uniform spin-precession current proportional to the gradient $\nabla \alpha$. The total free energy should now include the gradient energy of $^3$He-B  \cite{Fom-80}:
\begin{eqnarray}
F_\nabla=A(u){\nabla\alpha^2\over 2}+{\chi \over \gamma^2}\left[c_\parallel^2{\nabla\Phi^2\over 2}+c_\perp^2{\nabla u^2\over 2(1-u^2)}
\right],
     \label{grad}     \end{eqnarray}   
where 
\begin{eqnarray}
A(u)  ={\chi \over \gamma^2}[c_\parallel^2 (1-u) ^2+c_\perp^2 (1-u^2)], 
   \label{gr}       \end{eqnarray} 
$u=\cos\beta$,  and $c_\parallel$ and $c_\perp$ are longitudinal and transverse spin wave velocities. The expression for $F_\nabla$ assumes that all gradients are normal to the axis $z$ parallel to the dc magnetic field.

  An important feature of the dipole energy  in $^3$He-B is its independence of the precession angle $\alpha$. In accordance with Noether's theorem this means that the precession moment is strictly conserved. Thus the equations describing stationary precession with frequency $\omega_P$ are
  \begin{eqnarray}
-\omega_P=-\omega_L+\gamma{\delta (F_\nabla +V) \over \delta P}, 
   \label{prec1}
          \end{eqnarray}   
\begin{eqnarray}
\bm \nabla \cdot \bm  J =0, 
            \end{eqnarray}   
where 
 \begin{eqnarray}
\bm  J =-{\partial F\over \partial \bm \nabla \alpha}= -A(u) \bm \nabla \alpha
            \end{eqnarray}   
is the spin-precession current. 

Apart from $\nabla \alpha$, other gradients are absent: $\nabla u =\nabla \Phi =0$. Then equations~ (\ref{dip})  and  (\ref{grad})--(\ref{prec1})
yield the following equation for $u$:
 \begin{eqnarray}
(\omega_P-\omega_L)\omega_L  -[c_\parallel^2+(c_\perp^2-c_\parallel^2) u ]\nabla\alpha^2  =0                &\mbox{at}~\beta<104^\circ~(u>-1/4),
\nonumber \\
(\omega_P-\omega_L)\omega_L  -[c_\parallel^2+(c_\perp^2-c_\parallel^2) u ]\nabla\alpha^2+{4 \Omega^2(1+4u)\over  15}=0  &\mbox{at}~\beta>104^\circ~(u<-1/4).
\nonumber \\
 \label{curSt}  \end{eqnarray}

The proper way to check stability of the current state given by equation~(\ref{curSt}) is to use the Landau criterion \cite{ES-87}.  Since we check stability of the relative minimum at fixed averaged gradient $\nabla \alpha$, we must do a new Legendre transformation choosing a new Gibbs thermodynamic potential 
\begin{eqnarray}
\tilde G =G+\bm J \cdot \bm \nabla \alpha=F+\omega_ P {P\over \gamma} +\bm J \cdot \bm \nabla \alpha,
      \label{GibCur}    \end{eqnarray}   
which has a minimum at the specified values of the precession $P=M(1-u_0)$ and the gradient $\nabla \alpha_0$.  Now we must find the energy increase due to fluctuations. It is easy to check that fluctuations of $\Phi$ always increase $\tilde G$, so it suffices to retain in the fluctuation energy only terms quadratic in small deviations $u'=u-u_0$ and $\nabla \alpha'=\nabla \alpha -\nabla \alpha_0$ from the stationary values $u_0$ and $\nabla\alpha_0$ (we omit hereafter the subscript 0):
\begin{eqnarray}
\delta \tilde G =A(u){(\nabla\alpha')^2\over 2}+A'(u) u' \nabla\alpha'+ {1\over 2}\left[V''(u)+A''(u){\nabla \alpha^2\over 2}\right]{u'^2\over 2}.
          \end{eqnarray}   
The fluctuation energy is positive definite, i.e., the current state is stable, so long as $\nabla \alpha$ does not exceed the critical value
\begin{eqnarray}
\nabla \alpha_c =\left[A(u)V''(u)\over A'^2 (u) -A''(u)A(u)/2\right]^{1/2}={4\Omega\over 5\sqrt{3}}\left\{c_\parallel^2 (1-u) ^2+c_\perp^2 (1-u^2)\over [c_\parallel^2+(c_\perp^2-c_\parallel^2) u] ^2+c_\perp^4 /3\right\}^{1/2}.  
 \label{CrLan}                   \end{eqnarray}   
This expression yields the critical gradient on the order of the inverse dipole length $1/\xi_d = \Omega /c_\perp$, but it is valid only  at $u<-1/4$. At $u>-1/4$ ($\beta<104^\circ$) the dipole energy vanishes, $\nabla \alpha_c$, and the superfluid precession transport is impossible. 

Another definition of the critical gradient was suggested by Fomin \cite{Fom-87}: He  believed that the spin-current state can be stable as far as the gradient does not exceed the value  $\nabla \alpha_c=\sqrt{(\omega_P-\omega_L)\omega_L}/c_\perp$, which is  the maximum gradient at which equation~(\ref{curSt}) for $u$  has a solution. Fomin's theory allows stable supercurrents for $\beta <104^\circ$  when 
the dipole energy and the Landau critical gradient  vanish.  Arguing in favor of his  critical gradient,  Fomin \cite{Fom-88} stated that the Landau criterion is not necessary for the superfluid spin transport since emission of spin waves, which comes into play after exceeding the Landau critical gradient, is not essential in the experimental conditions (see also the similar conclusion after equation  (2.39) in the  review by  Bunkov \cite{Bun}). This argument is conceptually inconsistent. If  the experimentalists observed ``dissipationless'' spin transport simply  because dissipation were weak,  they would deal with ballistic  rather than superfluid transport. An example of ballistic spin transport will be considered in section~\ref{bal}. As was stressed in section \ref{intro}, the essence of the phenomenon of superfluidity is not the absence of sources of dissipation, but ineffectiveness of these sources due to energetic and topological reasons. The Landau criterion is an absolutely necessary condition for superfluidity.  Fortunately for the superfluidity scenario in the $^3$He-B, Fomin's estimation of the role of dissipation by spin-wave emission triggered by violation of the Landau criterion is not conclusive. He found that this dissipation  is weak compared to dissipation by spin diffusion.  But this is an argument in favor of importance rather than unimportance of the Landau criterion. Indeed, spin-diffusion, whatever high the diffusion coefficient could be, is ineffective in the subcritical regime, in which the gradient of  the ``chemical potential'' is absent. On the other hand, in the supercritical regime  the ``chemical potential'' is not constant anymore and this triggers the strong spin-diffusion mechanism of  dissipation.

\subsection{Experimental evidence of the superfluid spin-precession transport}

\begin{figure}
\begin{center}
\begin{minipage}{100mm}
{\resizebox*{7cm}{!}{\includegraphics{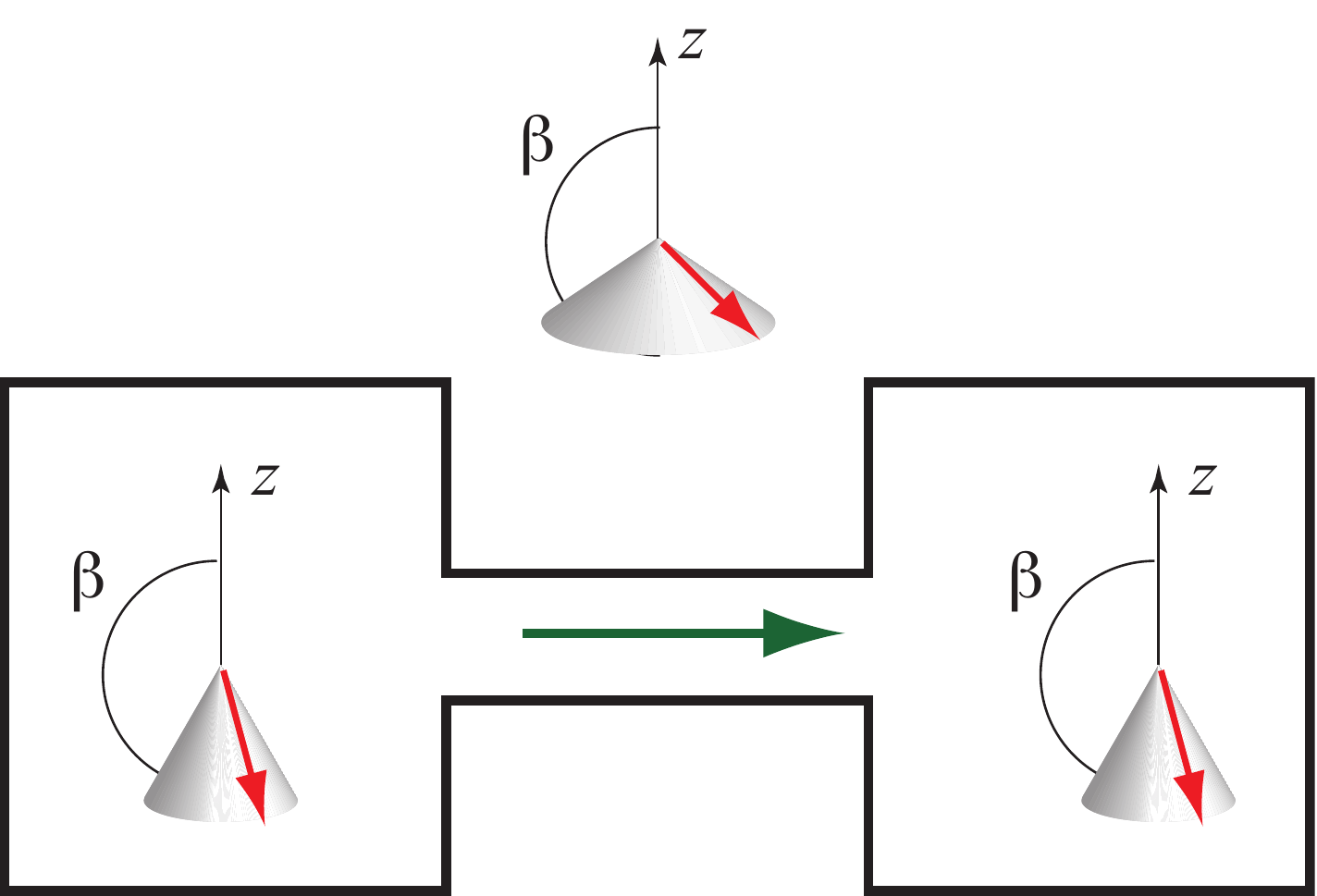}}}%
\caption{Spin-precession transport through a channel connecting  two cells filled by HPD. The horizontal  arrow shows the direction of the spin-precession current in the channel. The precession angle $\beta$ in the channel is less than in the cells (the analogue of the the Bernoulli effect, see the text). }%
\label{fig4.1}
\end{minipage}
\end{center}
\end{figure}

As was discussed above, the appearance of spin current itself is not yet a manifestation of spin superfluid transport. Supercurrents appear in spin waves or domain walls where they transport spin on distances of the order of wavelength or width of domain walls, but hardly it would be reasonable to call it superfluid transport. Similarly spin currents in the domain wall separating HPD from the bulk without precession cannot be a manifestation of superfluid transport. A convincing evidence of spin superfluidity would be spin transport on {\em long} distances. This evidence was presented by Borovik-Romanov et al. \cite{flux} studying spin current through a long channel connecting two cells filled by HPD. The schematic set up of their experiment is shown in figure~\ref{fig4.1}. There is a dc magnetic field parallel to the vertical axis $z$. 
The HPDs in the two cells were supported with independently rotating rf magnetic fields and with different precession phases as a result of it.  A small difference in the frequencies of the two rf fields leads to a linear growth of  difference of the precession phases $\alpha$ in the cells. This creates a phase gradient $\nabla \alpha$ in the channel accompanied by a spin-precession supercurrent. The rf coils can monitor precession phases in different parts of the set up. Due to a linear growth of $\nabla \alpha$ in time eventually it reaches the critical value at which a 2$\pi$ phase slip occurs. It is possible to register this event via its effect on NMR absorption \cite{Bun}.  Thus the critical gradient can be measured as a function of the precession frequency.

Despite aforementioned conceptual flaw of Fomin's theory, Borovik-Romanov et al. \cite{flux,crCur} found an agreement of this theory with  the experiment. Let us compare now the experiment with the theory based on the Landau criterion.  Equation~(\ref{CrLan}) for the Landau critical argument  contains the value $u=\cos \beta$ inside the channel, which is different from $u$ in the cells where there is no precession phase gradients $\nabla \alpha$ (see figure \ref{fig4.1}). This is an analogy with the  Bernoulli law in hydrodynamics (liquid density is less in areas with higher currents). The value of $u$ in the channel grows with  $\nabla \alpha$ according to equation~(\ref{curSt}) at fixed  precession frequency $\omega_P$. The latter is controlled in the experiment and in stationary states does not vary in space, exactly like  the chemical potential  in stationary states of superfluids.  If
\begin{eqnarray}
(\omega_P-\omega_L)\omega_L< {A'(-1/4)\over 2}{\gamma^2\over \chi } (\nabla \alpha_c)^2,
    \label{Lan}      \end{eqnarray}   
where $\nabla \alpha_c$ is the critical gradient from equation (\ref{CrLan})  at $u=-1/4$, $u$ reaches the value  -1/4 earlier than $\nabla \alpha$ reaches the Landau critical gradient. Since no stable supercurrent is possible at $u>-1/4$  the critical argument is determined as a solution of equation~(\ref{curSt}) at $u=-1/4$:
\begin{eqnarray}
\nabla \alpha_c =\left[{4\omega_L(\omega_P-\omega_L) \over 5c_\parallel^2-c_\perp^2}\right]^{1/2}.
    \label{CrLan'}       \end{eqnarray}   
The experiment was done at small $\omega_P-\omega_L$, and  its results must be compared with  equation~(\ref{CrLan'}). For the ratio $c_\parallel^2/c_\perp^2=4/3$ \cite{Fom-88} the latter gives the value of  $\nabla \alpha_c$ by the numerical factor $\sqrt{12/17}\approx 0.84$ smaller than Fomin's result, which was in about 1.5 times larger than the critical gradient in the experiment. Thus the theory based on the Landau criterion even better agrees  with the experiment \cite{flux,crCur}, and  an approximate  agreement of Fomin's result with the experiment cannot be used as an argument in its favor. At larger values of $\omega_L-\omega_P$,  when the condition (\ref{Lan}) is not satisfied, the critical argument is determined by (\ref{CrLan})  and not proportional  to $\sqrt{(\omega_P-\omega_L)\omega_L}$. So   the difference with Fomin's theory becomes more essential (for more details see reference \cite{ES-88}).

Further important development in experimental studies of spin-precession superfluidity was observation of a spin-current analog of the Josephson effect \cite{Joseph}. The weak link was formed by making a constriction of the channel  (orifice) connecting the two cells.

\subsection{Spin-precession vortex and its nucleation}

As was already discussed in section~\ref{stab}, at gradients less than the Landau critical gradient, the barrier, which impedes the current decay, is related to  vortex motion across the flow streamlines (phase slips). Similarly,  vortices called spin-precession vortices appear in the spin-precession flow, and the vortex core radius was estimated to be on the order of the dipole length $\xi_d  $ \cite{ES-87}.  The barrier  for vortex growth in the phase-slip process vanishes at phase gradients of the order of the inverse core radius. So the threshold for vortex instability agrees with the critical gradient from the Landau criterion [equation~(\ref{CrLan})]. This is usual in the conventional superfluidity theory \cite{ES-82}.

Later Fomin \cite{Fom-88} showed that the vortex core must be determined by another scale  $\xi_F=c_\perp/\sqrt{(\omega_P-\omega_L)\omega_L}$, where  $\omega_P$ and $\omega_L$ are 
 the precession and the Larmor frequencies.  This was supported by Misirpashaev and Volovik \cite{misir}  on the basis of the topological analysis. According to equation~(\ref{curSt}) in  the ground state without spin currents  $(\omega_P-\omega_L)\omega_L=16\Omega^2 |u+1/4| /15$. So if $u$ is not too close to -1/4 $\xi_d=c_\perp/ \Omega$ and $\xi_F$  are of the same order of magnitude. But if $u\to -1/4$, i.e. the precession angle $\beta$ approaches to the critical value $\beta_c=1.82$ rad (or $104^\circ$)  the core radius becomes  $r_c \sim \xi_F \sim \xi_d/(\beta-\beta_c)$, i.e., by the large factor $1/(\beta- \beta_c)$ differs  from the earlier  estimation $r_c \sim \xi_d$ \cite{ES-87}. So the latter is valid only far from the critical angle, where $\beta- \beta_c \sim 1$.  Since no barrier impedes vortex expansion across a channel if the gradient is on the order of $1/r_c$, the large core $r_c \sim \xi_d/(\beta-\beta_c)$ at  $\beta \to  \beta_c$   leads to the strange (from the point of view of the conventional superfluidity theory) conclusion: The instability with respect to vortex expansion occurs at the phase gradients $\sim 1/r_c$ essentially less than the Landau critical gradient $\sim 1/\xi_d$, obtained for {\em any} $\beta>\beta_c$.  Recently a resolution of this paradox was suggested \cite{ES-08}:  At precession angles close to 104$^\circ$ at phase gradients less than the Landau critical gradient but larger than the inverse core radius,  
no barrier impedes phase slips at the stage of vortex motion across streamlines, but there is a barrier, which blocks phase slips on the very early stage of nucleation of the vortex core. So for these gradients stability of current states is determined not by vortices but by vortex-core nuclei.

It should be stressed that, in contrast  to the previous subsection, where the growth of $\nabla \alpha$ was accompanied by the growth of $u$ at fixed $(\omega_P-\omega_L)\omega_L$, the present analysis is performed at $\beta=\arccos u$ fixed in the channel  excepting an area a vortex core or its nucleus.  It never exactly equal to $\beta_c$ though  $\beta-\beta_c$ could be whatever small. Thus $(\omega_P-\omega_L)\omega_L$ grows with $\nabla \alpha$. Vortex nucleation starts from a "protonucleus", which is a slight localized depression of the superfluid density [determined by $A(u)$ in our case]. The nucleus, which is related with a peak of a barrier, corresponds to an extremum (saddle point)  of the Gibbs potential given by equation~(\ref{GibCur}). 
Therefore, the nucleus  structure should be found from  solution of  the Euler-Lagrange equations for this Gibbs potential. The first step is to vary the Gibbs potential with respect to $\alpha$. Let us restrict ourselves with a 1D problem, when the distribution in the nucleus depends only on one coordinate $x$. Then the distribution of $\nabla \alpha$ is given by
$\nabla \alpha=-J/A(u)$, where $J$ is equal to the  spin-precession current $J= -A(u_\infty)\nabla  \alpha_0$, which  is determined by the gradient $\nabla  \alpha_0$ far from the nucleus center.  Expanding with respect to small deviation $g=u-u_\infty$ from the fixed value  $u_\infty$ at infinity one obtains 
\begin{eqnarray}
\tilde G=-{J^2\over 2A(u)}+{\chi c_\perp^2 \over  \gamma^2}\left[{(\nabla u)^2\over 2(1-u^2)} +{u\over \xi_F^2}\right]+V(u)
\nonumber \\
\approx {\chi c_\perp^2 \over  \gamma^2} {(\nabla g)^2\over 2(1-u_\infty^2)}+g\left\{{d\over du}\left[-{J^2\over 2A(u_\infty)}+V(u_\infty)\right]+{\chi c_\perp^2 \over  \gamma^2}{1\over \xi_F^2} \right\}
\nonumber \\ 
+{g^2\over 2}{d^2\over du^2}\left[-{J^2\over 2A(u_\infty)}+V(u_\infty)\right]
+{g^3\over 6}{J^2\over 2}{d^3A(u_\infty)^{-1}\over du^3},
        \end{eqnarray} 
where we took into account that $d^3V(u)/du^3=0$. The linear in $g$ terms must vanish at the stationary current state. The term quadratic in $g$ determines the stability of the current state: it vanishes at the Landau critical current
\begin{eqnarray}
J^2_c=2 {d^2V(u_\infty)\over du^2}\left\{{d^2[A(u_\infty)^{-1}]\over du^2}\right\}^{-1}.
        \end{eqnarray} 
This is exactly the Landau critical argument $\nabla\alpha_c$, which was determined in section~\ref{SPV} [equation~(\ref{CrLan})].  Considering the case of the current close to the critical value and using the Taylor expansion of $A^{-1}(u)$ around $u=-1/4$ one obtains
\begin{eqnarray}
\tilde G  = {16\chi c_\perp^2\over 15\gamma^2}\left[{(\nabla g)^2\over 2}  + a { g^2\over 2}-b{ g^3\over 6}\right],
        \end{eqnarray} 
where
\begin{eqnarray}
a=0.239\left({\gamma^2  \over    \chi c_\perp^2}\right)^2(J_c^2-J^2),~~b=0.577\left({\gamma^2  \over    \chi c_\perp^2}\right)^2J_c^2.
        \end{eqnarray} 
The Euler-Lagrange equation for this Gibbs potential, $-\Delta g=ag-bg^2/2=0$,  determines the distribution of $g$: 
\begin{eqnarray}
g=g_0 \left(1- \tanh^2{x\over r_p}\right), 
        \end{eqnarray} 
where $g_0= 3a/b=1.24 (J_c^2-J^2)/J_c^2$ is the value of $g$ in the nucleus center and $r_p=2/\sqrt{a}=4.1  \chi c_\perp^2/\gamma^2\sqrt{J_c^2-J^2}$ is the nucleus size. The energy of the nucleus,
\begin{eqnarray}
\epsilon= {16\chi c_\perp^2S\over 15\gamma^2} \int_0^{3a/b}\sqrt{ag^2- {b g^3\over 3}} dg={64\chi c_\perp^2\over 25\gamma^2}{a^{5/2} \over b^2}S
=0.214S{(J_c^2-J^2)^{5/2} \over J_c^4},
  \label{energ}      \end{eqnarray} 
 determines the barrier for the process of the vortex core nucleation. Here $S$ is the cross-section area of the channel. Since in the limit $J\to J_c$ the nucleus size $r_p$ diverges,  our 1D description is always valid  close enough to the critical point, where  $r_p \gg \sqrt{S}$. When $r_p$ becomes smaller than the transverse size of the channel, one should consider the 3D or 2D (in the case of a thin layer) nucleus.  The first stage of this problem is to find the distribution of $\nabla \alpha$ from the continuity equation  $\bm \nabla [g \bm \nabla \alpha]=0$. Its solution demonstrates that outside the nucleus the distribution of $\nabla \alpha$ is the same as around the vortex ring (3D case) or the vortex dipole (2D case). In particular, in the 2D case
\begin{eqnarray}
\bm \nabla \alpha =\bm \nabla \alpha_0 -\int_0^\infty g(r_1)r_1^2\,dr_1 \left[ {\bm \nabla \alpha_0\over r^2} -{2\bm r (\bm r\cdot \bm \nabla \alpha_0)\over r^4}\right]. 
        \end{eqnarray} 
In contrast to the 1D case, the relation between $\nabla \alpha$ and $g$ is not local, so the following variation of the Gibbs potential with respect to $g$ leads to an integro-differential  equation. However  using the scaling arguments one may conclude that the nucleus energy can be roughly estimated from the expression   (\ref{energ}) for the 1D case with replacing $S$ by $r_p^2$ or by $r_pL$ for the 3D and the 2D cases respectively ($L$ is the  thickness of the 2D layer). 

This analysis demonstrates an unusual feature of the superfluid spin-precession transport at the precession angle close to the critical angle 104$^\circ$: A bottleneck of the  phase slip process is not connected with  expansion of already formed (i.e., with sizes exceeding the core size) vortices  but with the early stage of vortex nucleation. The spin-precession vortex in $^3$He-B was detected experimentally \cite{sVor,Bun-08b}.

\section{Ballistic spin transport by magnons} \label{bal}

Another interesting case of spin transport in magnetically ordered system is connected with magnons \cite{ML,Tra}.  This is an analogue of the normal mass current in superfluids, which arises due to transport of mass by quasiparticles (e.g., phonons at low temperatures).  Let us find first what  contribution to the spin current comes from one magnon. The simplest case is a magnon in an isotropic ferromagnet. Metastable spin-current states are impossible in such a ferromagnet but it is not essential for now: We look for a  ``normal'' spin current. 

We consider an isotropic ferromagnet subject to a magnetic field $\bm H$ with the free energy density
\begin{eqnarray}
F=-\bm M\cdot \bm H+{\alpha \over 2} \nabla_i \bm M\cdot \nabla_i \bm M.
              \end{eqnarray}
The free energy is invariant with respect to rotations around the magnetic field $\bm H$ (the $z$ axis), so in accordance with Noether's theorem  the $z$ component of spin is conserved.

In the homogeneous ground state the spontaneous magnetization   $\bm M_0$ is parallel to $\bm H$.  Linearizing the Landau-Lifshitz equation (\ref{LLP}) with respect to small perturbation $\bm m=\bm M-\bm M_0$ ($\bm m \perp \bm M_0$) one obtains 
\begin{eqnarray}
{1\over \gamma}{d \bm m \over dt} =- \alpha \nabla_i [\bm M_0 \times \nabla_i \bm m]+ \left[\bm H \times \bm m\right] .
              \end{eqnarray}
This equation yields spin waves $\propto e^{i\bm k \cdot \bm r-i\omega t}$ with the spectrum
\begin{eqnarray}
\omega(\bm k,H) = \gamma (H +\alpha M_0  k^2),
              \end{eqnarray}
which in contrast to the spectrum (\ref{IFsp}) without magnetic field, has a gap $\gamma H$ equal to the frequency of the ferromagnetic resonance.

In the linear approximation the spin current $\bm J^z$, which is determined by equation~(\ref{spCur}), vanishes after averaging over the period of the wave. So we need the terms of the second order in $\bm m$, which yield the current
 \begin{eqnarray}
 J_j^z= \alpha \langle  \hat z \cdot [\bm m  \times \nabla_j \bm m]\rangle = \alpha k_j \langle m^2\rangle .
              \end{eqnarray}
The energy density in a single spin-wave mode is $\epsilon = (\omega /2\gamma M_0)\langle m^2\rangle V$, where $V$ is the sample volume. Contributions of a magnon  with the energy $\epsilon = \hbar \omega$ to the squared transverse magnetization and to the spin current are $\delta \langle m^2\rangle= 4 \mu_B M_0/V$ and $\delta \bm J^{z}={\hbar \bm v(\bm k)/ V}$ respectively.  Here  $\mu_B= \gamma \hbar/2$ is the Bohr magneton and $\bm v(\bm k)=d\omega/d\bm k$ is the magnon group velocity. If there is an ensemble of magnons with the distribution function $n(\bm k)$  the total spin current will be 
  \begin{eqnarray}
\bm J^{z}={\hbar\over V}\sum _{\bm k}  n(\bm k)\bm v(\bm k)={\hbar \over (2\pi)^3}\int d\bm k \,n(\bm k) \bm v(\bm k). 
        \label{M-n}      \end{eqnarray}
For  the axisymmetric equilibrium Planck distribution the  spin currents vanishes. But the spin current appears  if the magnon distribution is the Planck distribution with a drift velocity $\bm v_n$:
 \begin{eqnarray}
n_0( \bm k,H)=\frac{1}{e^{\hbar [\omega(\bm k,H) -\bm k\cdot \bm v_n]/T}-1} .
       \label{mag-dr}       \end{eqnarray}
The magnon drift velocity $\bm v_n$ is  an analogue of the normal velocity in a superfluid liquid. Expanding the right-hand side of equation~(\ref{M-n}) in small $\bm v_n$ one obtains the linear relation between the spin current and the drift velocity.  The Plank distribution with a drift is valid only if interaction between magnons is more effective than interaction of magnons with other quasiparticles (electron, phonons), or lattice defects. Otherwise the magnon distribution must be determined from Boltzmann's equation, and the spin current appears only if there is a gradient of the magnetic field $\bm H$, which plays a role of the chemical potential for spin since $\bm H=- \partial F/\partial \bm m$. The spin current proportional to the gradient of $\bm H$ is  accompanied by dissipation and is determined by ``spin-conductivity'' $J^z/\nabla H$.

Another way to obtain the spin current in the magnon system is to connect a quasi-one-dimensional ferromagnetic channel of finite length with two bulk ferromagnets (they act as reservoirs for spin) via ideal contacts \cite{ML}. Magnons cross the channel without scattering (ballistic regime). A spin current appears if there is a difference $\Delta H$ of the magnetic fields in the two leads. The physical picture is similar to that for electron ballistic transport analyzed in the framework of the Landauer-B\"uttiker approach \cite{Dat}:  The ``right-movers'' (magnons moving to right) and ``left-movers'' (magnons moving to left) are described by the equilibrium distribution inside the leads, which magnons come from. Bearing in mind that the wave vector has the only component along the channel the spin current is given by 
 \begin{eqnarray}
J^{z}={\hbar \over 2\pi}\left[\int_0^\infty d k  {d\omega \over dk}n_0(k, H+\Delta H) +\int_{-\infty}^0 d k  {d\omega \over dk}n_0(k, H)\right]={ \mu_B  \over \pi}{ \Delta H \over e^{\mu_B H/2T}-1}. 
\nonumber \\
            \end{eqnarray}
Here $n_0(k, H)$ is the Planck distribution given by equation (\ref{mag-dr}) at $\bm v_n=0$, taking into account the dependence of $\omega$ on $H$. This yields the spin conductance $J^{z} /\Delta H$ obtained by Meier and Loss \cite{ML}. The origin of dissipation is similar to that for ballistic charge transport along a 1D channel \cite{Dat}: There is no dissipation in the channel itself, but magnons arriving to the leads have the distribution different from the equilibrium distribution inside the lead. Its relaxation to the equilibrium leads to dissipation.

It is worthwhile to stress again (see also section~\ref{SPV}) that though dissipation  in the channel is absent in the ballistic regime, this is not superfluidity: In the ballistic regime dissipation is absent because there is no sources of dissipation. 

\section{Equilibrium spin currents in helimagnets} \label{heli}

Though the previous analysis addressed the case of ferromagnets it can be extended on anti- and ferrimagnets. The analysis is also relevant for spin transport in spinor Bose condensates of cold atoms, for which the stability of spin-current states (spiral structures) was also analyzed in the spirit of the Landau criterion \cite{FerCon}. The condition for superfluid spin transport is a proper topology of the magnetic order parameter: the magnetic  order parameter space can be mapped onto a circumference with only weakly broken symmetry (or no broken symmetry at all) with respect to  rotation around the circumference. However, spin  currents in helimagnets still require a special discussion.

The magnetic structure in helimagnets is a spatial rotation of the magnetization (spin) in the easy plane. This structure appears due to Dzyaloshinskii-Moria interaction linear in gradients of magnetization, which break invariance with respect to space inversion. Its energy is given by $\bm D \cdot [\bm M \times [\bm \nabla \times \bm M]]$ \cite{LL}. This energy is relativistically small and can affect only the direction of the magnetization $\bm M$, but not its absolute value $M$. In easy-plane magnets $\bm M$ is fully determined by its angle in the easy-plane, and the free energy is 
\begin{eqnarray}
{\cal F}=\int d^3\bm R\,F =\int d^3\bm R\left\{{ m_z^2\over 2\chi} + {A(\bm \nabla \varphi)^2\over 2}+K[1-\cos (n\varphi)] - D M^2\nabla_z \varphi \right\},
 \label{FEhel}  \end{eqnarray}
where it is supposed that the magnetic spiral is oriented along the axis $z$. This energy differs from the free energy in equation~(\ref{an}) with the term linear on the angle gradient $\nabla_z \varphi$.  

The term linear in the angle gradient does not affect the equations of motion but it is important for definition of the equilibrium magnetic structure: the phase gradient is present in the ground state. It also changes the expression for the spin current  replacing equation~(\ref{cur}) with
\begin{eqnarray}
\bm J^z=- {\partial F \over \partial \bm \nabla \varphi} =- A    (\bm \nabla \varphi -\bm k),
        \end{eqnarray}
where $\bm k=(DM^2/A) \hat z$. The ground state is determined by minimization of the free energy with respect to the average gradient $\langle \nabla \varphi \rangle$, which determines the density $n \langle \nabla \varphi \rangle/2\pi$ of domain walls. Focusing on the limit of low density of domain wall [see equation~(\ref{enDW})] the free energy density  is 
\begin{eqnarray}
F =A\langle \nabla \varphi \rangle\left({ 4\over \pi\sqrt{n}   l}- k\right).
           \label{enDWa}  
                              \end{eqnarray}
Thus at  $k < 4/ \pi\sqrt{n} l$ (strong anisotropy) there is a complete ``phase fixation'', spin being directed along some of the in-plane easy axes, and the ground state is uniform  ($ \nabla \varphi=0$). 
This means that the spin current $\bm J^z= A\bm k$  is present in the ground state. At $k= 4/ \pi\sqrt{n}$ there is the phase transition to the spiral structure with $\langle \nabla \varphi \rangle \neq 0$, which is a chain of solitons of the sine Gordon equation obtained by variation of the free energy.  Near the phase transition the equilibrium spin current is still close to $\bm J^z= A\bm k$. However in the limit $k \gg  1/ l$ the anisotropy energy is not essential and $\nabla \varphi\approx  k$. Then the equilibrium spin current is absent despite the presence of the spiral structure with $\nabla \varphi\neq 0$.  
 
The equilibrium spin currents in helimagnets were calculated by Heurich et al. \cite{McD-2} and by Bostrem et al. \cite{Bos}. In their calculations rotational  invariance in the easy plane was broken by an inplane magnetic field $H_\perp$. This corresponds to the free energy (\ref{FEhel}) with $K=M_0 H_\perp$ and $n=1$.
Our analysis  shows that an  equilibrium spin current is a generic feature of any helimagnet even without an in-plane field since one cannot imagine an easy-plane magnetic material without at least some finite in-plane anisotropy. The phase transition between the phase-fixed state and the the state with a helical structure is a typical example of the commensurate--incommensurate phase transition, which is common in condensed matter physics \cite{Bak}. In particular,  such a transition driven by a in-plane magnetic field was studied experimentally \cite{QHhel} and theoretically \cite{QHhel1} in the quantum Hall bilayers. 

In addition to equilibrium spin currents there are also possible metastable spin currents. Their stability is determined like it was done in section~\ref{stab}, but  in the Landau criterion, $\nabla \varphi < \sqrt{E_A/A}$, and in the expression, which determines the  barrier for vortex expansion [equation~(\ref{barr})], one should replace  the phase gradient  $\nabla \varphi$ with $|\bm\nabla \varphi-\bm k|$. 

The helical structure at the equilibrium (and possibly in the metastable state) is stationary and does not move. However, as discussed in section \ref{real}, spin injection to a medium, which does not support  superfluid spin transport, is impossible without some nonequilibrium magnetization. Then the  helical structure must move, but this motion can be blocked by pinning. The effect of pinning on moving  incommensurate structures is well known \cite{Bak}. It is similar to the  effect of coercivity on motion of domain walls. However, in the limit of high spin currents most effective for dissipationless spin transport pinning is suppressed due to strong overlapping of domain walls. If they overlap so strongly that the phase (angle) gradient does not vary  in space, the pinning force must vanish.

Equilibrium spin currents are possible not only in helimagnets. They were revealed also in spin-Heisenberg  rings  subject to inhomogeneous magnetic fields \cite{SSBS,Kop}. These currents are connected with the geometric phase quantization in ring geometry and, being inversely proportional to the length of the ring,  are possible only in mesoscopic rings. Their origin and properties are similar to mesoscopic currents arising in systems without magnetic order, which will be considered in section~\ref{1Dring}. 

Equilibrium spin currents were considered by a number of researchers as a bizarre  and unphysical outcome. They believed that they are background currents, which do not correspond to real spin transport \cite{R1,R1a,Kop2}. This stimulated attempts to redefine the definition of spin current in order to avoid spin currents in the thermodynamic equilibrium.  For Heisenberg magnets it was done by Sch\"utz et al. \cite{Kop2}. In many aspects confusion about a proper spin-current definition in magnetically order systems reflects the similar problem in systems without magnetic order, and we postpone its discussion for section~\ref{defSC}.

\section{Magnon Bose-Einstein condensation  vs spin superfluidity} \label{MBEC}

Connection of Bose-Einstein condensation (BEC) with superfluidity was debated from the very beginning of the theory of superfluidity. This connection is not straightforward. BEC not necessarily leads to superfluidity as the simple case of the ideal Bose-gas demonstrates. On the other hand, one may explain superfluidity without referring to the concept of BEC as Landau did creating his two-fluid theory. There are debates also what the term BEC really means and how widely one may use it. Nowadays there is a strong tendency to see manifestation of BEC in various cases of formation of coherent states in magnetically ordered systems. The BEC scenario was suggested for  the phase transition to the antiferromagnetic state from the saturated ferromagnetic state in a strong magnetic field in magnetic insulators TlCuClO$_3$ \cite{nook,rueg}, and Cs$_2$CuCl$_4$ \cite{radu}.    In yttrium-iron-garnet films Demokritov et al. \cite{Dem} [see also \cite{Dem2}] observed condensation of magnons in a single state interpreting it in terms of BEC. Finally, Bunkov and Volovik
\cite{Bun-07,Vol,Bun-09}) reinterpreted spin-precession superfluidity in $^3$He-B (see section~\ref{HeB}) in terms of BEC. Dealing with semantics it is difficult  to decree strict rules and to reach a broad consensus. 
Without trying to do it, we shall simply  follow recent debates on the question when one may and when one may not use the label ``magnon BEC''.

I start from a general argument why the term ``magnon BEC'' is at least problematic (but not forbidden!).
Let us remind what did BEC mean at the time of Bose, Einstein, and London when the concept was coined. The quantum theory showed that if identical particles are described by the Bose-statistics, at low temperatures they can condense at the state of the lowest energy. Since all Bose-condensed particles are described by the same wave-function, the BEC yields a new coherent classical field, which cannot be imagined in the classical theory of particles. 
So from this point of view the non-trivial essence of the BEC is appearance of a coherent classical field, which {\em does not exist} in the classical description of particles. Returning now to the magnon BEC, a magnon can be a result of quantization of a classical field (spin waves).   If one claims magnon BEC, this in fact means that ``particles'' magnons can be described by a classical fields. However, this is a logical circle: First we state that we need to quantize a classic field and to introduce quanta of this field, then we tell that we should describe condensed ``particles'' (quanta) by a classical field. So we return exactly to the same classical field, which we started from! In other words, in this special case quantization of spin waves was not necessary. Calling coherent spin waves by a Bose-condensate of magnons, one must allow also to call laser modes ``Bose-condensates of photons", or deformed states of solids ``Bose-condensates of phonons". Analogy between Bose-condensates and coherent states was well known long ago \cite{Haken}. But analogy does not mean identity. Some reasonable restriction should be imposed on using the term BEC. It seems that more careful is to use this term only referring to objects, which can be particles in the classical limit. This means that coherent sound (phonons), electromagnetic (photons) or spin (magnons) waves are not Bose-condensates, but excitons in semiconductors are. However, no clever classification can avoid ``gray areas'', where it fails to provide definite conclusions. Can one call coherent polariton states as Bose-condensates keeping in mind that the polariton is a superposition of an exciton and a photon? I leave this question without an answer.

Putting aside this argument as too rigorist let us discuss   various claims of ``magnon BEC''. 
They can be classified as equilibrium phenomena, or non-equilibrium ones, which require energy  pumping for their observation. Let us start from the equilibrium case of the phase transition to the antiferromagnetic state from the saturated ferromagnetic state in a strong magnetic field in magnetic insulators TlCuClO$_3$ \cite{nook,rueg}, BaCuSi$_2$O$_6$\cite{BaCu},  and Cs$_2$CuCl$_4$ \cite{radu}.  The theoretical basis for it was known from 80s \cite{BB} (see also \cite{Giam}): The transition from the ferromagnet (all spin parallel to the magnetic field) to the two-sublattice antiferromagnet with spins of sublattices in the plane can be described in terms of magnon  Bose-operators in the ferromagnetic state. This transition is one from numerous examples of orientational phase transitions in magnetism. If the BEC label is used for this case it could be used for any other. More generally, any transition described in terms of soft-mode instability  (e.g., structure transitions in solids) may be considered as BEC. Then the phenomenon of BEC loses its uniqueness. Reservations concerning using the term ``magnon BEC'' for equilibrium phenomena were expressed in a number of publications \cite{Mills,Vol}.

Another class of magnon BEC is non-equilibrium states, which require energy pumping for their support.   There are two ways to support a coherent non-equilibrium state. The first one  is to pump energy at the same frequency with which  the phase of the coherent state varies in time. An example of such non-equilibrium coherent states is the coherent precession in $^3$He-B discussed above in section~\ref{HeB}. In these cases there is a magnetic field rotating with the same frequency as the precession angular velocity. Snoke \cite{Snoke} called these examples of coherent states ``driven condensates''  suggesting that in such cases the word ``condensate''  was not appropriate and contradicted to the spirit of the term ``condensate''  as a spontaneous thermodynamic phenomenon. Putting aside this argument for a while and accepting the view of Bunkov and Volovik \cite{Bun-07,Vol,Bun-09}, who considered the coherent spin precession as a BEC phenomenon, the question arises why they do it 
only with respect to the coherent precession in $^3$He-B. The coherent precession takes place in the nonlinear ferromagnetic or antiferromagnetic resonances, which were studied more than 50 years. Already at that time it was well realized that precessing magnetization may be treated purely classically as a gigantic magnetic dipole, which emits coherent radiation similar to Dicke superradiance \cite{Blom}. Later experimental evidence of this radiation was reported \cite{RadDam}. Following the semantic criteria of  Bunkov and Volovik one may call it ``discovery of magnon BEC''.


Really spontaneous magnon condensation was observed by Demokritov et al. \cite{Dem} in yttrium-iron-garnet films: they pumped microwave energy at higher magnon modes and observed accumulation of magnons in the lowest-energy magnon state, which corresponded to nonzero wave vector. It is important that pumped magnons were incoherent \cite{Chum}, so ensuing accumulation in a single state was   spontaneous indeed.
Earlier the phenomenon was discussed theoretically by Kalafati and Safonov \cite{Kal}. Condensation requires some threshold pumping level. Accepting undisputed similarity of this phenomenon with BEC,  it is  worthwhile to draw attention to another analogy, which is not less ``intriguing'' than the BEC analogy: the analogy with lasers.  
Keldysh \cite{Keld} defined ``lasing'' in his review on the exciton BEC (see   p. 255 there): ``The accumulation of a macroscopic number of initially incoherent excitation quanta in a single-photon mode is lasing''. Replacing ``single-photon mode'' by ``single-magnon mode'' one can apply this definition to the experiment of Demokritov et al.  \cite{Dem}. On the basis of this analogy and since magnons are coupled with electromagnetic waves, one should expect coherent microwave radiation similar to laser radiation. This radiation was really observed recently by Dzyapko et al. \cite{Rad}. Radiation results from confluence of two magnons with opposite wave vectors (one-magnon radiation is forbidden by the momentum conservation law). It was demonstrated that the frequency of radiation is determined by the condensed magnon frequency but not the frequency of the pumping signal. Description of  magnon accumulation in a single state in terms of coherent magnon states was suggested many years ago  \cite{RezCoh} by analogy with coherent photon states in lasers. On the basis of this idea recently Rezende  \cite{Rez} developed the theory of coherent microwave radiation detected by Dzyapko et al. \cite{Rad}.
If it were possible to check coherence of this radiation, it would be a demonstration of {\em magnon laser}.

In order to distinguish between BEC and laser, it was suggested to use BEC trademark only if there is a quasi-equilibrium distribution of non-condensed magnons  with a well defined chemical potential that requires an effective magnon thermalization \cite{Vol} (in lasers  non-condensed photons are not in equilibrium). As one of possible semantic restrictions on using the term BEC, this criterion of BEC is formally legitimate. Demokritov et al. \cite{Dem3} experimentally demonstrated that magnon condensation in the momentum space  in yttrium-iron-garnet films was accompanied with the quasi-equilibrium Bose-Einstein distribution of magnons.  On the other hand, since the property of coherence  is  related only with condensed magnons  it is difficult to understand why  the distribution function of non-condensed magnons  is so crucial.

Up to now we discussed the property of coherence but not of superfluidity. It is necessary to stress that coherence does not lead automatically to superfluidity as sometimes assumed \cite{Bun-09}. An example of the coherent state without superfluidity was presented in section~\ref{stab}:  an  (anti)ferromagnet fully isotropic, or with uniaxial easy-axis anisotropy. There is long-range correlation of spins in these cases, but no metastable spin-current states. The property of superfluidity is determined by topology of the order parameter for a concrete type of a condensed state. Let us check it for various BEC cases considered in the present section. The equilibrium phase transition in strong magnetic fields from ferromagnetic to antiferromagnetic ordering certainly leads to potential possibility of superfluid spin transport: In the antiferromagnetic phase the strong magnetic field keeps the magnetizations of the sublattices in the plane normal to the field. So spiral structures with the antiferromagnetic vector rotating in the plane can be stable and provide dissipationless spin currents. The precession states in $^3$He-B also can be accompanied by stable spin-precession currents as was demonstrated theoretically and experimentally (section~\ref{HeB}). As for magnon condensation in yttrium-iron-garnet films \cite{Dem}, one cannot expect superfluid spin transport in this case since there is no easy plane for order-parameter rotation in the experimental geometry (the magnetic field was parallel to the film). However, if the modification of the experimental geometry (the magnetic field normal to the film) recently suggested  by Tupitsyn et al. \cite{Tup} were realized, the question on possible superfluid spin transport would become relevant.

\vspace{1cm}
\centerline{\bf \Large Part II:Spin currents without magnetic order}

\section{Definition of spin current} \label{defSC}
 
Up to now we considered spin currents, which are possible only in magnetically ordered media. Meanwhile, a lot of attention was devoted to other types  of dissipationless spin currents, which can appear without magnetic ordering. They are analogues of persistent charge currents in normal metals \cite{Imry} and can appear even in equilibrium. Certainly in this case they cannot relax and are genuinely persistent, in contrast to superfluid spin currents, which are only metastable.

We shall analyze the spin transport in normal systems using the simplest model: noninteracting electrons. Then one can start from a single electron in magnetic ($\bm H$) and  electric ($\bm E$) fields, which are not uniform in general. The electron wave function is a two-component spinor
\begin{eqnarray}
\mathbf{\Psi}=\left( \begin{array}{c} \psi_\uparrow\\ \psi_\downarrow\end{array}\right).
               \end{eqnarray}
We need the Hamiltonian with relativistic corrections \cite{BLP}:
  \begin{eqnarray}
\hat {\cal H}={\hbar^2 \over 2m}\left(-i \bm \nabla  -{2\pi \bm A \over \Phi_0}-\lambda [\bm \sigma \times \bm E] \right)^2 -\mu_B \bm H \cdot \bm\sigma ,
  \label{hamG}   \end{eqnarray}
where $\Phi_0=hc/e$ is the single-electron magnetic-flux quantum and $\bm\sigma $ is the vector of Pauli matrices. In vacuum the constant $\lambda$ is equal to $e/4mc^2$, but in crystals it can be much larger \cite{ERH}.  The electric field $\bm E$ may include not only an external field but also crystal fields. The crystal electric field is possible in 3D systems with broken space-inversion symmetry. In 2D electron systems on  semiconductor surfaces the effective electric field normal to the surface can result from asymmetry of the electrostatic potential confining the 2D system to the surface (structure inversion asymmetry). This leads to Rashba spin-orbit coupling (section   \ref{eqCur}). Since the Pauli matrices do not commute, the spin-orbit term proportional to $\bm E$ can be considered as a non-Abelian gauge field similar to the Yang-Mills fields \cite{Fro}. 

Writing down the Schr\"odinger equation for the Hamiltonian (\ref{hamG}),
  \begin{eqnarray}
i\hbar{\partial \mathbf{\Psi} \over \partial t}=\hat {\cal H} \mathbf{\Psi},
     \end{eqnarray}
one can derive the continuity equations for charge:
  \begin{eqnarray}
e{\partial (\mathbf{\Psi}^\dagger \mathbf{\Psi}) \over \partial t}=-\bm \nabla \cdot \bm j,
 \label{chBal}    \end{eqnarray}
and for spin:
  \begin{eqnarray}
{\partial S_\beta \over \partial t}={\hbar\over 2}{\partial (\mathbf{\Psi}^\dagger \sigma_\beta \mathbf{\Psi}) \over \partial t}=-\bm \nabla \cdot \bm j^\beta+G_\beta ,
  \label{SB}   \end{eqnarray}
with the following expressions for the currents:
  \begin{eqnarray}
\bm j={ e\hbar  \over m}\left\{-{i   \over 2}( \mathbf{\Psi}^\dagger \bm \nabla \mathbf{\Psi} -\bm \nabla \mathbf{\Psi}^\dagger  \mathbf{\Psi})-{2\pi \bm A  \over  \Phi_0 }(\mathbf{\Psi}^\dagger \mathbf{\Psi})-\lambda  [(\mathbf{\Psi}^\dagger \bm \sigma \mathbf{\Psi}) \times \bm E]\right\},
     \label{cur-ch}\end{eqnarray}
  \begin{eqnarray}
\bm j^\beta={ \hbar^2  \over 2m}\left\{-{i   \over 2}( \mathbf{\Psi}^\dagger\sigma_\beta \bm \nabla \mathbf{\Psi} -\bm \nabla \mathbf{\Psi}^\dagger \sigma_\beta \mathbf{\Psi})-{2\pi \bm A  \over \Phi_0 }(\mathbf{\Psi}^\dagger\sigma_\beta \mathbf{\Psi})
\right. \nonumber \\ \left.
-{\lambda\over 2}  [(\mathbf{\Psi}^\dagger (\sigma_\beta\bm \sigma+\bm \sigma  \sigma_\beta)\mathbf{\Psi}) \times \bm E]\right\}.
    \label{cur-sp}\end{eqnarray}
Since spin is not conserved, there are source terms (torques) in the spin-balance equations:
 \begin{eqnarray}
G_\beta= -{\gamma \hbar  \over 2}[\bm H \times (\mathbf{\Psi}^\dagger \bm \sigma \mathbf{\Psi})]_\alpha
-{i\hbar \lambda\over 2}\{\mathbf{\Psi}^\dagger [\bm \sigma \times [\bm E \times (\bm \nabla- 2\pi i\bm A/\Phi_0) ]]_\beta \mathbf{\Psi}
\nonumber \\
+[[\bm E \times (\bm \nabla+ 2\pi i\bm A/ \Phi_0)\mathbf{\Psi}^\dagger] \times \bm \sigma]_\beta\mathbf{\Psi}\}.
     \end{eqnarray}
 The spin current,  equation~(\ref{cur-sp}),  agrees with the current definition used by Rashba \cite{R} and many  others:
 \begin{eqnarray}
\bm j^\beta   ={\hbar \over 4}(\mathbf{\Psi}^\dagger\{ \sigma_\beta  \bm  v + \bm  v \sigma_\beta \}\mathbf{\Psi}),
      \label{spDef}       \end{eqnarray}   
where 
\begin{eqnarray}
\bm v  ={\hbar\over m}\left(-i \bm \nabla -{2\pi \bm A \over \Phi_0}- \lambda  [   \bm  \sigma  \times  \bm E ]\right) 
           \end{eqnarray}  
is the operator of the electron group velocity.

One may write down the spinor wave function $\mathbf{\Psi}$ as 
\begin{eqnarray}
\mathbf{\Psi}=\sqrt{n}\left( \begin{array}{c} \cos {\gamma \over 2}e^{i\phi_\uparrow} \\ \sin {\gamma \over 2}e^{i\phi_\downarrow}\end{array}\right),  
               \end{eqnarray}
where $n= (\mathbf{\Psi}^\dagger\mathbf{\Psi})$ is the electron density, $\gamma$ is the  tipping angle of the average spin with respect to the $z$-axis, $\phi_\uparrow$  and $\phi_\downarrow$ are the phases of the two spinor components.  Note that here we consider a single-electron wave function, and the electron density is normalized to unity: $\int n(\bm R)\,d\bm R=1$. 
Introducing the global phase $\phi={1\over 2}(\phi_\uparrow+\phi_\downarrow)$  and the relative phase $\phi_z= \phi_\downarrow-\phi_\uparrow$ the averaged spin depends on the latter, which is the angle of rotation around the $z$ axis in the spin space:
\begin{eqnarray}
\langle \sigma_x \rangle= {(\mathbf{\Psi}^\dagger\sigma_x\mathbf{\Psi})\over n}= \sin \gamma \cos \phi_z,~ 
\langle \sigma_y \rangle = {(\mathbf{\Psi}^\dagger\sigma_y\mathbf{\Psi})\over n}=\sin \gamma \sin \phi_z,
\nonumber \\   
\langle \sigma_z \rangle = {(\mathbf{\Psi}^\dagger\sigma_z\mathbf{\Psi})\over n}= \cos \gamma.
            \end{eqnarray}

In the variables $n$, $\gamma$, $\phi$, and $\phi_z$ the electron (free) energy is given by 
 \begin{eqnarray}
{\cal F}=\int d^3\bm R\,F
=\int d\bm R\,n\left\{ {\hbar^2 \over 2 m} \left[\cos^2{\gamma\over 2}\left(\bm\nabla\phi_\uparrow-{2\pi \bm A\over \Phi_0}\right)^2
\right.\right. \nonumber \\ \left. \left.
+\sin^2{\gamma\over 2}\left(\bm \nabla\phi_\downarrow -{2\pi \bm A\over \Phi_0}\right)^2\right] 
-\mu_B B \cos \gamma 
+{\hbar^2 \lambda \over 2m}[\hat z \times \bm E]\bm \nabla\phi_z
\right. \nonumber \\ \left.
-{\hbar^2 \lambda \over 2m}[\langle \bm \sigma\rangle  \times \bm E]\cdot (\bm\nabla\phi_\uparrow+\bm\nabla \phi_\downarrow)+{\hbar^2\lambda^2 \over 2 m}E^2+ \epsilon_\nabla(\nabla n, \nabla \gamma)\right\},
       \label{enGen}       \end{eqnarray}  
where the two last terms are not essential for the further analysis. 
The charge and the $z$ spin component currents are:
 \begin{eqnarray}
\bm j={e\over \hbar} {\partial F\over  \partial \bm \nabla\phi}={ne\hbar \over m } \left\{\bm \nabla\phi-{2\pi \bm A\over  \Phi_0}-{\cos \gamma\over 2}\bm \nabla\phi_z-\lambda [\langle\bm \sigma\rangle \times \bm E]\right\},
              \end{eqnarray}   
 \begin{eqnarray}
\bm j^z=-{\partial F\over  \partial\bm \nabla \phi_z}={n\hbar^2 \over 2 m }\left\{
\cos \gamma\left(\bm \nabla\phi-{2\pi \bm A\over  \Phi_0}\right)-{\bm \nabla\phi_z\over 2}-\lambda [\hat z \times \bm E]
\right\} ={\hbar \langle \sigma_z\rangle\over 2e}\bm j+\bm J^z.
\nonumber \\
              \end{eqnarray}   
Here we divided the spin current on two parts. The first one is a convection: the charge current is transferring also the average spin. The second one may be called a {\em pure} spin current, which is determined in the coordinate frame moving together with electrons, i.e., with the velocity $\bm v=\bm j/en$:
 \begin{eqnarray}
\bm J^z={\hbar^2 \over 2 m }n \left\{-{\sin^2\gamma\over 2}
\bm \nabla\phi_z-\lambda [(\hat z-\langle \bm \sigma \rangle \langle \sigma_z\rangle ) \times \bm E]\right\}.
              \end{eqnarray}   
Neglecting the term proportional to the electric field this spin current is proportional to the gradient of the angle $\phi_z$ of the spin rotation around the axis $z$ in accordance with Noether's theorem. The coefficient before the gradient (stiffness) is proportional to the squared inplane component of the spin ($\propto \sin^2 \gamma$) similarly to the  spin current in magnetically ordered media (see the first paragraph of section~\ref{stab}).

Let us discuss various terms in the expressions for charge and spin currents. The terms related to the global-phase gradient $\bm \nabla \phi$ and the electromagnetic vector $\bm A$ are common and do not require any special comment. The terms proportional to the spin-angle gradient $\bm \nabla \phi_z$ are related with the geometrical (Berry) phase   \cite{Bal,Stern,AL}, which results from the transport of spin over a closed path in  a  parametric space, which is the configurational space in our case. For spin ${1\over 2}$ the Berry phase is $ - {1\over 2} \Omega$, where $\Omega$ is the solid angle subtended by the varying spin during the cyclic process. So the Berry-phase ``gradient'' is  $\bm \nabla \Gamma=-{1\over 2}\bm \nabla \Omega=-{1\over 2} (1-\cos \gamma)\bm \nabla\phi_z$. The gradient is in quotation marks because it is a well defined gradient of a scalar function only at constant tipping angle $\gamma$. In general the Berry phase  depends on the path used for its definition. In order to separate the Berry-phase current from that connected with the global phase the latter must be redefined. Introducing $\tilde \phi =\phi_\uparrow$ instead of $\phi={1\over 2}(\phi_\uparrow+\phi_\downarrow)$ the charge current is
 \begin{eqnarray}
\bm j={e\hbar \over m }n \left\{\bm \nabla \tilde \phi-\bm \nabla \Gamma-{\bm A\over 2\pi \Phi_0}-\lambda [\langle\bm \sigma\rangle \times \bm E]\right\}.
              \end{eqnarray}   
It is worthwhile to stress that the Berry phase not only path-dependent but is not single-valued even for a given path. Indeed, one may redefine the solid angle $\Omega$ in the Berry phase $\Gamma =-{1\over 2}\Omega$  replacing $d\Omega =2\pi(1-\cos \gamma) d\phi_z $ by  $d\Omega =-2\pi(1+\cos \gamma) d\phi_z $. So the Berry phase depends on what pole of the sphere was included into the solid angle. The difference  between two values can be accounted for in the  phase $\tilde \phi$. Thus the Berry phase is a part of the global geometric phase, which the electron obtains after moving around the closed trajectory, and separation of the Berry phase from the rest  part of the global phase is not unique.   

The charge and the spin currents also contain a spin-orbit term proportional to an electric field. These terms are a manifestation of the Aharonov--Casher effect \cite{AC} related with the non-relativistic interaction of the magnetic moment (spin in our case) with the electric field. The Rashba spin-orbit interaction (section \ref{RashEq}) and the spin Hall effect (section \ref{shsec}) originate from  this term. It can also be attributed to the Berry phase for the parallel transport in the momentum space \cite{AL}.

We have discussed spin currents in the free-electron model, but the presence of various components of the current is determined by symmetry and topology without being restricted with a particular microscopic model. For example, the Aharonov-Casher effect was  considered in the tight-binding model \cite{BA} and in the presence of random spin-orbit interaction \cite{MGE}. As already stressed in the end of section~\ref{magn}, presumption that spin transport requires mobile carriers of spin \cite{Bun,Niu} is not justified: For phenomenology it does not matter whether spin is transported by itinerant carriers or via exchange interaction  between localized electrons.

There were worries in the literature about ambiguity of the spin-current definition \cite{R1}. Indeed, the definition of the spin current given above [equation~(\ref{spDef})]  is not the only possible choice. One can redefine the spin current by adding to it {\em any} current $\delta j_i^\beta$  ($j_i^\beta \to j_i^\beta +\delta j_i^\beta$), if it is accompanied by redefinition of the spin torque: $G_\beta \to G_\beta +\nabla_i\delta j_i^\beta$. This is a purely formal ambiguity of current definition (like freedom to choose various definitions of potentials in  electrodynamics), which must not lead to any ambiguity in physical predictions. This  only means that any definition of current is not complete without accompanying definition of spin torque. 
In principle, it is possible even to avoid the spin-current term at all, treating  the whole current divergence $\nabla_i j_i^\beta $ as a part of the spin torque.  This would exclude the word {\em spin current} from the scientific lexicon. As discussed in section~\ref{real}, one can avoid using such terms as {\em flow} or {\em current}, and describe the same phenomena using only concepts of deformation, spin stiffness, or torque \cite{ES-82,McD-2,torque}. However, it is not necessary: In many cases the ``current language'' provides a very transparent physical picture of processes in various spin systems. 


There were also attempts to redefine the spin-current so that to eliminate the torque term $G_\beta$ from the spin continuity equation making it to look as a continuity equation for a conserved quantity. For example,   Shi et al. \cite{Niu} (see also the discussion of their work by Bray-Ali and Nussinov \cite{Nuss}) used the definition of a torque dipole term by Culcer et al. \cite{torque} rewriting the torque density as a divergence of a torque dipole density $\bm P_\beta$:
\begin{equation}
G_\beta = -\bm \nabla \cdot \bm P^\beta.
\end{equation}
The torque dipole density was included into  the redefined spin current $\bm  {\cal T}^\beta=\bm j^\beta + \bm P^\beta$. Then the spin continuity equation (\ref{SB}) becomes
 \begin{eqnarray}
{\partial S_\beta \over \partial t}+\bm \nabla \cdot \bm  {\cal T}^\beta=0.
         \end{eqnarray}
However, there is a payoff for this formally legitimate operation: The spin current becomes non-local since it is determined by an integral over the torque distribution in the whole bulk.  Eventually any choice of the spin-current definition must yield the same predictions for observations if the whole procedure is correct. But using nonlocally defined spin currents it is more difficult to study effects related to local processes, which do not conserve the total spin. We shall return back to this issue discussing the observation of spin currents in the equilibrium Rashba medium (section \ref{mechM}).


\section{Equilibrium spin currents in one-dimensional rings}

\subsection{A 1D ring in a homogeneous magnetic field}

The simplest example of the equilibrium spin and charge currents is persistent currents in a one-dimensional ring \cite{Butt,Bal}. Let us consider first a ring of radius $R$ in a constant magnetic field $\bm H=H\hat z$ normal to the $xy$ plane of the ring, which provides the magnetic flux $\Phi= \pi R^2 H$ through the ring.  We neglect the Zeeman energy since its effect (splitting of the Fermi energies for spins up and down) becomes important for the magnetic fields $\sim \Phi_0 n/R$ much higher than the fields $\sim \Phi_0 /R^2$ when the Aharonov-Bohm phase shift $\propto  \Phi $ becomes essential.  Here $n=N/2\pi R$ is the electron density for $N$ electrons in the ring. We consider  zero temperature.  
Introducing the azimuthal angle  $\phi$ for the position of the electron in the ring, the Hamiltonian is
    \begin{eqnarray}
\hat{\cal H}={\hbar^2 \over 2m R^2}\left(-i {\partial \over \partial \phi}-\rho\right)^2 ,
     \end{eqnarray}
where $ \rho=\Phi/\Phi_0$. Taking into account the periodic boundary conditions, the electron spinor components are eigenfunctions of the angular momentum: $\psi_{\uparrow,\downarrow} (\phi)\propto e^{i s \phi}$, where $s$ is an integer. Let us consider electrons with the same spin (up or down).
 The total energy of the Fermi sea and the ground-state persistent currents depend on whether the number of electrons s even or odd. If the number $2p+1$  of electrons is odd their energy is:
\begin{eqnarray}
E ={\hbar^2 \over 2m R^2}\sum _{s=-p}^p (s-  \rho)^2.   
    \end{eqnarray}
This energy corresponds to the ground state only if $-1/2<\rho < 1/2$. In other cases the electron Fermi see should be shifted. For example if $\rho$ becomes larger than 1/2, one must sum $s$ from $-p+1$ to $p+1$. The particle current at $-1/2<\rho < 1/2$ is given by 
 \begin{eqnarray}
j=-{e \over 2\pi \hbar}  {\partial E \over \partial s}=-(2p +1)  \rho{e \hbar \over 2\pi m R^2} \approx -  \rho{ e n \hbar\over 2m R}.
 \label{odd} 
          \end{eqnarray}
This function must be periodically extended on any  $\rho$ beyond  the interval $(-1/2,1/2)$.  This yields the periodic sawtooth dependence of the current on $ \rho$ with the period 1. If the number  $2p$ of electrons is even one should shift this dependence with the half-period 1/2.       

Taking into account the both directions of spin, the result depends on whether the electron number $N$ is even or odd. If $N$ is even the numbers of electrons with spins up and down are  even or odd together. Then the charge current obtained for one value of spin is doubled, whereas the spin current, which is proportional to the difference of charge currents for two values of spin,  vanishes. On the other hand, if $N$ is odd the charge-current dependence  for one  spin direction is shifted with respect to another. This lead to the persistent spin current $j^z=\hbar (j_+-j_-)/2e$ with the period 1, where $j_\pm$ are charge currents for two values of spin. The spin current jumps between values $\pm {\hbar^2  n / 2m R}$ every half-period.

\begin{figure}
\begin{center}
\begin{minipage}{100mm}
{\resizebox*{7cm}{!}{\includegraphics{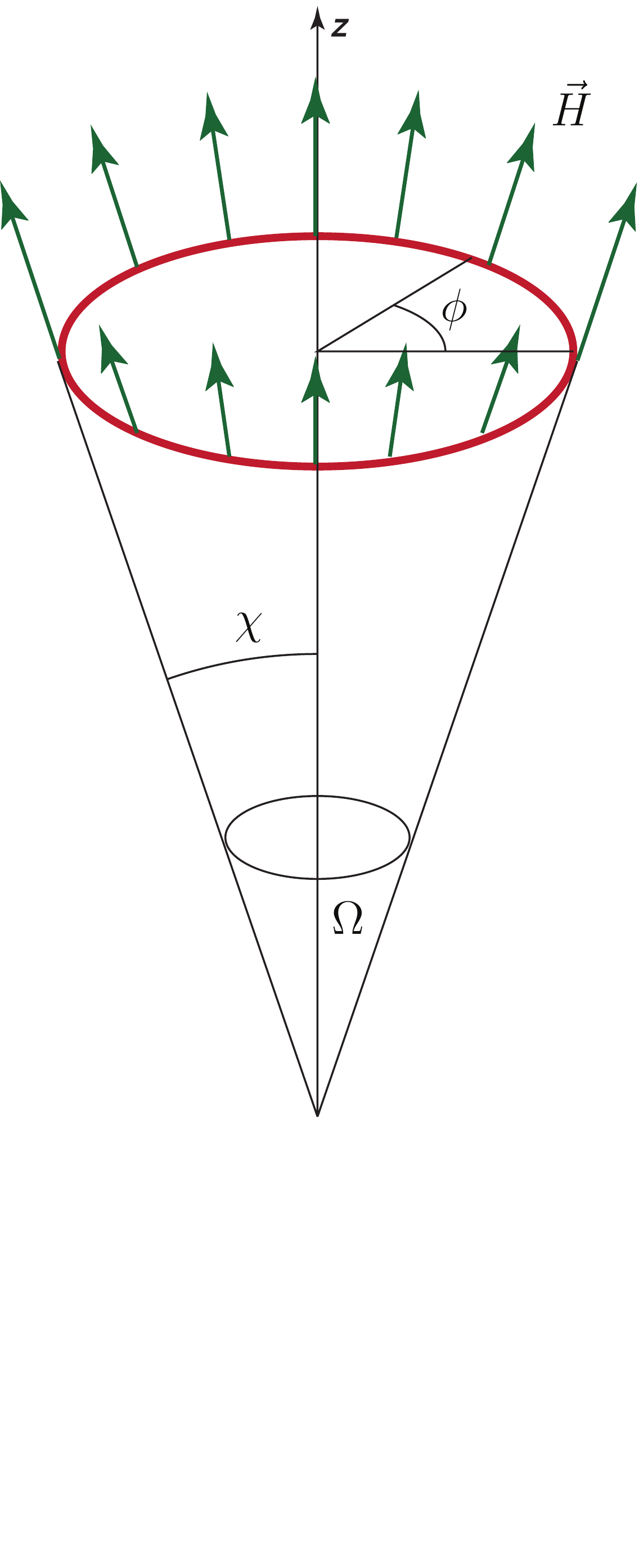}}}%
\caption{A 1D ring in a crown-shape  magnetic field. The solid angle $\Omega$ subtended by the spin parallel to $\bm H$ determines the Berry phase $\Gamma= -{1\over 2} \Omega$.}%
\label{fig4a}
\end{minipage}
\end{center}
\end{figure}

\subsection{A 1D ring in an inhomogeneous magnetic field: Berry-phase currents} \label{1Dring}

Let us consider now a ring in an inhomogeneous crown-shaped magnetic field distribution. This problem was considered in a number of papers starting from Loss et al. \cite{Bal}.  There is a constant $z$-component $H\cos \chi $ and a rotating transverse component:  $H_x= H\sin \chi  \cos \phi$ and $H_y= H\sin \chi  \sin \phi$ (crown-shaped field distribution). Here $\chi$ is the tipping angle of the magnetic field on the ring with respect to the $z$ axis (figure~\ref{fig4a}). The Hamiltonian for a single electron with a position characterized by the azimuthal angle $\phi$  is
    \begin{eqnarray}
\hat{\cal H}={\hbar^2 \over 2m R^2}\left(-i {\partial \over \partial \phi} -  \rho\right)^2 -\mu_B H \{ \sin \chi [\cos \phi \hat \sigma_x+\sin \phi \hat \sigma_y]+\cos \chi \hat \sigma_z\}. 
     \end{eqnarray}
This Hamiltonian can be exactly diagonalized \cite{Stern,QYS}. The eigenstates are spinors 
\begin{eqnarray}
|\mathbf{\Psi}_+(l)\rangle=\left( \begin{array}{c} \cos {\gamma \over 2}e^{il\phi} \\ \sin {\gamma \over 2}e^{i(l+1)\phi}\end{array}\right),
\nonumber \\
|\mathbf{\Psi}_-(l)\rangle=\left( \begin{array}{c} \cos {\gamma+\pi \over 2}e^{il\phi}\\ \sin {\gamma+\pi \over 2}e^{i(l+1)\phi} \end{array}\right)=\left( \begin{array}{c} -\sin {\gamma \over 2}e^{il\phi}\\ \cos {\gamma \over 2}e^{i(l+1)\phi} \end{array}\right)
              \end{eqnarray}
with the eigenvalues of the energy
 \begin{eqnarray}
\epsilon_\pm=E_\pm  \mp  \sqrt{{\Delta E^2\over 4}+(\mu_BH\sin \chi)^2}\left( 1- \cos \gamma \right).
              \end{eqnarray}
Here $l$ is the quantum number for the orbital moment, which should include also the Aharonov-Bohm phase shift,  and $\gamma$ is the tipping angles for the average spin with respect to the $z$-axis determined by
 \begin{eqnarray}
\cos \gamma={\Delta E\over \sqrt{\Delta E^2+4(\mu_BB\sin \chi)^2}}.
              \end{eqnarray}
Two states correspond to two opposite directions of spin with angles $\gamma$ and $\gamma+\pi$. The energies 
\begin{eqnarray}
E_+={\hbar ^2l^2\over 2 mR^2} - \mu_BH\cos \chi,
\nonumber \\
E_-={\hbar ^2(l+1)^2\over 2 mR^2} +\mu_BH\cos \chi,
              \end{eqnarray}
are obtained taking into account only the Zeeman energy in the longitudinal magnetic field $H_z=H\cos \chi$, and
 \begin{eqnarray}
\Delta E= E_--E_+={\hbar ^2(2l+1)\over 2 mR^2}+2\mu_BH\cos\chi .
              \end{eqnarray}
The signs + and - label the states with spin up and down in the limit of the longitudinal magnetic field ($\chi \to 0$). In the eigenstates the average spin rotates in space around the $z$ axis with the same speed as the magnetic field. Because of inhomogeneity of the  magnetic field the states are not eigenstates of the operator of the orbital moment but are mixtures of the  states with two different moments. However, they are eigenstates of the total moment (orbital moment + spin) with the quantum number $l+{1\over 2}$ since cylindrical symmetry is not broken. 

The charge and spin currents are obtained by summation over all $l$ and two possible spin states:
 \begin{eqnarray}
j= {e\hbar \over 2\pi m R^2}\left\{\sum_l \left\{ l+{1\over 2} [1-\cos \gamma(l)]\right\} 
+ \sum_l \left\{ l+1- {1\over 2} [1-\cos \gamma(l)]\right\} \right\},
              \end{eqnarray}
 \begin{eqnarray}
j^z= {\hbar^2 \over 4\pi m R^2}\left\{\sum_l \left\{ l\cos \gamma(l) -{1\over 2} [1-\cos \gamma(l)]\right\} 
\right. \nonumber \\ \left.
- \sum_l \left\{( l+1)\cos \gamma(l)+ {1\over 2} [1-\cos \gamma(l)]\right\} \right\}.
              \end{eqnarray}
The second terms $\propto (1-\cos \gamma)$ in the sums originate from  the geometrical (Berry) phase \cite{Bal}. The simplest case is the limit of strong magnetic field, when the spins in two states are parallel or antiparallel to the magnetic field, and $\gamma =\chi$ does not depend on $l$. Then the Berry contribution to the spin current $j^z$ is proportional to the total electron density, while the Berry contribution to the charge current $j$ is proportional to the differences of the densities in the two spin states, i.e., to the spin polarization. 

Similar effects exist in the presence  of a spatially rotating electric field (external or crystal), which generates an effective magnetic field via spin-orbit interaction \cite{AL,QYS}. The equilibrium spin  currents can appear without electron mobility when no charge current is possible. An example of it is the persistent spin current in a spin-1/2 Heisenberg  ring \cite{Kop,SSBS}: Similarly to the case considered above,  the   crown-shape effective magnetic field makes the spin  to subtend a solid angle, which leads to the Berry phase contribution to the spin current.

All effects discussed in the present section are purely mesoscopic: The derived charge and spin currents decrease with increasing size of the ring. The case of macroscopic equilibrium spin currents, which are finite in the thermodynamic limit, is discussed in the next section.

\section{Equilibrium spin currents in the 2D electron gas with spin-orbit interaction} \label{RashEq}

\subsection{Currents in a uniform 2D gas  with spin-orbit interaction} \label{eqCur}

Spin-orbit interaction allows to govern the spin transport by electric field, which is expected to be useful for applications  \cite{sptr}. This explains a great  interest to systems with spin-orbit interaction. A classical example of such a system is a 2D electron gas with the  Rashba spin-orbit term (let us call it {\em Rashba medium}). Rashba \cite{R} revealed that in the Rashba medium spin currents appear even at equilibrium. However, he qualified them as ``not real'', which cannot lead to transport and accumulation of spin. Their presence in the ground state was considered as an inherent problem in the spin current concept \cite{R1,R1a}.  In the present section I shall address equilibrium spin currents in the 2D electron gas with spin-orbit coupling and analyze whether these currents have something to do with spin transport. 

The Hamiltonian for a 2D electron gas with spin-orbit interaction is 
\begin{eqnarray}
{\cal H} ={\hbar^2\over 2 m}\left\{\bm \nabla \mathbf{\Psi}^\dagger \bm \nabla \mathbf{\Psi}
 +i\alpha(\bm r) (\mathbf{\Psi}^\dagger  [\bm \sigma \times \hat z]_i \bm\nabla_i\mathbf{\Psi} -\bm \nabla_i \mathbf{\Psi}^\dagger[\bm \sigma \times \hat z]_i\mathbf{\Psi} )
 \right. \nonumber \\ \left.
 +i\beta (\bm r) [\mathbf{\Psi}^\dagger  ( \sigma_x \nabla_x\mathbf{\Psi}-\sigma_y \nabla_y\mathbf{\Psi})  -( \sigma_x \nabla_x\mathbf{\Psi} ^\dagger-\sigma_y \nabla_y\mathbf{\Psi} ^\dagger)\mathbf{\Psi} )
 \right\},
       \label{Ham}          \end{eqnarray}
where $\alpha(\bm r)$ and $\beta(\bm r)$ are parameters of the Rashba and Dresselhaus spin-orbit interaction respectively. In general they may depend on the 2D position vector $\bm r$. For simplicity, we shall concentrate the further analysis on the  Rashba interaction ($\beta=0$) postponing its extension on Dresselhaus spin-orbit interaction  for the end of this subsection. The Rashba term $\propto \alpha$ is a particular case of the spin-orbit-interaction term in   the Hamiltonian equation~(\ref{hamG}), assuming that $\lambda \bm E=\alpha \hat z$.  The Schr\"odinger equations for components of the spinor $\mathbf{\Psi}$  are:
 \begin{eqnarray}
i\hbar \dot \psi_\uparrow 
 ={\hbar^2 \over m}\left(-{1 \over 2}\nabla^2 \psi_\uparrow+\alpha {\partial \psi_\downarrow\over \partial x}-i \alpha {\partial \psi_\downarrow\over \partial y} 
 +{1\over 2} {\partial \alpha\over \partial x}\psi_\downarrow-{i\over 2} {\partial \alpha\over \partial y}\psi_\downarrow\right)~,   \nonumber \\    
i\hbar\dot \psi_\downarrow 
 ={\hbar^2 \over m}\left(-{1 \over 2}\nabla^2 \psi_\downarrow-\alpha {\partial \psi_\uparrow\over \partial x}-i\alpha {\partial \psi_\uparrow\over \partial y}
  -{1\over 2} {\partial \alpha\over \partial x}\psi_\uparrow-{i\over 2} {\partial \alpha\over \partial y}\psi_\uparrow  \right).
    \label{inhom}       \end{eqnarray}           

The spin current and the torque in the spin-continuity equation (\ref{SB}) are 
\begin{eqnarray}
j_i^\beta=-{i\hbar^2 \over 4 m}( \mathbf{\Psi}^\dagger \sigma_\beta \nabla_i \mathbf{\Psi} -\nabla_i\mathbf{\Psi}^\dagger \sigma_\beta  \mathbf{\Psi}) 
-{\alpha\hbar^2\over 4m} (\mathbf{\Psi}^\dagger \{\sigma_\beta [\bm \sigma \times \hat z]_i +[\bm \sigma \times \hat z]_i \sigma_\beta\} \mathbf{\Psi}) ,
    \label{SC}         \end{eqnarray}
     \begin{eqnarray}
G_\beta=-  {i\alpha\hbar^2\over 2m}\left\{  \left(\mathbf{\Psi}^\dagger\{ [\bm \sigma \times [\hat z \times \bm  \nabla]]_\beta \mathbf{\Psi}\} \right)
 -\left(\{ [[\bm\nabla \times \hat z]\times \bm \sigma]_\beta \mathbf{\Psi}^\dagger\}   \mathbf{\Psi} \right)\right\}.
         \end{eqnarray}            
The Greek super(sub)script $\beta$ refers to three components $x,y,z$ in the 3D spin space, whereas the Latin super(sub)script $i$ is related with the two coordinates $x,y$ in the 2D electron layer.

\begin{figure}
\begin{center}
\begin{minipage}{100mm}
{\resizebox*{7cm}{!}{\includegraphics{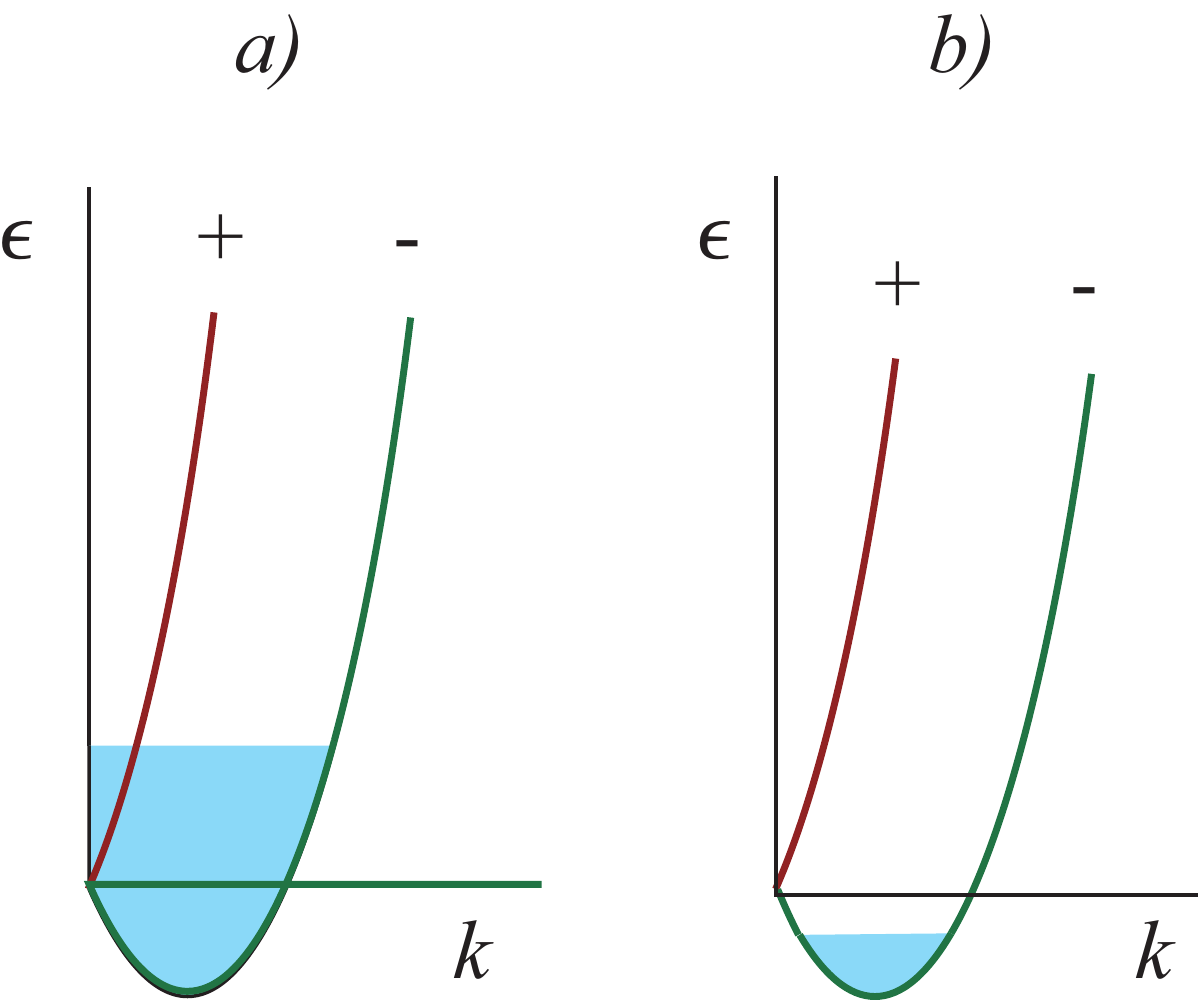}}}%
\caption{The ground state of the Rashba medium. a) The case $k_m >\alpha$, the Fermi sea (shaded blue) fills the upper (+) and the lower (-) band. b) The case $k_m<\alpha$, the Fermi sea fills only the lower band.}%
\label{fig5}
\end{minipage}
\end{center}
\end{figure}

    In the uniform Rashba medium eigenstates are plane waves given by spinors 
 \begin{eqnarray}
{1\over \sqrt{2}}\left( \begin{array}{c} 1 \\ \kappa \end{array}\right)e^{i\bm k  \bm r},
        \label{eqSp}      \end{eqnarray}
where $\kappa = \mp ie^{i\varphi}$,  $\varphi$ is the angle between the wave vector $\bm k$ and the axis $x$ ($k_x=k\cos \varphi$, $k_y=k\sin \varphi$), and the upper (lower) sign corresponds to the upper (lower) branch of the spectrum (band) with the energies (see figure \ref{fig5})      
 \begin{eqnarray}
\epsilon  ={\hbar^2\over m}\left({k^2\over 2}\pm \alpha k\right)={\hbar^2(k_0^2-\alpha ^2)\over 2 m}.
           \label{ener}  \end{eqnarray}   
The energy is parametrized by the wave number  $k_0$, which is connected with absolute values of wave vectors in two bands as $k=|k_0 \mp \alpha|$.  
 The eigenstates are spin-polarized in the plane with spins 
  \begin{eqnarray}
 \bm  s =\pm {\hbar\over 2}{[\bm k \times \hat z] \over k}
             \end{eqnarray}   
parallel or antiparallel to the effective spin-orbit magnetic field. There is no spin component normal to the plane ($z$ axis). The group velocities in two bands are given by
 \begin{eqnarray}
\bm v(\bm k)  ={\hbar \bm k\over m}+{2 \alpha \over m}  [\hat z \times   \bm  s  ] ={\hbar k_0\over m} {\bm k \over k}.
         \label{gv}    \end{eqnarray}  
Spin torque in the eigenstates is absent, but  there are inplane spin currents:
 \begin{eqnarray}
j^i_{j\pm}(\bm k) ={\hbar^2 \over 2m}\left(\pm {\varepsilon_{is} k_s \over k}k_j +\alpha \varepsilon_{ij}\right),
             \end{eqnarray}   
where $\varepsilon_{ij}$ is a 2D antisymmetric tensor with components $\varepsilon_{xy}=1$ and $\varepsilon_{yx}=-1$. The spin current does not vary in space since there is no precession of spin oriented along  the effective spin-orbit magnetic field. The latter is constant for a plane wave, and there is no torque on the spin violating its conservation.  

Though any eigenstate is spin-polarized,  after averaging over the equilibrium Fermi sea (we consider the $T=0$ case) the total spin vanishes. But  the total spin currents do remain. The Fermi energy is $\epsilon_F=\hbar^2(k_m^2-\alpha^2)/2$, where  $k_m$ is the maximum value of $k_0$. 
In the case $k_m>\alpha$, when the both electron bands are filled (figure \ref{fig5}a), the inplane spin currents are sums of contributions from two bands \cite{BR}: 
\begin{eqnarray} 
j^i_j={ \hbar^2 \over 4\pi m} \varepsilon_{ij}\left[\int_0^{k_m-\alpha} \left({k\over 2} +\alpha\right)k\,dk 
+ \int_0^{k_m+\alpha}\left(-{k\over 2} +\alpha\right)k\,dk\right]  
={\alpha^3\hbar^2\over 6\pi m}\varepsilon_{ij} .
   \label{maxCur}         \end{eqnarray}   
At $k_m<\alpha$ only the lower band is filled (figure \ref{fig5}b), and
\begin{eqnarray}
j^i_j= { \hbar^2 \over 4\pi m}\varepsilon_{ij} \int_{-k_m+\alpha}^{k_m+\alpha}\left(-{k\over 2} +\alpha\right)k\,dk
 ={ \hbar^2 \over 4\pi m}\varepsilon_{ij} \left(-{k_m^3\over 3} + k_m\alpha^2\right).
    \label{PWcur}         \end{eqnarray}   

Here we presented the calculation of equilibrium spin currents in the 2D gas with Rashba spin-orbit interaction at zero temperature.   Recently  Bencheikh and Vignale \cite{BenVig} performed a more general calculation taking into account temperature effects and Dresselhaus spin-orbit coupling. In the presence of  Dresselhaus spin-orbit coupling equation (\ref{maxCur}) is generalized to 
\begin{eqnarray} 
j^i_j={\hbar^2\over 6\pi m}[\alpha(\alpha^2-\beta^2)\varepsilon_{ij} + \beta(\alpha^2-\beta^2)(\sigma_z)_{ij}] .
            \end{eqnarray}   
The existence of equilibrium spin currents is related with broken space inversion symmetry and is a generic phenomenon in systems with spin-orbit interaction related to the non-Abelian gauge invariance \cite{Tok}.  In order to demonstrate that  an equilibrium spin current is able to transport spin, we should consider nonuniform media.

\subsection{Spin currents in a nonuniform Rashba medium}

Let us consider a slightly modulated Rashba medium with the Rashba parameter varying in space as \cite{BR}:
 \begin{eqnarray} 
\alpha(\bm r)=\alpha_0+\alpha_1 \cos(\bm p\cdot\bm r) . 
             \end{eqnarray}   
The eigenstates found above must be corrected using the perturbation theory with respect to $\alpha_1$: $\mathbf{\Psi}=\mathbf{\Psi}_0+\mathbf{\Psi}'$. Here $\mathbf{\Psi}_0$ is the spinor for a uniform medium with $\alpha=\alpha_0$ given by equation~(\ref{eqSp}). The equations for the first order correction $\mathbf{\Psi}'$ are (in components):
 \begin{eqnarray}
{m\over \hbar^2}\Delta\epsilon\psi_\uparrow'- \alpha_0 [k \kappa^*(\bm k) +p \kappa^*(\bm p) ]\psi_\downarrow'
={\alpha_1 \kappa(\bm k)\over \sqrt{2}}\left[k \kappa^*(\bm k) +{p \kappa^*(\bm p)\over 2}\right] e^{i\bm k\cdot\bm r} \cos(\bm p\cdot\bm r),   
\nonumber \\    
-\alpha_0 [k \kappa(\bm k) +p \kappa(\bm p) ] \psi_\uparrow'+ {m\over \hbar^2}\Delta\epsilon\psi_\downarrow' 
={\alpha_1\over \sqrt{2}}\left[k \kappa^*(\bm k) +{p \kappa^*(\bm p)\over 2}\right] e^{i\bm k\cdot\bm r} \cos(\bm p\cdot\bm r),
\nonumber \\
            \end{eqnarray}     
where  
\begin{eqnarray}
\Delta\epsilon=-{\hbar^2\over m}\left({p^2 \over 2}+\bm p\cdot \bm k \mp \alpha_0 k\right) .
            \end{eqnarray}  
The solution of linear equations for  $\mathbf{\Psi}'$ should be used for derivation of all relevant physical quantities (densities, torques, and currents). The general expressions are rather cumbersome. Moreover, the perturbation theory fails in the limit $p\to 0$. Therefore, we restrict ourselves with the limit $p \gg k, \alpha_0$.  The torques and currents for inplane spin components  are (only  linear in $\alpha_1$ terms are kept) 
\begin{eqnarray}
G_{i\pm}(\bm k) =\pm {2  \alpha_1\alpha_0\hbar^2\over m} \varepsilon_{ij}{p_jk\over p^2}\left[1-{(\bm p\cdot\bm k)^2\over p^2 k^2 }\right]\sin(\bm p\cdot\bm r),
            \end{eqnarray}     
\begin{eqnarray}
j_{j\pm}^{i}(\bm k)= { \alpha_1\hbar^2\over 2 m}\left\{- p_j {  \varepsilon_{is}p_s\over p^2}+\varepsilon_{ij} 
\mp{4\varepsilon_{ij} \alpha_0 k\over p^2} \left[1-{(\bm p\cdot\bm k)^2\over p^2k^2}\right]\right\} \cos (\bm p\cdot\bm r)\, .
            \end{eqnarray}    
The torque and the current for the $z$-component of spin are given by terms of higher order in $1/p$ and  vanish after integration over the Fermi sea. The integration of the torque and the current for inplane spin over the Fermi sea yields for the case $k_m >\alpha_0$:
\begin{eqnarray}
G_{i}={1\over 4\pi^2} \int G_{i+}(\bm k)\,d\bm k +{1\over 4\pi^2} \int G_{i-}(\bm k)\,d\bm k
\nonumber \\
 =- {  \alpha_1\alpha_0\hbar^2\over \pi m} {\varepsilon_{ij}p_j\over p^2}\left(k_m^2\alpha_0+{\alpha_0^3\over 3}  \right)\sin(\bm p\cdot\bm r),
            \end{eqnarray}     
\begin{eqnarray}
j_{j}^{i}={1\over 4\pi^2}\int j_{j+}^{i}(\bm k)\,d\bm k +{1\over 4\pi^2} \int j_{j-}^{i}(\bm k)\,d\bm k
\nonumber \\
= { \alpha_1\hbar^2\over 8\pi^2 m}\left[ \left(\varepsilon_{ij}-p_j {  \varepsilon_{is}p_s\over p^2}\right)n 
+{8\pi\varepsilon_{ij} \alpha_0 \over p^2} \left(k_m^2\alpha_0+{\alpha_0^3\over 3}  \right)\right] \cos (\bm p\cdot\bm r).
            \end{eqnarray}    
If $k_m<\alpha_0$:
\begin{eqnarray}
G_{i}={1\over 4\pi^2} \int G_{i-}(\bm k)\,d\bm k
 =- {  \alpha_1\alpha_0\hbar^2\over \pi m} {\varepsilon_{ij}p_j\over p^2}\left(k_m\alpha_0^2+{k_m^3\over 3}  \right)\sin(\bm p\cdot\bm r),
            \end{eqnarray}     
\begin{eqnarray}
j_{j}^{i}= {1\over 4\pi^2}\int j_{j-}^{i}(\bm k)\,d\bm k
\nonumber \\ 
= { \alpha_1\hbar^2\over 8\pi  m}\left[  \left(\varepsilon_{ij}-p_j {  \varepsilon_{is}p_s\over p^2}\right)n 
+{8\pi\varepsilon_{ij} \alpha_0 \over p^2}\left(k_m\alpha_0^2+{k_m^3\over 3}  \right)\right] \cos (\bm p\cdot\bm r).
            \end{eqnarray}  
The first term in the spin current, which is proportional to electron density $n$, is divergence-free, whereas the  divergence of the second term does not vanish and compensates the spin torque in  the spin balance. Thus the second term is responsible for spin transport from areas, where spin is produced  ($G_i >0$) to areas where spin is absorbed ($G_i<0$). One may consider this as a manifestation of spin transport even though it does not result in spin accumulation. Thus equilibrium spin currents can transport spin, and an attempt to distinguish equilibrium (persistent) currents from transport spin currents  \cite{Sun1} hardly would be reasonable.

It is worthwhile to note that the spin current in a modulated Rashba medium is linear in the spin-orbit coupling constant $\alpha_0$,  whereas the dependence of the current on $\alpha $ in the uniform Rashba medium is cubic.  Apparently the linear dependence is a general property of nonuniform media. In particular, the same dependence was predicted by Sablikov et al. \cite{Sab}, who considered equilibrium spin currents along the interface between the media with and without spin-orbit coupling.

\subsection{Interference and torque at  edges of the Rashba medium} \label{edge}


\begin{figure}
\begin{center}
   \leavevmode
  \includegraphics[width=0.7\linewidth]{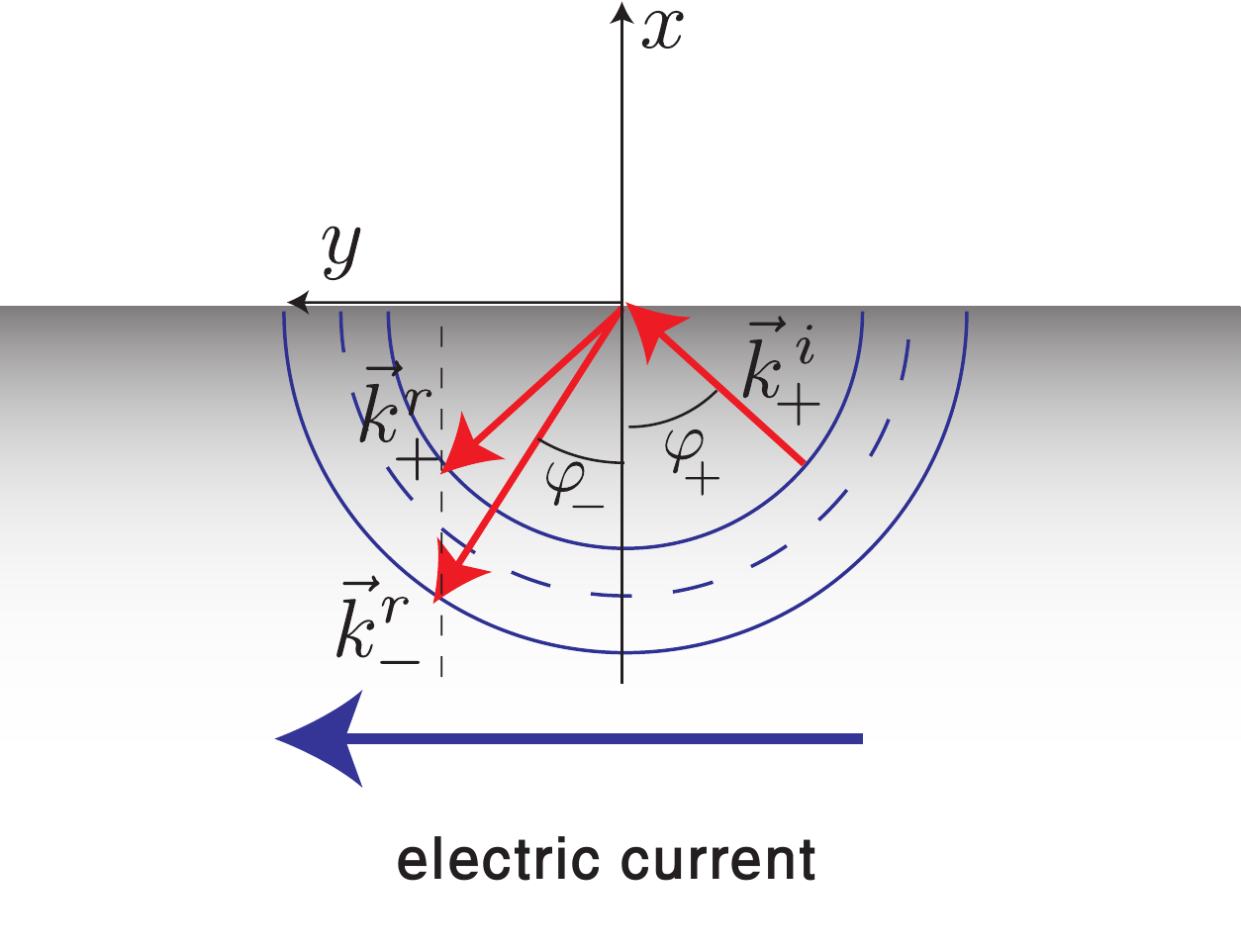}
 \caption{Spin-dependent reflection of electrons from an ideal impenetrable wall. The electron from the upper band ($\bm k_+^i$) is reflected either as  an electron from the same band ($\bm k_+^r$), or as  an electron from the lower band ($\bm k_-^r$).}
 \label{fig1r}
 \end{center}
\end{figure}

One cannot understand the physical meaning of  spin currents without a clear picture of what is going on at borders of a system with a bulk spin current.
Suppose that the Rashba medium occupies the semispace $x<0$ while at $x>0$ spin-orbit interaction is absent. The two semispaces have also different potentials, so the Hamiltonian is
\begin{eqnarray}
{\cal H} ={\hbar^2\over 2 m}\left\{\bm \nabla \mathbf{\Psi}^\dagger \bm \nabla \mathbf{\Psi}
 +i\alpha(\bm r) (\mathbf{\Psi}^\dagger  [\bm \sigma \times \hat z]_i \bm\nabla_i\mathbf{\Psi} -\bm \nabla_i \mathbf{\Psi}^\dagger[\bm \sigma \times \hat z]_i\mathbf{\Psi} )\right\}
 \nonumber \\
 +V(\bm r)\mathbf{\Psi}^\dagger  \mathbf{\Psi},
                   \end{eqnarray}
where
 \begin{eqnarray}
V(\bm r) =\left\{ \begin{array}{cc} 0 &  \mbox{at}~x<0 \\ U  & \mbox{at}~x>0 \end{array}\right. ~,~~
\alpha(\bm r) =\left\{ \begin{array}{cc} \alpha &  \mbox{at}~x<0 \\ 0  & \mbox{at}~x>0 \end{array}\right. ~.
              \end{eqnarray}
The electron states near the interface between two regions with different spin-orbit  constants  have already been analyzed  earlier \cite{Marig,Fink}. The spin currents in a hybrid ring consisting of parts with and without Rashba spin-orbit coupling were also analyzed by Sun et al. \cite{Sun}. 

At $x<0$ one should look for a superposition of plane waves:  one incident wave, which is coming from $x =-\infty$, and two reflected waves  (figure \ref{fig1r}). 
 For high-energy  electrons with $k_0 >\alpha $  and the incident electron in the upper band, the superposition is
\begin{eqnarray}
{\mathbf \Psi}={e^{ik_y y}\over \sqrt{2}}\left[\left( \begin{array}{c} 1\\ \kappa_+ \end{array}\right)e^{ik_{+ x} x} 
+r_1\left( \begin{array}{c} 1\\   \kappa_+^* \end{array}\right)e^{-ik_{+ x} x}
+r_2\left( \begin{array}{c} 1\\  \kappa_-^* \end{array}\right) e^{-ik_{- x} x} 
\right],  
     \label{x<<}         \end{eqnarray} 
where   $ \kappa_\pm=\mp ie^{i\varphi_\pm}$, $\varphi_\pm= \arctan(k_y/k_{\pm x})$ are the azimuthal angles of the wave vectors in the plane $xy$,   and  $k_{\pm x} =\sqrt{(k_0 \mp \alpha)^2-k_y^2}$ are the $x$ components of the wave vectors corresponding to states of the same energy in the upper (+) and the lower (-) band.  At $x>0$ the wave function is evanescent: $\mathbf{\Psi}=\left( \begin{array}{c} t_\uparrow \\ t_\downarrow\end{array}\right)
e^{ik_y y}  e^{-k_b x}$,  where $k_b =\sqrt{2mU/\hbar^2 -k_0^2+\alpha^2}$. 

The wave superposition should satisfy the boundary conditions, which include continuity of the both components of the spinor and jumps of  first derivatives of these components \cite{Marig,Fink} related with derivatives of $\alpha$ in equation (\ref{inhom}): 
 \begin{eqnarray}
\left.{\partial \mathbf{\Psi} \over \partial x }\right |_{+0}-\left.{\partial \mathbf{\Psi}\over \partial x }\right |_{-0}=-  i\alpha  \sigma_ y \mathbf{\Psi} .
              \end{eqnarray}

The expressions for the reflection coefficients are rather cumbersome in general, and we restrict ourselves with the case  of an infinite potential step at the  $x=0$ ($U,k_b \to \infty$). Then the spinor wave function  at $x=0$ vanishes, and  expressions for the  reflection coefficients become simple:
\begin{eqnarray}
r_1={\kappa_+ -\kappa^*_- \over \kappa^*_--\bar \kappa_+ }={e^{i(\varphi_+ +\varphi_-)} -1 \over e^{i(\varphi_- -\varphi_+)}+1},~~r_2={  \kappa^*_+ -  \kappa_+\over  \kappa^*_-- \kappa^*_+}=-{2 e^{i\varphi_- }\cos \varphi_+\over  e^{i(\varphi_- -\varphi_+)}+1}.
             \end{eqnarray} 
The relation between the angles $\varphi_+$ and $\varphi_-$  is determined from the condition that scattering does not change the component $k_y=(k_0-\alpha) \sin \varphi_+ =(k_0+\alpha) \sin \varphi_- $.

For low-energy electrons $k_0<\alpha$ all three waves in the superposition belong to the two parts of the lower band, either to the right ($k>\alpha$) or to the left ($k<\alpha$) from the energy minimum (figure~\ref{fig5}). If the incident wave corresponds to the state with $k>\alpha$, the superposition is
\begin{eqnarray}
{\mathbf \Psi}=e^{ik_y y}\left[\left( \begin{array}{c} 1\\ \kappa_+\end{array}\right)e^{ik_+ x} +r_1\left( \begin{array}{c} 1\\  \kappa^*_+\end{array}\right)e^{-ik_+ x}
+r_2\left( \begin{array}{c} 1\\ \kappa_-\end{array}\right) e^{ik_- x} \right],  
     \label{x<}         \end{eqnarray} 
where   $ \kappa_\pm= ie^{i\varphi_\pm}$, $\varphi_\pm= \arctan(k_y/k_{\pm x})$, and   $k_{\pm x} =\sqrt{(\alpha \pm k_0)^2-k_y^2}$. The positive sign before $k_-$ in the exponent of the second reflected wave was chosen because the negative group velocity of this wave. Since the electron transport is determined by the group velocity,  the latter should be directed from the boundary into the bulk even though the wave vector is directed to the boundary. The reflection coefficient are
\begin{eqnarray}
r_1={\kappa_+ -\kappa_- \over \kappa_-- \kappa^*_+ }={e^{i(\varphi_+ -\varphi_-)} -1 \over e^{-i(\varphi_- +\varphi_+)}+1},~~~r_2={ \kappa^*_+ -  \kappa_+\over  \kappa_-- \kappa^*_+}=-{2 e^{-i\varphi_- }\cos \varphi_+\over  e^{-i(\varphi_- +\varphi_+)}+1}.
             \end{eqnarray}

Whereas in plane-wave eigenstates of the Rashba Hamiltonian the spin has no $z$ component, the interference between the waves in the superposition leads to partial spin polarization along the $z$ axis. For $k_0>\alpha$ the oscillating spin density $s_z=(\hbar/2) {\mathbf \Psi}^\dagger \hat \sigma_z {\mathbf \Psi}$ (Friedel-like oscillation)  is given by
 \begin{eqnarray}
s_{+z}(\bm k) ={\hbar \over 4}\left\{r_1(e^{-2i\varphi_+}
+1)e^{-2ik_1x} 
+r_2[e^{-i(\varphi_+ +\varphi_-)}
\right. \nonumber \\ \left.
+1]e^{-i(k_{+ x}+k_{- x})x}
+r_1^* r_2[e^{i(\varphi_+ -\varphi_-)}+1]e^{i(k_{+ x}-k_{- x}) x}\right\}
+\mbox{c.c.} 
\nonumber \\
={\hbar  (\sin\varphi_+ + \sin\varphi_-)\cos \varphi_+\over 1+\cos (\varphi_+-\varphi_-)}[\sin (2k_{+ x}x)
\nonumber \\
-\sin (k_{+ x}x+k_{- x}x)-\sin (k_{+ x}x-k_{- x}x)].~~
             \end{eqnarray}   
Exchanging $+$ and $-$ one obtains the spin density $s_{-z}(\bm k)$ for the incident electron from the lower band.
 Similar expressions can be derived for  the low-energy  case $k_0 <\alpha $, when all waves belong to the lower band.
 
The expressions given above are valid only if $k_y < k_+$, or $\sin \varphi_- <| k_0-\alpha|/(k_0+\alpha)$.  At $k_-> k_y > k_+$ the reflection of the incident electron from the lower band to the upper one is forbidden by the conservation law. But the contribution of the upper band into the wave superposition is still present in the form of the evanescent mode. The wave superposition in this case is 
\begin{eqnarray}
{\mathbf \Psi}={e^{ik_y y}\over \sqrt{2}}\left[\left( \begin{array}{c} 1\\ \kappa_- \end{array}\right)e^{ik_{- x} x} 
+r_1\left( \begin{array}{c} 1\\  \kappa^*_-  \end{array}\right)e^{-ik_{- x} x}
+g\left( \begin{array}{c} 1\\ s \end{array}\right) e^{p x} \right],  
      \end{eqnarray} 
where $\kappa_-=ie^{i\varphi_-} $ and
\begin{eqnarray}
p=\sqrt{(k_0+\alpha)^2\sin^2\varphi_- -(k_0-\alpha)^2},~~s={k_y-p\over k_0 -\alpha}
\nonumber \\
r_1=-{s-ie^{i\varphi_-}\over s+ie^{-i\varphi_-}},~~
g=-{2i\cos \varphi_-\over s+ie^{-i\varphi_-}}.
        \end{eqnarray} 
The $z$ spin density for this wave superposition contains not only the interference contributions but also the contribution from the evanescent component $\propto e^{px}$:
 \begin{eqnarray}
s_{-z}(\bm k) =  {p\cos^2 \varphi_-\over k_0\sin\varphi_-}  [e^{2px}+\cos (2k_{-x}x)-2e^{px}\cos(k_{-x}x)]
\nonumber \\
+ {\cos\varphi_-\over k_0\sin\varphi_-}[2k_0-(k_0+\alpha)\cos^2\varphi_-]
 [\sin(2k_{-x}x)
 -2e^{px}\sin(_{-x}kx)]. ~~
          \label{evans}    \end{eqnarray}   
This expression is valid independently of whether the electron energy is high ($k_0>\alpha$) or low ($k_0<\alpha$).

All contributions to the $z$ spin density are odd with respect to the sign of $k_y$ and vanish in the equilibrium state. But in the presence of the voltage bias along the $y$ axis the distribution function also has  an  odd component, and spin polarization becomes possible. This leads to the edge spin accumulation (polarization), which is important for investigation of the intrinsic spin Hall effect (section \ref{intrin}).

The wave interference near the edge leads not only to $z$ spin polarization but also to the spin torque. The existence of this torque is required by the spin balance (\ref{SB}):  If there is no current in the vacuum and there is a bulk current normal to the boundary,  the presence of an edge torque is inevitable and its total value (integral over the whole edge area) must be equal to the spin current from the bulk independently  of particular properties of the edge (an ideally reflecting wall in our case). Moreover, the integral edge torque should compensate the bulk spin current not only at the border with the vacuum but also at the interface between the Rashba medium and a medium, which does not allow dissipationless spin currents. Indeed, in the latter case spin diffusion accompanied by spin accumulation is the only mechanism of spin transport, but in equilibrium  no dissipative process is possible and the spin current must vanish at the interface.

But the type of the edge does influence the spatial distribution of the torque. We shall derive this distribution for a simpler case $k_m \ll \alpha$ (figure \ref{fig5}b) when all expressions can be expanded in $k_0$. Here $k_m$ is the maximum value of $k_0$ corresponding to the Fermi level. In this limit the main contribution to the torque originates from interference of the incident wave with the second reflected wave in the superposition  (\ref{x<}): 
  \begin{eqnarray}
G_{y+}(\bm k)=-{\alpha \hbar^2 k_y\over m} \mbox{Re} \left\{ e^{-ik_+ x +i k_- x}\left(1-\kappa^*_+  \kappa_- \right)r_2 
\right\}.
     \label{torq}      \end{eqnarray}   
A similar contribution $G_{y-}$ comes from the conjugate superposition,  in which the incident plane wave $\propto e^{-i k_- x}$ corresponds to the wave number $k<\alpha$ with the negative group velocity. The subsequent integration over the whole Fermi sea in the lower band yields 
\begin{eqnarray}
G_y(x)
= - { \alpha^2 \hbar^2 k_m^2\over \pi m}\left[\, _1F_2\left(-{1\over 2};1,{3\over 2};-k_m^2 x^2\right)
\right. \nonumber \\ \left.
-{1\over 2} {_1F}_2\left(-{1\over 2};{3\over 2},3;-k_m^2 x^2\right)+{2\over 3}k_mx \right] ,
            \end{eqnarray} 
where 
$_pF_q(a_1,...,a_p; b_1,...,b_q;z)$ is the generalized hypergeometric function \cite{TF}.  The total torque over the whole bulk   $\int_{-\infty}^0G_y(x)dx= \hbar^2 \alpha^2 k_m/4\pi m$   exactly compensates the bulk spin current [see equation (\ref{PWcur}) in the limit $k_m \ll \alpha$].

At large distances from the border the torques for single modes oscillate fast, so the asymptotic behavior of the torque can be analyzed using the steepest-descent method. This yields the asymptotic torque at $x\to -\infty$:
\begin{eqnarray}
G_y =-\sqrt{\pi\over k_m }{\alpha^2\hbar^2\over 4\pi^2 m |x|^{5/2}}  \sin\left(2k_m x-{\pi\over 4}\right)\, .
             \end{eqnarray}              
This Friedel-like oscillation may be suppressed by disorder or electron-electron interaction, which were neglected in our analysis.

In  summary, we have obtained the following picture of spin currents and torques in the restricted Rashba medium. There is no spin torque inside the medium far from medium edges, but there is a constant spin current there. On the other hand, interference of incident and reflected plane waves leads to  spin torques of opposite signs (source and drain of spin) near the two edges. The role of the bulk equilibrium spin current is to transport spin from the spin source near one edge to the spin drain near the opposite edge.

\begin{figure}
\begin{center}
\begin{minipage}{100mm}
{\resizebox*{7cm}{!}{\includegraphics{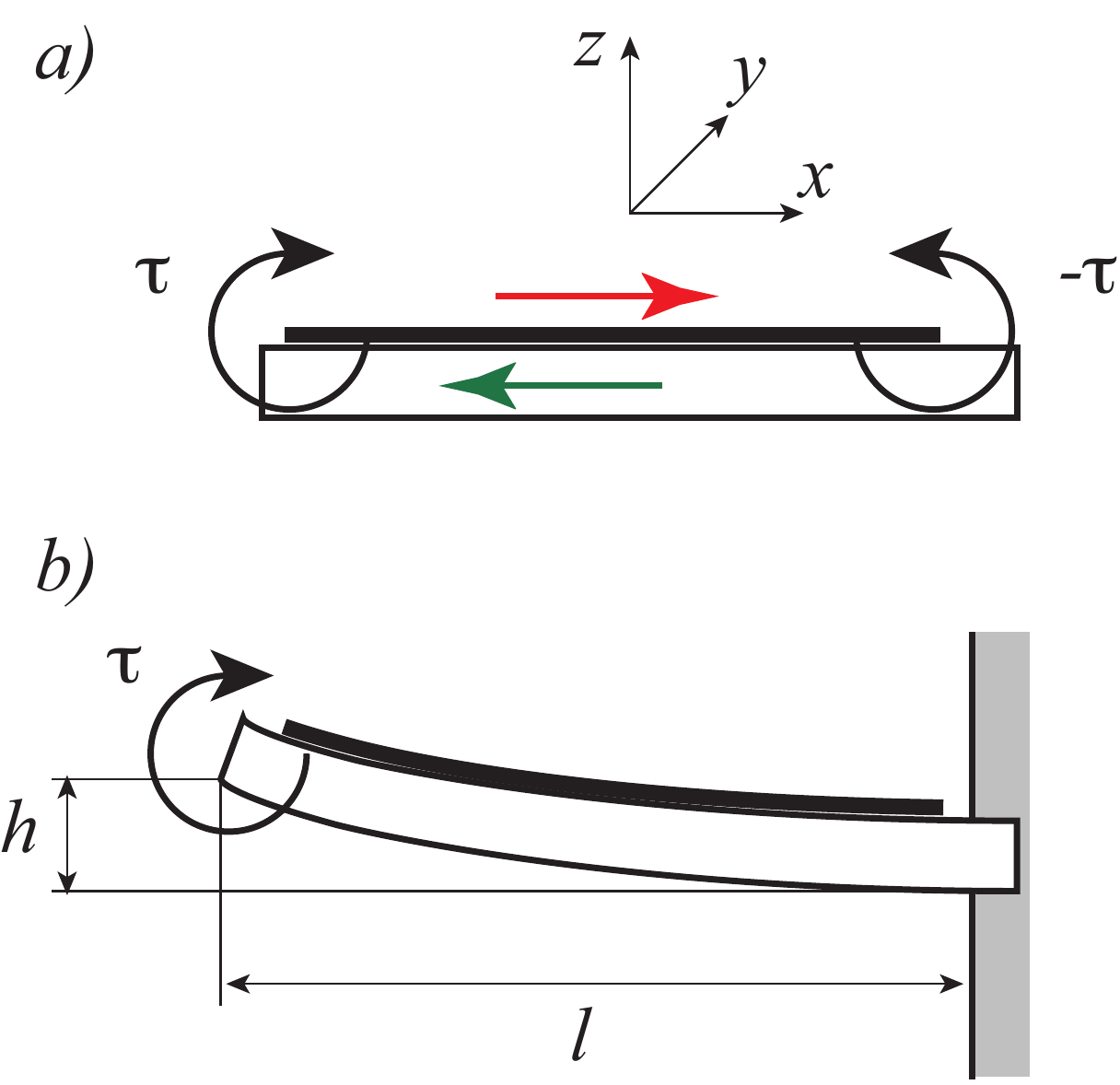}}}%
\caption{The cantilever with  the integrated Rashba medium (thick solid line) on it. (a) A rigid substrate. The red arrow (above the Rashba medium) shows the direction of the spin current. The green arrow (inside the substrate) show the direction of the orbital-moment current. The currents result in mechanical torques $\pm \tau$ at the edges of the substrate (b) The substrate is now a flexible cantilever. The edge torque at its free end leads to its displacement $h$. }%
\label{fig6}
\end{minipage}
\end{center}
\end{figure}

\subsection{Experimental detection of equilibrium spin currents} \label{mechM}

The central question for understanding the physical sense of the spin current is  {\em how is it  possible, if possible at all, to  detect the existence of spin currents experimentally}.  This question is especially acute for equilibrium spin currents since they do not lead to any spin accumulation. However, spin current leads to electric polarization, which might be detected via electric fields produced by it. Indeed, the spin currents is a flow of magnetic moments related with spin. According to classical electrodynamics  \cite{Griff} a magnetic moment $\bm m$ moving with velocity $\bm v$ creates an electric dipole ${\bm p}=[{\bm v} \times {\bm m}]/2 c$. The terms  $m_i v_j$ in this expression are proportional to the spin currents $ j_j^i$:  $m_i v_j=\gamma j_j^i$.  So the dipole moment is $p_i = \varepsilon_{ijk} \gamma j_k^j/2c=\varepsilon_{ijk} e j_k^j/2mc^2$. But this relation does not take into account crystal effect, which can strongly amplify spin-orbit interaction comparing with vacuum electrodynamics. A more reliable definition of spin-orbit dipole moment is using   
the thermodynamic relation ${\bm p} =\partial \langle \hat H \rangle /\partial {\bm E}$ applied to the  Hamiltonian (\ref{hamG}) \cite{R1}.  This yields $p_i = \varepsilon_{ijk}\lambda j_k^j$. For the Rashba medium with the spin-orbit constant $\alpha$ the spin current  $ j_j^i$ is determined by  equations (\ref{maxCur})  and (\ref{PWcur}), $\lambda =\partial \alpha/ \partial E_z$, and the electric dipole  is parallel to the $z$ axis.

The electric fields induced by stationary spin currents in conducting media were discussed by Hirsch \cite{Hirsch2} and Sun et al. \cite{Sun2}.   These fields were also discussed for magnetically order systems \cite{ML,Kop,SCD,SCDa}, where there are no itinerant carriers.  This phenomenon is an effect inverse to the spin Hall effect (generation of spin current by an electric field), which will be discussed in the next section \ref{shsec}.
The inverse spin Hall effect was observed experimentally by Valenzuela and Tinkham \cite{VT}.  Though the experiment was realized for spin-diffusion currents but not equilibrium spin current discussed here, there is no evident reason why the origin of the spin current would be essential for the existence of the effect. 
In any case, this suggests at least  a {\em Gedanken experiment} for detection of equilibrium spin currents: as well as the charge currents  in currents loops, which do not lead to any charge accumulation, are detected by magnetic-field  measurement,  one may detect equilibrium spin currents by electric-field  measurement even in the absence of any spin accumulation. 

Recently it was suggested \cite{Mech} to detect an equilibrium spin current in the Rashba medium by measuring a mechanical torque on a substrate at edges of the Rashba medium caused by the spin current.  If the substrate is flexible, the edge torques should deform it (figure~\ref{fig6}),  and measurement of  this deformation would provide a method to detect equilibrium spin currents experimentally.  An appropriate   experimental technique for such a measurement is already known: a  mechanical cantilever magnetometer with an integrated 2D electron system \cite{Gamb}. Earlier mechanical detectors  were suggested for detection of non-equilibrium diffusion spin currents \cite{Malsh}. A mechanical  stress produced by spin currents in mesoscopic structures with collinear magnetic order was also studied by Dugaev and Bruno \cite{MME}. They called it the {\em magneto mechanical effect}.

Derivation of the mechanical torque produced by bulk spin currents is based on the conservation law for the total angular momentum (the spin + the orbital moment) in the system ``2D electron gas + substrate''. The  continuity equation (\ref{SB})  for spin must be supplemented by the continuity equation for orbital moment:
\begin{eqnarray}
{\partial L_\beta \over \partial t}+\nabla_\gamma \tilde J_\gamma^\beta  =-G_\beta.
                \end{eqnarray} 
Here $L_\beta$ are $\beta$  components of  the orbital-moment densities and $\tilde J_\gamma^\beta$ is  the orbital-moment flux-tensor. The torque  $G_\beta$ in this equation  is the same  as in the continuity equation (\ref{SB})  for spin, but appears with an opposite sign. This provides the conservation of  the total angular momentum $\bm S+\bm L$.  We consider currents of $y$ components along the  axis $x$, so $\beta=y$ and $\gamma=x$.  At the equilibrium the time derivatives of  momenta are absent. In section~\ref{edge}  we saw that inside the Rashba medium there is a spin current but no torque, while at the edge there is an edge spin torque, which compensates the bulk current. According to the total angular momentum conservation law this leads to an edge orbital torque and to  a flux of the orbital  moment with a sign opposite to that of the spin current,  as shown in figure~\ref{fig6}. Since the 2D electron gas has no $y$ component of the orbital moment, the whole orbital torque must be applied to an edge of the  substrate. Now if the substrate is a cantilever rigidly fixed at one end (figure \ref{fig6}), the mechanical torque $\tau =\tilde J_x^y=-j_x^y = \int G_y(x) dx$ will deform the cantilever, and the displacement of its free end can be measured.

The fact that the spin current must be accompanied by an opposite orbital-moment current was already noticed by Sheng and Chang \cite{ShCh} and Zhang and Yang \cite{Zhang}.  Transformation of spin to angular momentum at an edge of the Rashba medium was recently discussed  by Teodorescu  and Winkler  \cite{TW}. Since the counterflow of the spin and the orbital moment does not lead to any flow of the {\em total} moment, Zhang and Yang \cite{Zhang} have concluded that the spin current is not observable and cannot induce electric fields discussed in the beginning of this subsection. However, this conclusion ignores the fact that the spin and the orbital moments have different gyromagnetic ratios. Therefore though the flow of the total {\em mechanical} moment really vanishes,  the flow of the total {\em magnetic} moment does not. In fact, the orbital moment current, which compensates the spin current, is not an obstacle but an instrument for spin-current detection as is shown here.

For numerical estimation of the effect we shall use the maximal value of the spin-orbit coupling $\alpha \hbar^2/m =6 \times 10^{-9}~\mbox{eV cm}=10^{-20}~\mbox{erg cm} =10^{-11}~\mbox{J m}$ quoted by Rashba \cite{R2} for InAs based quantum wells. When calculating the spin torque,  it was simpler to consider the case  $k_m \ll \alpha$. But the  mechanical torque reaches its maximum at $k_m > \alpha$  [see equation~(\ref{maxCur})] when the spin current is $j_x^y=-\alpha ^3 \hbar^2/6\pi m\sim 10^{-8}$  erg/cm=$10^{-3}$  J/m.  In order to estimate the displacement $h$ of the cantilever end  (see figure \ref{fig6}), we use the cantilever parameters from reference \cite{CL}:  the length $l=120~\mu$m and the spring constant $k=F/h=86~\mu$N/m, where $F$ is the force on the cantilever end.  Using the theory of elastic plates \cite{LLe}, one obtains that the torque $\tau=-j_x^y$ produces the displacement $h=3 \tau /2 k l$. This yields $h \sim 0.45~\mu$m. Less optimistic estimations of the spin-orbit coupling ($\alpha \hbar^2/m =3 \times 10^{-9}~\mbox{eV cm}$ for InAs, or $1.4  \times 10^{-9}~\mbox{eV cm}$ for GaSb \cite{semiRev}) predict an order or more smaller displacements,  but
certainly measurable with the modern micromechanical technique. The torque can be enhanced and tuned by an external magnetic field. 

In the presence of the external magnetic field  one should add the Zeeman energy $-\mu_B \bm \sigma \cdot \bm H$ to  the Hamilton equation~(\ref{Ham}). The spin current in a single-electron state is determined by the spin, which is parallel or antiparallel to  the ``effective'' magnetic field $\bm H-\alpha \Phi_0 [\bm k \times \hat z]/\pi$ acting on the electron.
Here $\Phi_0=hc/e$ is the single-electron flux quantum.
Integrating the single-state spin current over the whole $\bm k$ space and assuming that the Zeeman energy is much larger  than the spin-orbit energy, one obtains the bulk spin current 
\begin{eqnarray}
j^y_x  = \mp { \alpha\over 8\pi} { \epsilon_F ^2-\mu_B^2  H^2\over  \mu_B  H^3}(H_z^2+H_x^2),
       \label{h-z}       \end{eqnarray}   
where $\epsilon_F$ is the Fermi energy. In the interval $-\mu_B H<\epsilon_F<\mu_B H$  electrons fill only the lower band [the lower sign in equation (\ref{h-z})]. Then in terms of the electron density $n=m(\epsilon_F +\mu_BH)/2\pi \hbar^2$
\begin{eqnarray}
j^y_x  =-{\alpha \hbar ^2\Phi_0\over 2 m H}n\left( {H \over  \Phi_0}-n \right) {H_z^2+H_x^2\over H^3}.
       \end{eqnarray}   
If $\epsilon_F>\mu_B H$, electrons fill the both bands,  the contributions from two bands to the spin current cancel each other, and spin current vanishes  in our approximation.  The present analysis ignores the effect of the electromagnetic vector potential on the electron momentum, but this effect (which deserves a special analysis) is absent for the inplane magnetic field $H_x$.

It is worthwhile to comment that ambiguity of spin-current definition, which was intensively discussed  in the literature, has no impact on the effect considered here. As discussed in the end of section \ref{defSC}, one may redefine the spin current $j_j^i$ by adding to it an arbitrary term ($j_j^i \to j_j^i +\delta j_j^i $) but at the same time it is necessary to compensate it by redefinition of the spin torque ($G_i \to G_i +\nabla_j \delta j_j^i $). If the balance of the orbital part of the angular momentum is also considered, the definitions of the orbital torque and flux must be compatible with those of the spin part, in order not to violate the conservation law of the total angular momentum. Eventually  whatever definition was used any {\em correct} calculation must predict the same observable effect (displacement of the cantilever). 
After we defined the spin current by equation~(\ref{SC}),   we are not free anymore in the choice of the definition of the torque and the current  of the orbital angular momentum: The flux of the orbital angular momentum in the elastic cantilever should be defined as $\tilde J_j^i =\varepsilon_{imn} x_m T_j^n$ where $T_j^n$ is the elastic stress tensor. This choice looks most natural since it defines the mutual torque between the  spin and the orbital moment as a derivative of the spin-orbit energy with respect to the rotation angle.

Let us look at an alternative definition of spin current \cite{Niu}, which was discussed in the end of section~\ref{defSC}. The ``spin-conserving''  current $\bm {\cal T}^\beta(x)=\bm j^z(x)+\bm P^\beta(x)=\bm j^z(x)+\int_x^0 G_\beta(x')dx'$, which includes the dipole torque term $\bm P^\beta(x)$,  is constant, and it is difficult to notice in this picture that there is a process of angular-momentum transfer  between spin and orbital degrees of freedom. This is due to a nonlocal character of the current $\bm {\cal T}^\beta$: it controls only the global balance of spin. Globally no change of spin occurs in the sample, spin being generated at one edge and absorbed  at another. Meanwhile, exactly {\em local} torques, which do not violate the {\em global} spin balance, are responsible for the angular-momentum transfer to the orbital subsystem, which leads to the mechanical deformation discussed in this section.

\section{Spin Hall effect} \label{shsec}

\subsection{Phenomenology of spin Hall effect} \label{phen}

If an electrical current flows through a conductor with spin-orbit coupling,  this can give rise to a spin current flowing normally to the direction of the electrical current. This effect predicted by Dyakonov and Perel \cite{DP} was later called {\em spin Hall effect} \cite{Hirsch}. In contrast to the usual Hall effect, the spin Hall effect originates from the effective magnetic field produced by spin-orbit interaction and does not require an external magnetic field for its existence. Experimental evidences of the effect  in bulk semiconductors \cite{Kato}, in a  hole \cite{Wu} and an electron \cite{Sih} 2D gas  have already been reported. Observation of the inverse spin Hall effect  \cite{VT}  also  provides evidence for the spin Hall effect since the direct and the reverse effects are connected with the Onsager relations.

There are two possible origins for the spin Hall effect. The first one, which was discussed by Dyakonov and Perel  \cite{DP}, is due to spin-dependent scattering on impurities and is called {\em extrinsic} spin Hall effect. But broken space inversion symmetry also can allow  an {\em intrinsic} effect, which is not connected with impurities directly, though can be strongly affected by them (see below). 

In general spin currents in the spin Hall effect are not equilibrium currents considered in the previous section. They are accompanied by dissipation and require an energy input for their existence, despite that dissipation is related not with the spin current itself but with the longitudinal charge current. What does unite them with the spin currents discussed through the present review, is common controversies about their definitions and the impact of the absence of spin conservation. We consider the 2D electron gas assuming the presence of spin-orbit interaction and a weak electric field parallel to the axis $y$. Independently of the microscopic origin of the spin Hall effect, the broken space inversion symmetry allows the current of the $z$ spin component along the axis $x$ given by
\begin{equation}
j_x^z=\sigma_{SH} E,
  \label{SHE} \end{equation}
where $\sigma_{SH}$ is the Hall spin conductivity. In a uniform system with constant $\sigma_{SH}$ this current is constant by definition. On the other hand, at the sample border the spin current must vanish. Thus one should look for a spin current of another nature or an edge torque, which would compensate the spin current of electric origin. According to Dyakonov and Perel \cite{DP} (see also the recent phenomenological analysis by Dyakonov \cite{Dy})  another current is a dissipative spin diffusion current, i.e. the total spin current is $j_x^z+j_D^z$, where
\begin{equation}
j_D^z=-D_s \nabla_x S_z.
                 \end{equation}
This current is not constant because of inevitable longitudinal spin relaxation determined by the time $T_1$. Thus the continuity equation for the $z$ spin component is
\begin{equation}
{\partial S_z\over \partial t}=-\nabla_x     j_D^z-{S_z\over T_1}=D_s \nabla_x^2 S_z -{S_z\over T_1}. 
     \label{SEcont}            \end{equation}
For the stationary process with $\partial S_z/\partial t=0$ and under the condition that the total current vanishes at the sample border $x=0$ one obtains 
\begin{equation}
j_D^z=-j_x^z e^{-x/L_s},~~~S_z= -{j_x^z T_1\over L_s}e^{-x/L_s}, 
                 \end{equation}
where $L_s=\sqrt{D_s T_1}$ is the spin-diffusion length, which has already  been introduced in section~\ref{real}. Thus the spin Hall effect leads to accumulation of the $z$ spin component at the sample edge (called also spin orientation). The degree of spin accumulation is governed by dissipation parameters $T_1$ and $D_s$. A reader can find a detailed discussion of physical mechanisms responsible for dissipation in the review by \v Zuti\'c et al. \cite{sptr}.

\subsection{Extrinsic spin Hall effect}

The extrinsic spin Hall effect originates from the spin-orbit interaction related to the local electric fields induced by impurities or defects. A well accepted model for this interaction \cite{ERH}  is described by the Hamiltonian term
\begin{equation}
H_{extr}=\lambda \bm  \sigma \cdot  [\bm p \times \bm \nabla V_i(r)],
    \label{ex}  \end{equation}
where $\bm p$ is the electron momentum and $V_i(r)$ is the impurity axisymmetric potential, which depends on the distance $r$ from the impurity. The effect of the spin-orbit interaction on elastic scattering was known long time ago \cite{Mott}.
The interaction leads to asymmetry of scattering (skew scattering), and the differential cross section depends on spin of an incident electron:
\begin{equation}
\sigma_\pm(\theta)=\sigma_0(\theta)[1 \pm S(\theta)],
      \end{equation}
where $\sigma_0(\theta)$ is an even function of the scattering angle $\theta$ and determines scattering of unpolarized electron beams, while $S(\theta)$ is an odd function of $\theta$ called Sherman function. The Sherman function determines the skew scattering. The upper and the lower signs correspond to spins $+1/2$ and $-1/2$. 

Various types of impurities and of 2D quantum wells were discussed in the literature (see, e.g., \cite{HVL} and references therein). Here we present the simple but general discussion in terms of effective cross sections similar to that by Engel et. al. \cite{Eng}. 
A straightforward and physically transparent method to calculate a bulk spin current  is solution of the Boltzmann equation. Studying the spin Hall effect they usually used the quantum Boltzmann equation, in which the distribution function was a matrix 2 $\times$ 2 in spin indices  \cite{DK,Kha2,Shyt,Dy08}.  However, for the extrinsic effect under consideration the $z$ spin component is a good quantum number, which is affected neither by external electric field, nor by spin-orbit interaction determined by equation (\ref{ex}).  Indeed, the effective spin-orbit magnetic field is normal to the 2D layer plane and scattering cannot lead to any spin-flop. This means that off-diagonal elements of the spin-density matrix vanish. Two diagonal elements correspond to two distribution functions $f_\pm(\bm k)=f_0(k) +f'_\pm(\bm k)$  for electrons with positive (+) and negative (-) $z$ spins.  Here $f_0(k)$ is the Fermi equilibrium distribution function, which does not depend on spin and direction of the electron wave vector $\bm k$. The Boltzmann equation for the  non-equilibrium distribution function $f'_\pm(\bm k)$ generated by a weak inplane electric field $\bm E$ is 
\begin{eqnarray}
{e\over \hbar}\bm E {\partial f_0(k)\over \partial \bm k}=n_i v_F \int[ f_\pm (\phi) -f_\pm (\phi')]\sigma_\pm (\theta)\,d\theta , 
        \end{eqnarray} 
where $v_F$ is the Fermi velocity, $n_i$ is the impurity density, and  $\phi$ and $\phi'=\phi+\theta$ are angles between $\bm k$ and $\bm E$ before and after a collision. 
The solution of this equation, as one may check by substitution, is 

\begin{equation}
f'_\pm(\bm k)={e\hbar\tau\over m}{\partial f_0(\epsilon)\over \partial \epsilon}\left(\bm E \cdot \bm k\pm {\sigma_s\over \sigma_t} \bm E \cdot [\hat z\times \bm k]\right).
      \end{equation}
Here $\epsilon=\hbar^2k^2 /2m$ is the electron energy, and the relaxation time 
\begin{equation}
\tau ={\sigma_t \over n_i v_F (\sigma_t^2+\sigma_s^2)}
      \end{equation}
is determined by the transport cross section, 
\begin{equation}
\sigma_t =\int I(\theta) (1-\cos\theta)\,d\theta, 
      \end{equation}
and the effective cross section related to scattering asymmetry,
\begin{equation}
\sigma_s =\int  I(\theta)S(\theta) \sin\theta\,d\theta.
      \end{equation}
The parameter $\sigma_s/\sigma_t$ is called {\em transport skewness} \cite{Eng}.

Knowing the non-equilibrium distribution function one can calculate the charge current parallel to $\bm E$ and the $z$ spin current transverse to $\bm E$ by integration over the momentum space and summation over two spin values. Let us consider the zero temperature limit when $\partial f_0(\epsilon)\/ \partial \epsilon =\delta(\epsilon-\epsilon_F)$. The charge current is given by the usual Drude formula,
\begin{eqnarray}
j=\sigma E=   {e^2\over \pi} E{ \tau \varepsilon_F \over \hbar^2 },
        \end{eqnarray} 
while the spin current $j_z=\sigma_{SH} E$ is determined by spin Hall conductivity \cite{Eng} 
\begin{equation}
\sigma_{SH}= {e\over 2\pi}{ \tau \varepsilon_F \over \hbar }{\sigma_s\over \sigma_t}={\hbar \over 2e} {\sigma_s\over \sigma_t} \sigma .
      \end{equation}
Here $\sigma$ is the ohmic electron conductivity. 

It is believed that the extrinsic spin Hall effect was detected by Kato et al. \cite{Kato} who observed spin accumulation  on edges of $n$-GaAs layers. Engel et al. \cite{Eng} have found a rough quantitative agreement of the experiment with the theory presented in this subsection, though the signs of the effects were opposite. This disagreement remains unresolved (see discussion in  \cite{Eng}).

\subsection{Intrinsic spin Hall effect and edge spin accumulation} \label{intrin}

The intrinsic spin Hall effect does not rely on spin-dependent effects in scattering but is related entirely to the uniform bulk spin-orbit interaction, which leads to spin-orbit-split band structure. The effect was proposed by Murakami et al. \cite{Mu1} in bulk $p$-type semiconductors using the effective Luttinger Hamiltonian for holes and by  Sinova et al.  \cite{Sinova} in 2D electron systems with the Rashba spin-orbit interaction. 
In the literature there were debates on the strength of the intrinsic spin Hall effect in the 2D electron gas. Relating the  spin current to the acceleration of electrons by the electric field in a pure material, Sinova et al.   \cite{Sinova} concluded that the intrinsic  spin Hall effect  is characterized by the ``universal''  spin Hall conductivity $\sigma_{SH}=e/8\pi$. Originally it was believed that this result is insensitive to a small concentration of impurities, despite the fact that a steady state of a conductor in an electric field is impossible without collisions, which compensate the monotonous acceleration of electrons by the electric field. Later on it turned out that in the Rashba medium even rare collisions should not be ignored and eventually  the intrinsic spin-Hall conductivity at zero frequency vanishes in an infinite system  (see discussion and references in the review by Engel  et al. \cite{ERH}). But this conclusion is valid only for Rashba spin-orbit coupling linear in the electron wave vector $\bm k$, and the intrinsic spin Hall effect for Rashba coupling nonlinear in $k$, or in the Luttinger model for spin-orbit coupled systems \cite{Bern}  is not ruled out.  The intrinsic effect in the medium with nonlinear in $k$ Rashba coupling  follows from the theory using the quantum Boltzmann equation for the spin density matrix \cite{Shyt,Kha}.  In contrast to the extrinsic spin Hall effect, the intrinsic spin Hall effect is not possible without off-diagonal matrix elements since the $z$ spin current is proportional to them. This is because the bulk spin-orbit interaction keeps the spin inside the $xy$ plane, and only correlation between eigenstates of the Rashba Hamiltonian (like interference between plane-wave states near edges) leads to $z$ spin polarization and $z$ spin currents. 

However, the experimental detection of the intrinsic spin Hall is rather problematic. The experimental evidences of the spin Hall effect reported in the literature were based on optical measurement of the spin accumulated on the sample edges presumably resulting from the spin Hall effect.  Meanwhile, spin accumulation (polarization) near boundaries might not be so simply related to spin currents in the bulk as repeatedly pointed out \cite{Shyt,Gal,Dy08}.  As was shown above (section \ref{eqCur}), the bulk spin current is not necessarily accompanied by spin accumulation.  Moreover,  spin accumulation at sample edges is possible even without bulk spin current. It was demonstrated for the ballistic spin Hall effect \cite{Niko,Usaj,Zu}, when the electron mean-free path exceeds the sample sizes. Edge spin accumulation without bulk spin currents takes place also in the standard collisional regime when the electron mean-free path is much shorter than the sizes of the sample but still longer than the distance where interference of incident and reflected waves responsible for edge accumulation takes place \cite{edge}. Let us discuss edge accumulation without bulk spin currents in more details.

Since our goal is to analyze the case without bulk spin current, we may use the standard linear-in-momentum Rashba Hamiltonian. Moreover, we do not need to deal with the quantum Boltzmann equation and restrict ourselves with the classic Boltzmann equation for two scalar  distribution functions corresponding to two diagonal elements of the spin density matrix. Indeed, according to references \cite{Kha2,Shyt},  for  the linear-in-momentum Rashba interaction  bulk spin currents vanish together with off-diagonal matrix elements, to which they are proportional. 

 In section \ref{edge} it was shown that the eigenstates of the Rashba Hamiltonian are partially $z$ spin polarized  near the boundaries, though in the equilibrium  there is no total spin polarization after integration over the Fermi sea. But in the spin Hall effect there is a charge current    (along the $y$ axis at our choice of the coordinate frame, see figure \ref{fig1r}), and the non-equilibrium distribution function has a component odd with respect  to the  wave vector component $k_y$. First we shall consider the ballistic regime when the voltage drop $V$ occurs at the contacts, and there is no electric field inside the sample. In the narrow interval of energies $\epsilon_F+eV>\epsilon>\epsilon_F$ around the Fermi surface only left-moving electrons with $k_y>0$ are present. They are responsible for the edge accumulation of the $z$ spin.
Bearing in mind that $eV =d\epsilon = (\hbar^2/m)k_{\pm F}dk$, the two band contributions to the total 
spin density $s_z(x)=s_{+z}(x)+s_{-z}(x)$ are determined by integrals over the Fermi circumferences  of the two bands  with the Fermi wave vectors $k_{\pm F}=|k_m \mp \alpha|$:
 \begin{eqnarray}
 s_{\pm z}(x)={meVk_{\pm F}\over 4\pi^2 \hbar^2 k_m}\int_0^{\pi/2} s_{\pm z}(\bm k)d\varphi _\pm.
      \label{SD}       \end{eqnarray}   
Here $k_m$ is the value of $k_0$ at the Fermi circumferences.
The spin densities $s_{\pm z}(\bm k)$ for the eigenstates, which correspond to the wave vectors $\bm k$ of incident waves, were found in   section \ref{edge}.  
The asymptotic behavior of the spin density is determined by the evanescent-mode contribution [see equation (\ref{evans})]  and at $x \to -\infty$ is given by
\begin{eqnarray}
s_z(x)={meV\over 8\pi^2 \hbar } \sqrt{\alpha\over  k_m} { 1\over k_{+F}^2k_{-F}}{1\over |x|^3}.
             \end{eqnarray} 
The total accumulated spin   is given by \cite{Zu}
\begin{eqnarray}
S_z=\int_{-\infty}^0 s_z(x)dx={meV\over 8\pi^2 \hbar  \alpha }   \left(\ln{k_m+\alpha\over |k_m-\alpha|}   -{2\alpha\over  k_m} \right). 
           \label{rho}
                 \end{eqnarray} 
For further comparison with the collisional regime it is convenient to connect the total spin not with the voltage $V$  but with the electric current,  
\begin{eqnarray}
j= {e^2n V\over \pi \hbar  k_m }\times \left\{ \begin{array}{cc} {2k_m^2\over k_m^2 +\alpha^2}& \mbox{at}~k_m>\alpha\\ 1& \mbox{at}~k_m<\alpha\end{array}\right. ,
      \end{eqnarray} 
where the 2D electron density is $n=(k_m^2+\alpha^2)/2\pi $ at $k_m>\alpha$ and $n=\alpha k_m /\pi $ at $k_m<\alpha$. Then  
\begin{eqnarray}
S_z= {mj\over 8 \pi en \alpha } \left(\ln{k_m+\alpha\over |k_m-\alpha|}-{2\alpha\over  k_m}    \right)
\times \left\{ \begin{array}{cc} {k_m^2 +\alpha^2  \over  2k_m}& \mbox{at}~k_m>\alpha\\k_m& \mbox{at}~k_m<\alpha\end{array}\right.. 
          \label{balCur}
                 \end{eqnarray} 
In the limits of weak ($\alpha \to 0$) and strong ($\alpha \to \infty$) spin-orbit interaction this yields $S_z=(mj/24 \pi en)(\alpha^2/k_m^2)$ and $S_z=- mj/4 \pi en$ respectively. At $k_m=\alpha$ there is a logarithmic divergence, which can be cut either by the sample size or by nonlinear effects. 

Let us switch now to the collisional regime. As was explained above, since the bulk spin current is absent, one may use the standard Boltzmann equation for two scalar distribution functions $f_\pm(\bm k)= f_0(\epsilon)+f'_\pm(\bm k)$ for the two bands, where $ f_0(\epsilon)$ is the equilibrium Fermi distribution function, which depends only on energy $\epsilon$. The stationary solution of the Boltzmann equation for the non-equilibrium distribution function $f'$ in a weak electric field  along the $y$ axis  is  \cite{Abrikos}
\begin{eqnarray}
f'_\pm = {e\tau \bm E \over \hbar}{\partial f_0(\bm k)\over \partial \bm k} 
=eE\tau {\hbar k_m \over  m}\sin \varphi_\pm \delta (\epsilon-\epsilon_F).
             \end{eqnarray} 
Here the relaxation time $\tau$ for elastic scattering on defects is determined by the transport cross section and  in general is a function of $k$.  In principle $\tau$ should differ for two bands. But the difference vanishes for weak spin-orbit coupling and further will be neglected. The functions $f'_\pm$ determine the electric current equal to $j=e^2 E\tau k_m^2/2\pi m$ for $k_m>\alpha$ and to $j=e^2 E\tau \alpha k_m/2\pi m$  for $k_m<\alpha$. 
The $z$ spin densities for the two bands  instead of (\ref{SD}) are given by 
\begin{eqnarray}
 s_{\pm z}(x)={eE\tau k_{\pm F}\over 4\pi^2 \hbar }\int_{-\pi/2}^{\pi/2}\sin\varphi _\pm  s_{\pm z}(\bm k)d\varphi _\pm.
          \end{eqnarray}   
Tedious but straightforward integrations similar to those for the ballistic regime yield the total edge spin:
\begin{eqnarray}
S_z= -{mj \over 32\pi^2 e n }  {k_m^2 +\alpha^2\over k_m^4}
\left[3 (k_m^2-\alpha^2)\arctan {2\sqrt{\alpha k_m}\over k_m-\alpha}
 \right. \nonumber  \\ \left.
-{2 \sqrt{\alpha k_m}(3 k_m^2 +2k_m\alpha+3\alpha^2 )\over k_m+\alpha}+\pi (k_m^2+3\alpha^2)\right]~~
\label{colCur}
         \end{eqnarray} 
for the high-energy case $k_m>\alpha$, and 
\begin{eqnarray}
S_z = -{mj \over 16\pi^2 en  \alpha k_m}  
\left [3 (\alpha^2 -k_m^2)\arctan  {2\sqrt{\alpha k_m}\over \alpha-k_m}
   \right. \nonumber  \\ \left.
  -{\sqrt{\alpha k_m}(6\alpha^2 +6k_m^2 +4 \alpha k_m)\over \alpha +k_m}
 +4\pi \alpha k_m\right]~~
          \end{eqnarray} 
for the low-energy case $\alpha>k_m$.

\begin{figure}
\begin{center}
   \leavevmode
  \includegraphics[width=0.9\linewidth]{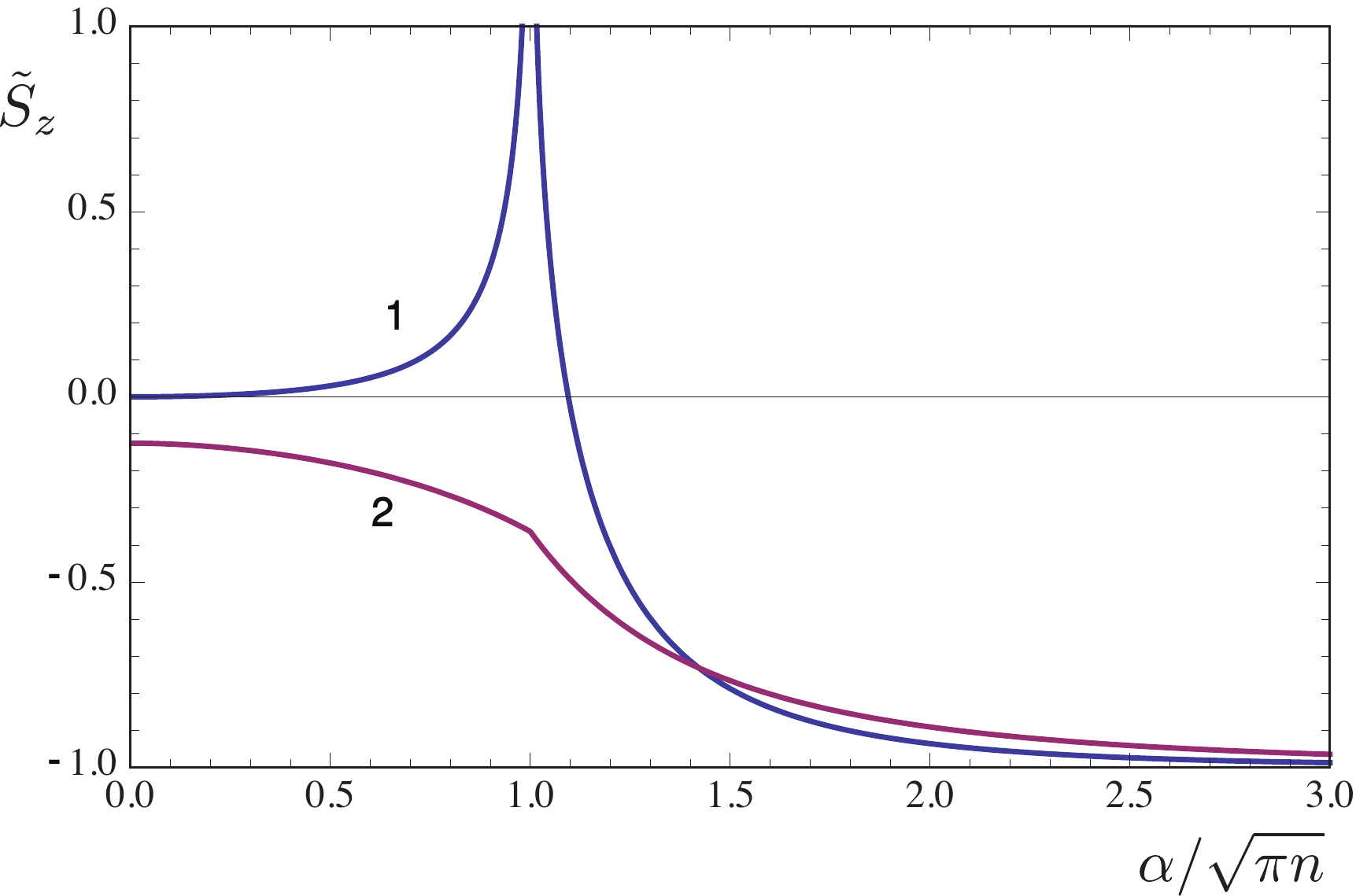} 
 \caption{ The plot of the reduced total spin $\tilde S_z=4\pi en S_z/mj$ as functions of $\alpha/\sqrt{\pi n}$. {\sl 1} -- the ballistic regime. {\sl 2} -- the collisional  regime.}
 \label{fig3r}
 \end{center}
\end{figure}

When $\alpha \to \infty$ the difference between the ballistic and collisional regime vanishes. On the other hand, in contrast to the ballistic regime, in the collisional regime the accumulated spin remains finite even in the limit of zero spin-orbit coupling $\alpha \to 0$ being equal to 
\begin{eqnarray}
S_z= -{mj \over 32\pi e n }. 
      \label{fina}         \end{eqnarray} 
This paradoxical result is explained by the divergence of the width $\sim 1/(k_{-x}-k_{+x})$ of the spin accumulation area in this limit. In the ballistic regime this divergence is canceled after summation over the two bands. However, our analysis is valid only if all relevant scales including $ 1/(k_{-x}-k_{+x})$ are less than the electron mean-free path. When this condition is violated the spin accumulation should go down. Figure \ref{fig3r} shows the reduced total accumulated spin $\tilde S_z=4\pi en S_z/mj$ for the ballistic (curve 1) and the collisional (curve 2) regimes as functions of the density-dependent parameter $\alpha/\sqrt{\pi n}$. Edge spin accumulation without bulk spin currents is possible for other types of spin-orbit interaction.  In particular, Bokes and Horv\'{a}th \cite{BH} considered spin-orbit interaction related to the edge potential $V(x)$ confining the electron gas: $H_{SO} =\alpha_E   \bm  \sigma \cdot  [\bm k \times \bm \nabla V(x)] $.  They obtained the edge accumulation linear in the spin-orbit constant $\alpha_E$, which is insensitive to the details of the edge potential. This allows to expect that the assumption of the hard-wall potential made in our calculation for the Rashba spin-orbit interaction also is not so crucial for the final outcome of the calculation.

Originally the spin Hall effect was defined as an effect related to bulk spin currents. The present discussion shows that
spin accumulation is not really a probe of the bulk spin current: the former can be absent in the presence of the bulk current and can appear in the absence of the latter. 
Therefore the question arises whether edge accumulation without bulk currents may be called  the spin Hall effect. A choice of terminology usually  is a matter of convention, taste, or tradition. Edge spin accumulation and spin currents require the same symmetry, and 
one may call  the edge spin accumulation without bulk currents the {\em edge} spin Hall effect. 

In order to compare the edge and the bulk spin Hall effects, we scale the latter using the ``universal'' spin conductivity $\sigma_{SH} = j^z/E=e/8\pi $, though in reality this is far from being universal \cite{ERH}. Here $j^z$ is the bulk  current  of the $z$ spin. Assuming that at the edge the bulk spin current is fully compensated with spin diffusion current  (section \ref{phen}), the total accumulated spin is $j_z T_1=eET_1 /8\pi$, where $T_1$ is the longitudinal spin relaxation time. So ratio of the edge to the bulk spin  Hall effect  is $\sim \tau /T_1$.  

For comparison with the spin Hall effect observed in the 2D hole gas \cite{Wu,Nomura} on may use  $\tau \sim 10\hbar /E_F=20/k_m v_F$, $n= 2\times 10^{12}$ cm$^{-1}$, and  the accumulation area width 10 nm given by Nomura {\em et al.} \cite{Nomura}. Then  the total spin accumulated  due to the edge spin Hall effect  at $\alpha \to 0$ is about 70 \% of the experimental value. So the interpretation of this experiment in the terms of the bulk spin currents probably must be reconsidered even if the spin-orbit interaction for these materials should be described by a model more general  than used for the present analysis.

There are other cases of electrically generated edge spin accumulation  without bulk spin currents inside the sample.    Adagideli and  Bauer \cite{edDif} discussed edge spin accumulation governed by spin diffusion, which occurred within the distance of the order of the spin-diffusion length $L_s$ from the  interface between  the media with and without spin-orbit coupling. In contrast, interference spin accumulation  discussed above, is not related to any dissipative process and occurs  at the distance on the order spin-orbit length, which was assumed to be much shorter than the mean-free path. It is interesting that the diffusion governed accumulation provides the total accumulated spin $\sim eE\tau$ \cite{R1a} of the same order of magnitude  as the interference mechanism. Since $L_s$ must essentially exceed the mean-free path, the  interference mechanism provides much higher spin density but in a much narrower layer. Another example of edge spin accumulation without bulk spin currents will be discussed in section \ref{QSH} addressing the quantum spin Hall effect .

\subsection{Spin currents in spin Hall insulators}
 
 As was already mentioned, though transverse spin currents induced by the spin Hall effect are dissipationless themselves, the whole process is accompanied by dissipation in the longitudinal channel and therefore requires an energy source. Murakami et al. \cite{Mu2} (see also reference \cite{Mu}) proposed the totally dissipationless spin-Hall effect in insulators with spin-orbit interaction. 
 In a band insulator at zero temperature and the Fermi level within the gap between conductance and the valence bands a weak electric field cannot generate an electric current, but may result in a transverse spin current, i.e., the spin conductivity $\sigma_{SH}$ in the spin-Hall relation  (\ref{SHE}) is finite being determined by topological invariants of the band structure. Materials with this property are called {\em spin Hall insulators}. The   dissipationless spin currents in  insulators, when the Fermi level is inside  the forbidden gap, were revealed by numerical calculations using realistic parameters of semiconductor band structures \cite{YF}.  The spin Hall effect is possible not only in band insulators. Meier and Loss \cite{ML} considered the spin Hall effect in a two-dimensional Heisenberg model consisting of localized spin, in contrast to itinerant electrons in band insulators. 

The spin Hall effect in insulators has a very important feature discerning it from the spin Hall effect in conductors. Since there is no charge current in the former case, there is no Joule heating and no energy input \cite{Mu}. As a result, no dissipation process is possible, and the problem reduces to  the equilibrium problem of an insulator in an electric field. The bulk spin current cannot be compensated by the dissipative mechanism of Dyakonov and Perel \cite{DP}, and one should look for a Hamiltonian mechanism of absorption (generation) of spin near sample edges. So there is a close analogy between spin currents in the spin Hall insulator and  equilibrium currents in the Rashba medium analyzed in section \ref{RashEq}. Indeed, the Rashba spin-orbit interaction is connected with an internal or external electric field normal to the electron-gas plane. The electric  field is not able produce a current since electrons are confined in a 2D layer. So the 2D gas is an insulator in the $z$ direction, but this does not rule out a Hall spin current along the layer.

The equilibrium character of spin currents in spin Hall insulator impose a serious restriction on methods of spin-current detection. In particular, spin injection into a non-magnetic material without spin-orbit interaction is impossible, since in this material spin can be transported only by a dissipative (diffusion) current, which must be supported by energy pumping.  On the other hand, it is possible to extract spin from  a spin Hall insulator putting it into a contact with a magnetically ordered medium supporting dissipationless spin transport.

 \subsection{Spin accumulation and the quantum spin Hall effect in topological insulators} \label{QSH}

\begin{figure}
\begin{center}
   \leavevmode
  \includegraphics[width=0.6\linewidth]{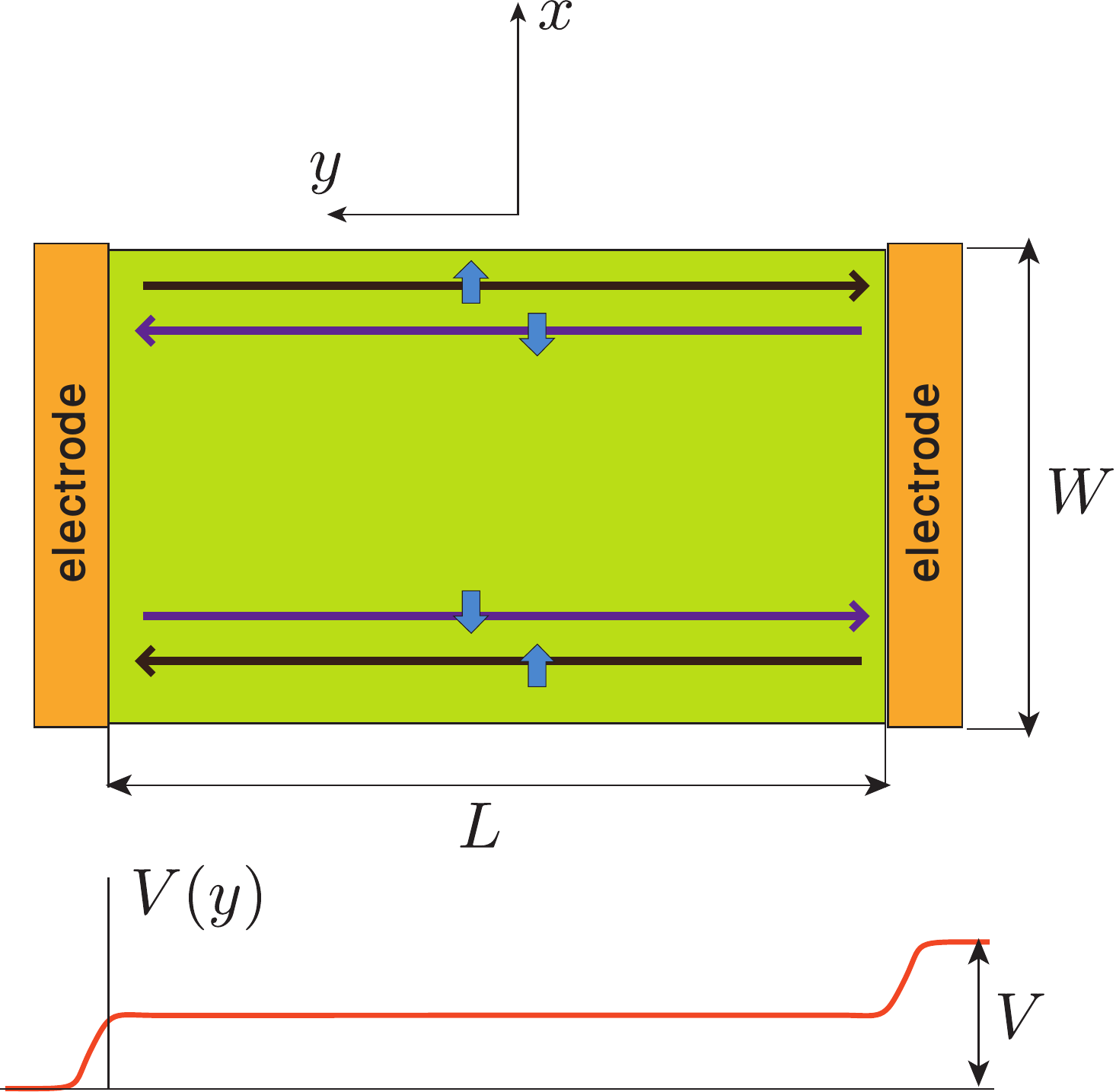} 
 \caption{ Edge states in a topological insulator. Wide blue arrows show spin direction (spin quantization axis is not necessarily in the plane as in the figure). At the upper edge spins up are rightmovers while spins down are leftmovers. At the lower edge directions of motion are opposite. The ballistic edge channels short-circuit the bulk insulator with infinite resistance. Therefore the whole sheet is globally a ballistic conductor without electric field inside. The external voltage bias $V$ drops inside electrodes (see voltage distribution in the lower part of the figure).}
 \label{figTI}
 \end{center}
\end{figure}

 Recently great attention was attracted to remarkable properties of {\em topological insulators}, in which the quantum spin Hall effect was predicted \cite{KM,BHZ}. The hallmark of a topological insulator is a forbidden gap originated from spin-orbit coupling and the presence of topologically stable helical edge states.  Topological criteria and classification for these materials have already been carefully analyzed \cite{Koenig,ZhangT,Hasan}.  The outcome of this analysis most important for the goals of the present review  is illustrated in figure \ref{figTI}. At two edges of the sample parallel to the electric field (we address the 2D case) each spin is able to move only in one direction, which is opposite for two spin directions. The edge states cross the whole forbidden gap  connecting the valence-band and the conduction-band bulk continua. The system is time-reversal invariant, and two states  with opposite directions of the spin and the wave vectors form a Kramers degenerate pair. Because of helicity (one-way motion) the edge states are robust against elastic backscattering and therefore may be treated as ballistic. This means that the ballistic edge channels short-circuit the bulk insulator with infinite resistance, and the whole sample is globally a ballistic conductor without electric field inside. 

Thus, whatever the bulk spin conductivity $\sigma_{SH} = j^z/E$ could be,  bulk spin currents are {\em absent} simply because the bulk electric field is absent. Nevertheless, the voltage drop between two leads (see the voltage distribution along the sample in figure  \ref{figTI}) leads to edge spin accumulation  similar to that considered in section \ref{intrin} for bulk conductors. However, mechanisms of the edge accumulation in two case are different. Whereas in bulk conductors the spin accumulation arose from interference of electron plane waves reflected from the boundary, in topological insulators spin polarization appears because the leftmoving and rightmoving edge states with oppositely directed spins have different densities proportional to the voltage bias. The latter directly follows from the Landauer-B\"uttiker approach. The leftmovers occupy states below the Fermi level in the right electrode, whereas the rightmovers occupy states below the Fermi level in the left electrode (figure \ref{figTI}). The difference between the two Fermi energies is $eV$, and the edge spin density (spin per unit length along the edge) is determined by the number of states in this energy interval:
 \begin{equation}
S_z={e s _e\over h v_{Fe}}V,
 \label{acc}   \end{equation}
where $  v_{Fe} =d \varepsilon (k_y)/\hbar\, d k_y$ is the Fermi velocity of the edge mode with the spectrum $\varepsilon (k_y)$ and $s_e$ is the effective ``spin'' of the edge states, which may be different from $\pm \hbar/2$ in general. The  parameters of edge modes were calculated numerically and analytically using simple but reliable models \cite{Koenig,ZhangT}.  The word ``spin'' is in quotation marks since it is actually a combination of the electron spin (which is $\pm \hbar/2$ of course) and the orbital moment of the electron in the band. Though it is dangerous to jump to general conclusions on which angular momentum eventually is relevant for various experiments, one may definitely expect that   electromagnetic experiments (Kerr  or Faraday effect, induced electric fields) address the magnetic moment in  the edge mode.

For estimation of the effective spin related to the magnetic moment one can consider the model usually used for the topological insulator \cite{Koenig,ZhangT}:  the edge state crosses the forbidden gap separating the conduction and the valence bands, which originate from $s$-type ($l=0$) and   $p$-type ($l=1$) atomic  orbitals. The $p$-type orbital corresponds  to quantum numbers  $j=3 /2$ for  the total moment and $m_j=\pm 1/2$ for its  projection on the quantization axis. The  Lande factor,
\[
g_L=1+\frac{j(j+1) -l(l+1)+s(s+1)}{2j(+1)}, \]
is equal to the electron $g$ factor $g_e=2$ for the $s$-type orbital and to the factor $g_v =4/3$ for the $p$-type orbital. The edge mode is a superposition of the two states with the weights $t_e$ and $t_v$ respectively ($t_e+t_v=1$). Then the effective spin in equation (\ref{acc}) is 
 \begin{equation}
s _e={ \hbar \over 2} \left( t_c +{g_v \over g_e}t_v\right)={ \hbar \over 2} \left( t_c +{2 t_v\over 3}\right).
    \end{equation}
Using this value of the effective spin the accumulated magnetization is obtained from expression (\ref{acc}) by  multiplying with the electron gyromagnetic relation.

It is interesting to compare the edge spin accumulation due to the quantum spin Hall effect with the accumulation due to the intrinsic spin Hall effect in conductors (section \ref{intrin}). Introducing the average electric field $E=V/L$, where $L$ is the length of the sample, equation (\ref{acc}) at $s_e \sim h$ yields $S_z \sim e E L/v_{Fe}$. Comparing this with $S_z \sim e E \tau$ for the accumulation in conductors one sees that this transforms to  the accumulated spin in topological insulators after replacing the relaxation time $\tau$ by the time of flight $ L/v_{Fe}$ through ballistic edge channels.
  
Soon after the theoretical prediction the topological insulators were experimentally detected in the HgTe quantum well \cite{KoenigE} by studying charge transport. It was demonstrated that at the quantum well thickness exceeding the critical value 6.3 nm there was an interval of gate voltages where the conductance reaches the value $2e^2 /h$ independently of the sample width $W$ (see figure  \ref{figTI}). This is a clear evidence of the ballistic transport through edge states while the main bulk is not conducting. The topological insulators states were also detected in BiSb \cite{Hsi},  BiSe \cite{Xia}, and  BiTe \cite{Chen} compounds by the methods of angle-resolved photoemission spectroscopy (APRES). In the literature the topological insulator state is called also the quantum spin Hall state. It is worthwhile of mentioning that despite impressive experimental demonstration of the quantum spin Hall  {\em state} the quantum spin Hall {\em effect} itself is still wants its experimental confirmation: A ``smoking gun'' of edge spin accumulation have not yet been reported.

\section{Conclusions}

The present review focused on four types of dissipationless spin transport: (1) Superfluid transport, when the spin-current state is a metastable state (a local but not the absolute minimum in the parameter space). (2) Ballistic spin transport, when spin is  transported without losses simply because sources of dissipation are very weak.  (3) Equilibrium spin currents, i.e., genuine persistent currents. (4) Spin currents in the spin Hall effect.   The dissipationless spin transport was a matter of debates during decades, though sometimes they were to some extent semantic. Therefore it was important to analyze what physical phenomenon was hidden under this or that name remembering that any choice of terminology  is inevitably subjective and is a  matter of taste and convention.  The various hurdles on the way of using the concept of spin current (absence of the spin-conservation law, ambiguity of spin current definition,  etc.) were analyzed. The final conclusion is that the spin-current concept can be developed in a fully consistent manner, though this is not an obligatory language of description: Spin currents are equivalent to deformations of the spin structure, and one may describe the spin transport also in terms of deformations and spin stiffness.

The recent revival of interest to spin transport is motivated by emerging of spintronics and high expectations of new applications based on spin manipulation. This is far beyond the scope of the present review, but hopefully the review could justify  using of  the spin-current language in numerous investigations of spin-dynamics problems, an important example of which is the spin Hall effect. 

\section*{Acknowledgments}

I thank D. Bercioux, S.O. Demokritov, V.K. Dugaev, A. J. Leggett, Z. Nussinov, E.V. Thuneberg, and G.E. Volovik for useful comments. I have been benefited from discussion of the spin Hall effect in common and topological insulators  with B.A. Bernevig.
The work was supported by the grant of the Israel Academy of Sciences and Humanities.



\end{document}